\newcommand{\CLASSINPUTtoptextmargin}{1.88cm}
\newcommand{\CLASSINPUTbottomtextmargin}{1.8cm}
\documentclass[journal,twoside]{IEEEtran}

\usepackage{amssymb}

\usepackage{cite}
\ifCLASSINFOpdf
  \usepackage[pdftex]{graphicx}
\else
\fi
\usepackage{epstopdf}

\usepackage{tikz}
\usetikzlibrary{trees}
\usepackage{varwidth}

\usepackage[cmex10]{amsmath}
\usepackage{multirow}
\usepackage{booktabs}
\usepackage{makecell}

\newcolumntype{L}[1]{>{\raggedright\let\newline\\\arraybackslash\hspace{0pt}}m{#1}}
\newcolumntype{C}[1]{>{\centering\let\newline\\\arraybackslash\hspace{0pt}}m{#1}}
\newcolumntype{R}[1]{>{\raggedleft\let\newline\\\arraybackslash\hspace{0pt}}m{#1}}

\usepackage{algorithmic}
\usepackage{algorithm}

\usepackage{setspace}
\let\Algorithm\algorithm
\renewcommand\algorithm[1][]{\Algorithm[#1]\setstretch{1.3}}

\usepackage{subcaption}
\newcommand{\MONTH}{%
  \ifcase\the\month
  \or January% 1
  \or February% 2
  \or March% 3
  \or April% 4
  \or May% 5
  \or June% 6
  \or July% 7
  \or August% 8
  \or September% 9
  \or October% 10
  \or November% 11
  \or December% 12
  \fi}
\makeatletter
\newcommand{\YEAR}{\the\year}
\makeatother

\usepackage{color}

\begin{document}

%\title{GPU-Accelerated Decoding of Quasi-Cyclic Multi-Edge LDPC Codes for Multi-Dimensional Reconciliation in Long-Distance CV-QKD}
%\title{Quasi-Cyclic Multi-Edge LDPC Codes for Long-Distance CV-QKD with Multi-Dimensional Reconciliation}
%\title{Quasi-Cyclic Multi-Edge LDPC Codes for Long-Distance Quantum Cryptography}
%\title{LDPC Codes for Long-Distance Quantum Cryptography with Multi-Dimensional Reconciliation}

%\title{Multi-Dimensional Key Reconciliation with LDPC Codes for Long-Distance Quantum Cryptography}
\title{Key Reconciliation with Low-Density Parity-Check Codes for Long-Distance Quantum Cryptography}


%\author{Mario~Milicevic, Chen~Feng, Lei~M.~Zhang, and P.~Glenn~Gulak
%\thanks{This footnote is intentionally left blank}% <-this % stops a space
%\thanks{This footnote is intentionally left blank}
%}

\author{
    \IEEEauthorblockN{Mario~Milicevic\IEEEauthorrefmark{1}, Chen~Feng\IEEEauthorrefmark{2}, Lei~M.~Zhang\IEEEauthorrefmark{1}, P.~Glenn~Gulak\IEEEauthorrefmark{1}}
    
    \IEEEauthorblockA{\IEEEauthorrefmark{1}Department of Electrical and Computer Engineering, University of Toronto, Toronto, ON, Canada}  
    
    \IEEEauthorblockA{\IEEEauthorrefmark{2}School of Engineering, University of British Columbia, Kelowna, BC, Canada}
    
	\IEEEauthorblockA{Email:\ \ mario.milicevic@utoronto.ca\ \ chen.feng@ubc.ca\ \ leizhang@comm.utoronto.ca\ \ gulak@eecg.toronto.edu}   
}

%\markboth{Journal Name,~Vol.~1, No.~1, October~2016}%
\markboth{\MONTH \ \YEAR} %Draft 3
{Milicevic \MakeLowercase{\textit{et al.}}: Key Reconciliation with Low-Density Parity-Check Codes for Long-Distance Quantum Cryptography}




\maketitle

\begin{abstract}
%\boldmath

The speed at which two remote parties can exchange secret keys over a fixed-length fiber-optic cable in continuous-variable quantum key distribution (CV-QKD) is currently limited by the computational complexity of post-processing algorithms for key reconciliation. Multi-edge low-density parity-check (LDPC) codes with low code rates and long block lengths were proposed for CV-QKD, in order to extend the maximum reconciliation distance between the two remote parties. Key reconciliation over multiple dimensions has been shown to further improve the error-correction performance of multi-edge LDPC codes in CV-QKD, thereby increasing both the secret key rate and distance. However, the computational complexity of LDPC decoding for long block lengths on the order of $\mathbf{10^6}$ bits remains a challenge. This work introduces a quasi-cyclic (QC) code construction for multi-edge LDPC codes that is highly suitable for hardware-accelerated decoding on a modern graphics processing unit (GPU). When combined with an 8-dimensional reconciliation scheme, the LDPC decoder achieves a raw decoding throughput of 1.72Mbit/s and an information throughput of 7.16Kbit/s using an NVIDIA GeForce GTX 1080 GPU at a maximum distance of 160km with a secret key rate of 4.10$\mathbf{\times 10^{-7}}$ bits/pulse for a rate 0.02 multi-edge code with block length of $\mathbf{10^6}$ bits when finite-size effects are considered. This work extends the previous maximum CV-QKD distance of 100km to 160km, while delivering between 1.07$\times$ and 8.03$\times$ higher decoded information throughput over the upper bound on the secret key rate for a lossy channel. For distances beyond 130km, the GPU decoder delivers an information throughput between $\mathbf{1868\times}$ and $\mathbf{18790\times}$ higher than the achievable secret key rates with a 1MHz light source. The GPU-based QC-LDPC decoder achieves a 1.29$\times$ improvement in throughput over the best existing GPU decoder implementation for a rate 1/10 multi-edge LDPC code with block length of 2$^\mathbf{20}$ bits. These results show that LDPC decoding is no longer the computational bottleneck in long-distance CV-QKD, and that the secret key rate remains limited only by the physical parameters of the quantum channel and the latency of privacy amplification.

\end{abstract}


% \fi
\IEEEpeerreviewmaketitle

% -----------------------------------------------------------------------------------
% INTRODUCTION
% Why is this a new and important problem?
% What has been done before? How does your research bring
% significant new understanding to the field? The reader should
% find enough information to understand why your research was
% necessary, without having to refer to other source material or
% published works. The introduction should be concise, no
% more than one or two pages. It is written in the present tense. 
% -----------------------------------------------------------------------------------

\section{Introduction}
\label{Introduction}

Quantum key distribution (QKD), also referred to as quantum cryptography, offers unconditional security between two remote parties that employ one-time pad encryption to encrypt and decrypt messages using a shared secret key, even in the presence of an eavesdropper with infinite computing power and mathematical genius~\cite{Bennett1984, Gisin2002, Alleaume2014, Diamanti2016}. Unlike classical cryptography, quantum cryptography allows the two remote parties, Alice and Bob, to detect the presence of an eavesdropper, Eve, while also providing future-proof security against brute force, key distillation attacks that may be enabled through quantum computing~\cite{Morris2014}. Today's public key exchange schemes such as Diffie-Hellman and encryption algorithms like RSA respectively rely on the computational hardness of solving the discrete log problem and factoring large primes\cite{Rivest1978,kollmitzer2010}. Both of these problems, however, can be solved in polynomial time by applying Shor's algorithm on a quantum computer~\cite{Shor1997, Adrian2015, Lo2014}.

\begin{figure*}[htbp]
\centering
\includegraphics[trim=0in 0in 0in 0in, width=0.98\textwidth]{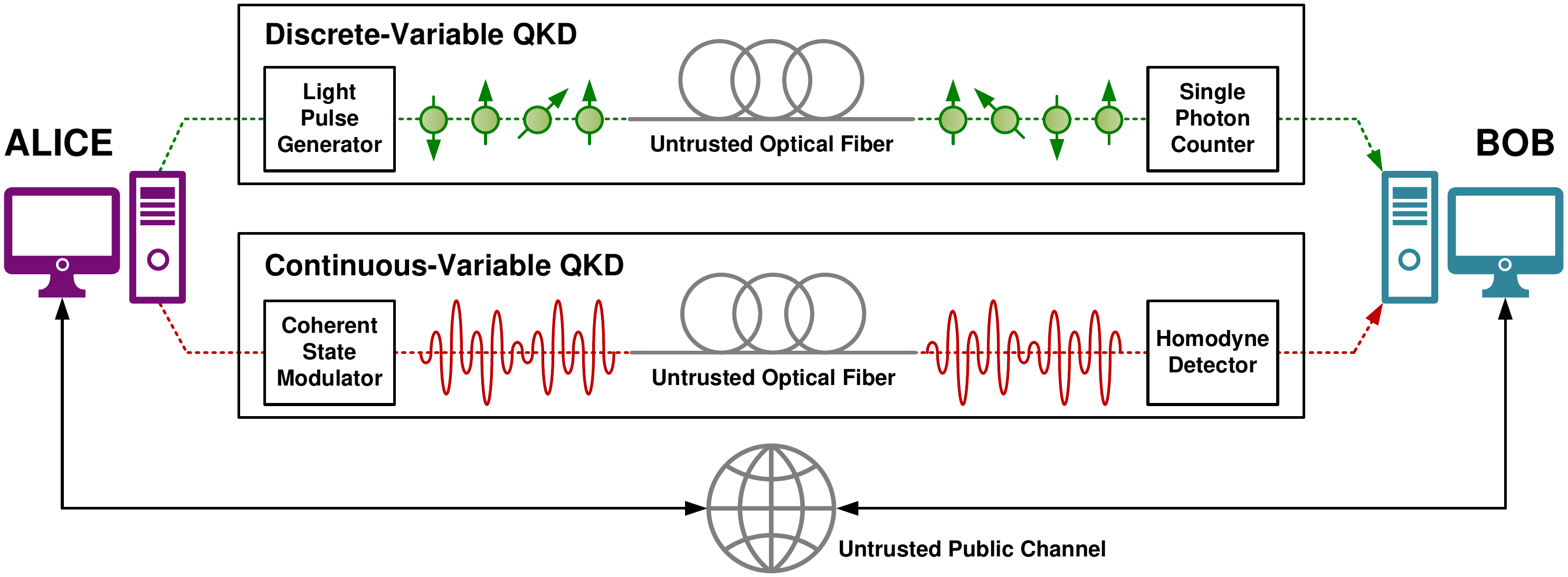}
\caption{Information transmission over untrusted quantum channel and authenticated public channel between Alice and Bob for CV- and DV-QKD.}
\label{fig:CVDVQKDAliceBob}
\end{figure*}

While quantum computing remains speculative, QKD systems have already been realized in several commercial and research settings worldwide~\cite{Peev2009,Sasaki2011,Jouguet2012OSA,Wang2014OSA}. Figure~\ref{fig:CVDVQKDAliceBob} presents two different protocols for generating a symmetric key over a quantum channel: (1) discrete-variable QKD (DV-QKD) where Alice encodes her information in the polarization of single-photon states that she sends to Bob, or (2) continuous-variable QKD (CV-QKD) where Alice encodes her information in the amplitude and phase quadratures of coherent states~\cite{Diamanti2016}. In DV-QKD, Bob uses a single-photon detector to measure each received quantum state, while in CV-QKD, Bob uses homodyne or heterodyne detection techniques to measure the quadratures of light~\cite{Diamanti2016}. While DV-QKD has been experimentally demonstrated up to a distance of 404km~\cite{Yin2016}, the cryogenic temperatures required for single-photon detection at such extreme distances present a challenge for widespread implementation~\cite{Diamanti2016}. CV-QKD systems on the other hand can be implemented using standard, cost-effective detectors that are routinely deployed in classical telecommunications equipment that operates at room temperature~\cite{Diamanti2016}. The majority of QKD research focuses on applications over optical fiber, since quantum signals for both CV- and DV-QKD can be multiplexed over classical telecommunications traffic in existing fiber-optical networks~\cite{Qi2010,Patel2012,Kumar2015}. Nevertheless, there has been recent progress in chip-based QKD, as well as free-space and Earth-to-satellite QKD~\cite{Vest2015,Vallone2015,Sibson2017}. It is noted here that quantum cryptography, i.e. QKD, differs from post-quantum cryptography, which is an evolving area of research that studies public-key encryption algorithms believed to be secure against an attack by a quantum computer~\cite{Chen2016NIST}. The discussion of post-quantum cryptography is beyond the scope of this work.

\begin{figure}[htbp]
\centering
\includegraphics[trim=0in 0in 0in 0in, width=0.485\textwidth]{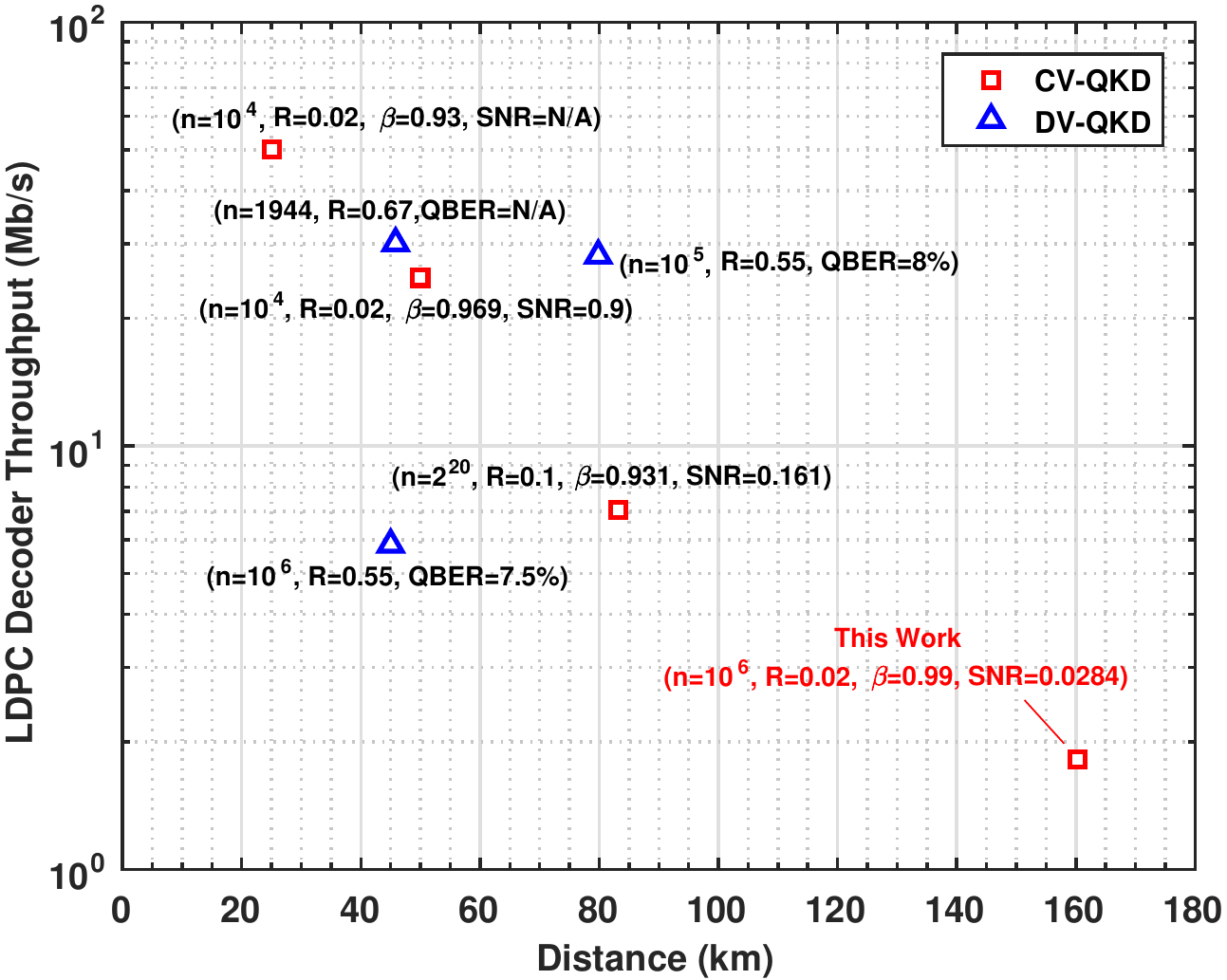}
\caption{Throughput vs. distance of GPU-based LDPC decoders for CV- and DV-QKD. The reported throughput is the raw GPU throughput without code- or error-rate scaling. For CV-QKD implementations~\cite{Huang2015,Wang2015,Jouguet2014}, the annotated values in parentheses indicate the LDPC code code block length $n$, the code rate $R$, the reconciliation efficiency $\beta$, and SNR of the quantum channel. For DV-QKD implementations, the annotated values indicate the block length $n$, code rate $R$, and QBER~\cite{Sasaki2011,Martinez-Mateo2013,Dixon2014}.}
\label{fig:CVDVQKD-Decoder-Comparison}
\end{figure}

The motivation of this work is to address the two key challenges that remain in the practical implementation of CV-QKD over optical fiber: (1) to extend the distance of secure communication beyond 100km with protection against collective Gaussian attacks, and (2) to increase the computational throughput of the key-reconciliation (error correction) algorithm in the post-processing step such that the maximum achievable secret key rate remains limited only by the fundamental physical parameters of the optical equipment at long distances~\cite{Jouguet2013,Jouguet2014,Huang2015}. Jouguet and Kunz-Jacques showed that Mbit/s error-correction decoding of multi-edge low-density parity-check (LDPC) codes is achievable for distances up to 80km~\cite{Jouguet2014}, while Huang~et~al. recently showed that the distance could be extended to 100km by controlling excess system noise~\cite{HuangHuangLinZeng2016}. This work explores high-speed LDPC decoding for CV-QKD beyond 100km.

%A particular challenge in designing LDPC codes for such long distances is the low signal-to-noise ratio (SNR) of the optical quantum channel, which typically operates below -15dB. At such low SNR, high-efficiency key reconciliation can be achieved only using low-rate codes with large block lengths on the order of 10$^{6}$ bits~\cite{Leverrier2009, Becir2012}, where approximately 98\% of the bits are redundant parity bits that must be discarded after error-correction decoding. Such codes require hundreds of LDPC decoding iterations to achieve the asymptotic, near-Shannon limit error-correction performance in order to maximize the secret key rate~\cite{Jouguet2011, Jouguet2014}. This is in contrast to LDPC codes employed in modern communication standards, such as IEEE 802.11ac (Wi-Fi) and ETSI DVB-S2X, where the target SNR is above 0dB and block lengths range from 648 bits to 64,800 bits~\cite{IEEE-80211ac, ETSI-DVBS2X}. In these standards, the LDPC decoder typically operates at 10 iterations to deliver Gbit/s decoding throughput~\cite{Blanksby2002, Zhang2009, Park2014}. In addition to long block-length codes, multi-dimensional reconciliation schemes have also been shown to improve error-correction performance of multi-edge codes at low SNR~\cite{Jouguet2013}. 

A particular challenge in designing LDPC codes for such long distances is the low signal-to-noise ratio (SNR) of the optical quantum channel, which typically operates below $-15\text{dB}$. At such low SNR, high-efficiency key reconciliation can be achieved only using low-rate codes with large block lengths on the order of 10$^{6}$ bits~\cite{Leverrier2009, Becir2012}, where approximately 98\% of the bits are redundant parity bits that must be discarded after error-correction decoding. Such codes require hundreds of LDPC decoding iterations to achieve the asymptotic, near-Shannon limit error-correction performance in order to maximize the secret key rate~\cite{Jouguet2011, Jouguet2014}. This is in contrast to LDPC codes employed in modern communication standards, such as IEEE 802.11ac (Wi-Fi) and ETSI DVB-S2X, where the target SNR is above 0dB and block lengths range from 648 bits to 64,800 bits~\cite{IEEE-80211ac, ETSI-DVBS2X}. In these standards, the LDPC decoder typically operates at 10 iterations to deliver Gbit/s decoding throughput~\cite{Blanksby2002, Zhang2009, Park2014}. Long block lengths allow Alice and Bob to generate longer secret keys, which can be used to provide unconditional security by employing the one-time pad encryption scheme. Shorter codes with block lengths of 10$^{5}$ bits, for instance, would not be suitable for low-SNR channels beyond 100km due to their less robust error-correction performance~\cite{Lodewyck2007, Jouguet2011}. In addition to long block-length codes, key reconciliation over multiple dimensions has also been shown to improve error-correction performance of multi-edge codes at low SNR~\cite{Jouguet2013}, thereby increasing both the secret key rate and distance. However, the computational complexity and latency of LDPC decoding for long block lengths on the order of $10^6$ bits remains a challenge. Figure~\ref{fig:CVDVQKD-Decoder-Comparison} presents a comparison of LDPC decoding throughput versus distance for several state-of-the-art CV- and DV-QKD implementations, illustrating that high-throughput reconciliation at long distances is achievable only with large block-length codes that approach the Shannon limit with more than 90\% efficiency for CV-QKD or less than 10\% quantum bit error rate (QBER) for DV-QKD. 

This work introduces a new, quasi-cyclic (QC) code construction for multi-edge LDPC codes with block lengths on the order of 10$^{6}$ bits~\cite{Richardson2002, Fossorier2004}. Computational acceleration is achieved through an optimized LDPC decoder design implemented on a state-of-the-art graphics processing unit (GPU). When combined with an 8-dimensional reconciliation scheme, the LDPC decoder achieves a raw decoding throughput of 1.72Mbit/s and an information throughput of 7.16Kbit/s using an NVIDIA GeForce GTX 1080 GPU at a maximum distance of 160km with a secret key rate of 4.10$\times 10^{-7}$ bits/pulse when finite-size effects are considered. The performance of this work in comparison to previous GPU-based decoders for QKD is plotted in Fig.~\ref{fig:CVDVQKD-Decoder-Comparison}, and discussed in greater detail in Section~\ref{sec:GPU}. This work extends the previous maximum CV-QKD distance of 100km to 160km, while delivering between 1.07$\times$ and 8.03$\times$ higher decoded information throughput over the upper bound on the secret key rate for a lossy channel~\cite{Pirandola2015}. These results show that LDPC decoding is no longer the computational bottleneck in long-distance CV-QKD, and that the secret key rate remains limited only by the physical parameters of the quantum channel and the latency of privacy amplification.

CV-QKD provides a new use case for hardware-based LDPC codes. Over the past 15 years, research in LDPC decoder design has primarily focused on application-specific integrated circuit (ASIC) implementations for wireless, wireline, optical, and non-volatile memory systems, due to the widespread adoption of LDPC codes in modern communication standards~\cite{Blanksby2002, Zhang2009, Park2014}. Although highly-customizable ASICs provide excellent energy efficiency, the silicon implementation of an LDPC decoder for long-distance CV-QKD with an LDPC code block length of 10$^{6}$ bits would require significant silicon die area, which may be prohibitively expensive to fabricate in a modern CMOS technology node~\cite{IRDS2016}. 

The high availability of on-chip memory and floating-point computational precision make GPUs a highly suitable platform for LDPC decoder implementation in long-distance CV-QKD systems~\cite{Falcao2011, Ji2009}, as opposed to ASICs or field-programmable gate arrays (FPGAs), which suffer from limited memory, fixed-point computational precision, high-complexity routing, and silicon area constraints~\cite{Mohsenin2010, Kim2011}. Since Alice and Bob are stationary and their communication occurs over a fixed-length fiber-optic cable, the traditional optimization parameters of energy efficiency and silicon chip area do not necessarily apply since the LDPC decoder does not need to assume an integrated circuit form factor, as in the case of a mobile hand-held device, where power consumption is a primary concern. Furthermore, GPUs seamlessly integrate into a post-processing computer system, and provide increasing computational performance at low cost with each successive architecture generation~\cite{Keckler2011}. The information throughput results presented in this work were measured using a single GPU, however, further computational speedup can be achieved with multiple GPUs.   

The remainder of this paper focuses on the design and implementation of high-efficiency LDPC codes for reverse reconciliation in CV-QKD systems that operate in the low-SNR regime at long distances. Section~\ref{sec:Background} presents the background on CV-QKD. Section~\ref{sec:LDPCforRecon} introduces the application of LDPC codes for information reconciliation in CV-QKD. Section~\ref{sec:GPU} presents the GPU-based LDPC decoder design and achievable secret key rate results with multi-dimensional reconciliation, as well as a comparison of this work to a recently published CV-QKD work that addresses the computational bottleneck of post-processing algorithms.

\section{Background}
\label{sec:Background}

\begin{figure}[t!]
\centering
\includegraphics[trim=0.
0in 0in 0in 0in, width=0.48\textwidth]{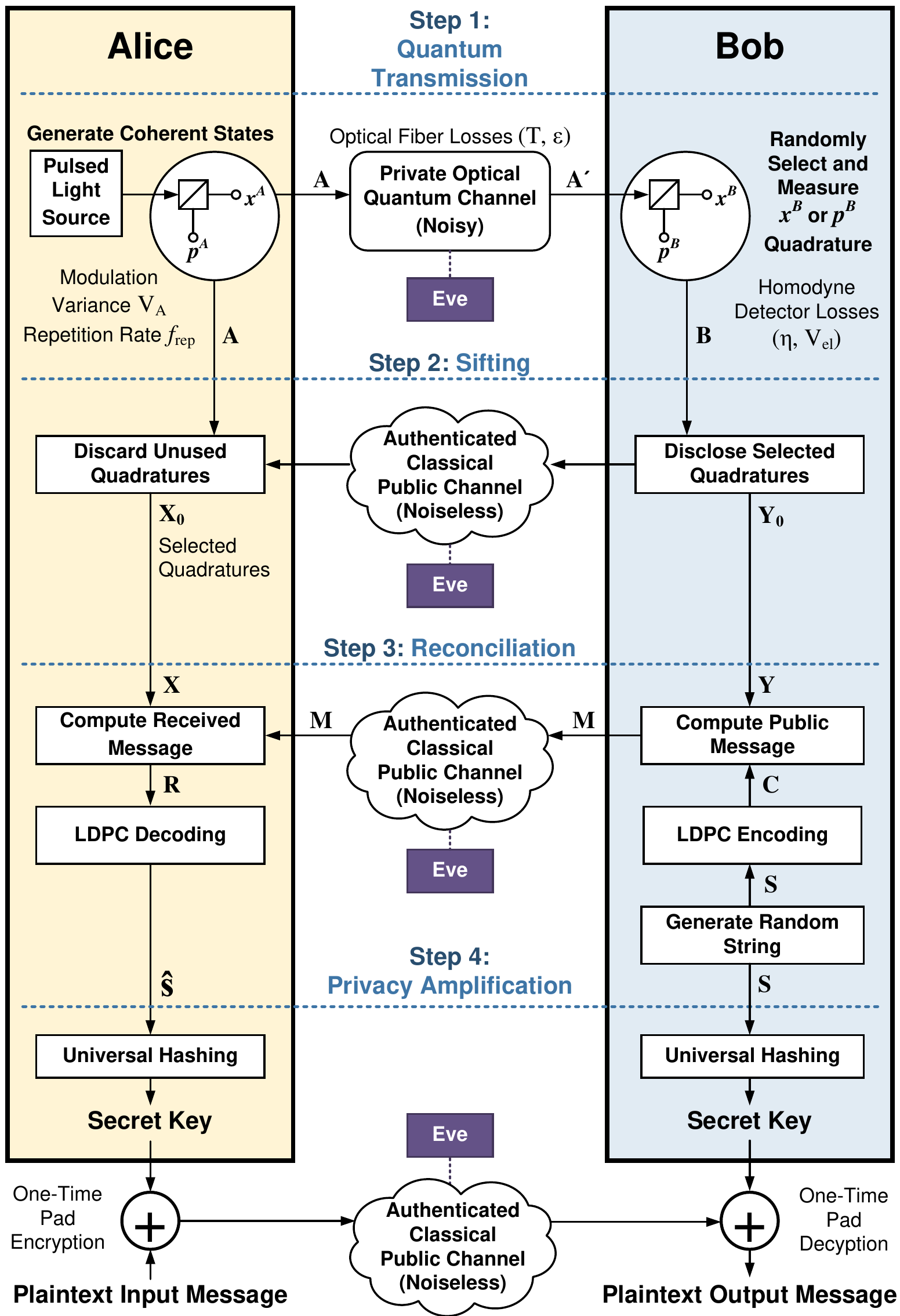}
\caption{CV-QKD model for secret key distillation with reverse reconciliation between Alice and Bob over a private quantum channel and public classical channel.}
\label{fig:RRCVQKDmodel}
\end{figure}

In a QKD system, two remote parties, Alice and Bob, communicate over a private optical quantum channel, as well as an authenticated classical public channel to generate a shared secret key in the presence of an adversary or eavesdropper, Eve, who may have access to both channels~\cite{Gisin2002}. The security of QKD stems from the no-cloning theorem of quantum mechanics, which states that any observation or measurement of the quantum channel by Eve would disturb the coherent states transmitted from Alice to Bob~\cite{GG02, Lodewyck2007}. Since Alice and Bob can calibrate their expected channel noise threshold for a fixed fiber-optic transmission distance prior to being deployed in the field, any quantum measurement by Eve would result in a channel noise increase, at which point, the reconciliation error rate would increase, and Alice and Bob could choose to terminate their communication if they suspect a man-in-the-middle attack~\cite{Jouguet2013}. A typical prepare-and-measure CV-QKD system is based on the Grosshans-Grangier 2002 (GG02) protocol~\cite{GG02}, which defines the following four steps presented in Fig.~\ref{fig:RRCVQKDmodel}: quantum transmission, sifting, reconciliation, and privacy amplification. Fully secure QKD networks can be built by designating intermediate trusted nodes~\cite{Alleaume2014, Diamanti2016}, or through measurement-device-independent QKD (MDI-QKD) using untrusted relay nodes in both CV- and DV-QKD~\cite{Lo2012,Pirandola2015high,Yin2016}. MDI-QKD is beyond the scope of this work, however, it does provide a viable solution to the quantum hacking problem by removing all detector side channels~\cite{Lo2012}. This section first provides a fundamental overview of QKD, and then presents the mathematical framework for key reconciliation using LDPC codes over multiple dimensions, with the consideration of finite-size effects on the secret key rate.

\subsection{Quantum Transmission and Measurement}

To construct a secret key using the prepare-and-measure CV-QKD protocol, Alice first transmits $N_\text{quantum}$ coherent states to Bob over an optical fiber. Each coherent state is comprised of a pair of amplitude and phase quadrature operators, $x$ and $p$, of the form $\left| x + jp \right\rangle$, $j=\sqrt{-1}$. Using a quantum random number generator, Alice prepares each coherent state by randomly selecting her $x^A$ and $p^A$ quadrature values according to a zero-mean Gaussian distribution with adjustable modulation variance $\sigma_{A}^2 = V_{A}N_{0}$, where $N_{0}$ represents the shot noise variance defined by the Heisenberg inequality $\Delta x \Delta p \geq N_{0}$~\cite{GG02, Lodewyck2007}. Alice transmits her train of $N_\text{quantum}$ coherent states to Bob by modulating a light source with a pulse repetition rate $f_\text{rep}$. She also records her selections of $x^A$ and $p^A$ for the next sifting step, by constructing a vector, $\mathbf{A}$, of length $2N_\text{quantum}$, from her $N_\text{quantum}$ coherent state quadrature operator pairs $(x^A,p^A)$, such that $A_{2i-1}=x_i^A$ and $A_{2i}=p_i^A$ for $i=1,2,\dots,N_\text{quantum}$. As such, $\mathbf{A}=(x_1^A,p_1^A,x_2^A,p_2^A,\dots,x_{N_\text{quantum}}^A,p_{N_\text{quantum}}^A)$ and $A \sim \mathcal{N}(0, \sigma_{A}^2)$. Bob randomly selects and measures either the $x$ or $p$ quadrature for each incoming pulse using an unbiased homodyne detector. Bob constructs his own vector, $\mathbf{B}$, of length $N_\text{quantum}$, comprised of the observed modulated quadrature measurements, where $B_{i} \in \{x^B_i,p^B_i\}$ with equal probability. Despite the losses in the optical fiber, and the added noise from the Bob and Eve's detection equipment, the $x^{B}$ and $p^{B}$ quadrature measurements can still be used to distill a secret key following the sifting and reconciliation (error correction) steps.

Without considering the presence of the eavesdropper (Eve), the quantum transmission is subject to path loss, excess noise in the single-mode fiber between Alice and Bob, the inefficiency of Bob's homodyne detection, as well as added electronic (thermal) noise~\cite{Jouguet2013}. The optical experimental setup is beyond the scope of this work, thus experimental values from previously published works have been used to characterize the quantum channel~\cite{Lodewyck2007}. The excess channel noise expressed in shot noise units is assumed to be $\epsilon = 0.005$, Bob's added electronic noise in shot noise units is chosen as $V_\text{el}=0.041$, Bob's homodyne detector efficiency is set to $\eta = 0.606$, and the single-mode fiber transmission loss is assumed to be 0.2dB/km, such that the transmittance of the quantum channel is given by $T = 10^{-\alpha d/10}$, where $d$ is the transmission distance in kilometers and $\alpha = 0.2$dB/km. The total noise between Alice and Bob is given by $\chi_\text{total} = \chi_\text{line} + \frac{\chi_\text{hom}}{T}$, where $\chi_\text{line} = ( \frac{1}{T}-1) + \epsilon$ is the total channel added noise referred to the channel input, and $\chi_\text{hom} = \frac{1 + V_\text{el}}{\eta}-1$ is the noise introduced by the homodyne detector. The variance of Bob's measurement is given by $\sigma_{B}^2 = V_{B}N_{0} = \eta T (V+ \chi_\text{total})N_{0}$. Although the adversary (Eve) may have access to the quantum channel, her presence is not considered in the channel characterization. Instead, the information leaked to Eve will be considered in the secret key rate calculation~\cite{GG02}. 

In the remaining post-processing steps of the QKD protocol, Alice and Bob communicate over an authenticated classical public channel, which is assumed to be noiseless and error-free. Eve may have access to this channel, however, her eavesdropping does not introduce additional errors~\cite{GG02}.

\subsection{Sifting}

Following the quantum transmission step, Alice's original transmission vector $\mathbf{A}$ contains twice as many elements as Bob's measurement vector $\mathbf{B}$. In the sifting step, Bob informs Alice which  of the $x^{B}$ or $p^{B}$ quadratures he randomly selected for each of his $N_\text{quantum}$ element measurements, such that Alice may respectively discard her $N_\text{quantum}$ unused $x^{A}$ and $p^{A}$ quadrature values~\cite{GG02}. After sifting, Alice and Bob share correlated random sequences of length $N_\text{quantum}$, herein defined as $\mathbf{X_{0}} = (X_{0_{1}},X_{0_{2}},\dots,X_{0_{N_\text{quantum}}})$ and $\mathbf{Y_{0}} = (Y_{0_{1}},Y_{0_{2}},\dots,Y_{0_{N_\text{quantum}}})$, respectively, where ($X_{0_{i}},Y_{0_{i}}$), $i=1,2,\dots,N_\text{quantum}$, are independent and identically distributed realizations of some jointly Gaussian random variables ($X_{0},Y_{0}$). For example, Alice and Bob may have the following random sequences after sifting: $\mathbf{X_{0}}=(x^A_1,p^A_2,p^A_3,x^A_4,\dots,p^A_{N_\text{quantum}})$ and $\mathbf{Y_{0}}=(x^B_1,p^B_2,p^B_3,x^B_4,\dots,p^B_{N_\text{quantum}})$. In the following reconciliation and privacy amplification steps, Alice and Bob apply error correction and hashing techniques to build a secret key using their sifted sequences of correlated quadrature measurements.

\subsection{Reconciliation}

During information reconciliation, Alice and Bob perform the first step in building a unique secret key by: (1) encoding a randomly-generated message using the sifted quadrature measurements, (2) transmitting the encoded message over an authenticated classical channel, and then (3) applying an error-correction scheme to decode the original message~\cite{Lodewyck2007, Leverrier2008}. In the direct reconciliation scheme, Alice generates and transmits a random message to Bob, who then performs the error-correction decoding based on his measured quadratures. However, previous works have shown that the transmission distance with direct reconciliation is limited to about 15km~\cite{Weedbrook2010, Jouguet2014directrecon, gehring2015imp}, and is thus not suitable for long-distance CV-QKD targeting transmission distances beyond 100km~\cite{Jouguet2011}. 

The long distance problem drives the need for an alternate, robust scheme that is capable of operating under the low-SNR conditions of the optical channel, even in the presence of excess noise introduced by an eavesdropper. In the reverse reconciliation scheme, the direction of classical communication between Alice and Bob is reversed. Reverse reconciliation achieves a higher secret key rate at longer distances in comparison to direct reconciliation, however, powerful error-correction codes are still required to combat the high channel noise at long distances without revealing unnecessary information to Eve during the reconciliation process~\cite{Leverrier2008, Jouguet2011, Jouguet2013}. 

Two-way interactive error-correction protocols such as Cascade or Winnow are not practical for long-distance QKD due to the large latency and communication overhead required to theoretically minimize the information leakage to Eve~\cite{Grosshans2003, Yan2009, Elkouss2010, Benletaief2014}. 
Blind reconciliation using short block-length codes on the order of 10$^3$ bits with low interactivity was proposed to reduce decoding latency~\cite{Martinez-Mateo2013}, however, the short block length is not suitable for error-correction at low SNR. Instead, one-way forward error-correction implemented using long block-length codes with iterative soft-decision decoding is required to achieve efficient error-correction at low SNR~\cite{Leverrier2008, Leverrier2009}. Jouguet~et~al.\ recently showed that multi-edge LDPC codes combined with a multi-dimensional reverse reconciliation scheme can achieve near-Shannon limit error-correction performance at long distances~\cite{Jouguet2011, Jouguet2013}. However, the computational complexity of LDPC decoding remains a limitation to the maximum achievable secret key rate in a practical QKD implementation~\cite{Jouguet2014}. Sections \ref{sec:LDPCforRecon} and \ref{sec:GPU} of this work present hardware-oriented optimization techniques to alleviate the time-intensive bottleneck of LDPC decoding for long distance CV-QKD systems, while the remainder of this section outlines the mathematical framework for long-distance reverse reconciliation.

\subsubsection{Reconciliation at Long Distances}

Strong error-correction schemes do not exist for systems with both a Gaussian input and Gaussian channel, as in the case of CV-QKD. However, at low SNR, the maximum theoretical secret key rate is less than 1 bit/pulse per channel use, and the Shannon limit of the additive white Gaussian noise (AWGN) channel approaches the limit of a binary-input AWGN channel (BIAWGNC)~\cite{Leverrier2008}. This makes binary codes highly suitable for error correction in the low-SNR regime~\cite{Richardson2001, Richardson2001_Shokrollahi}, as opposed to non-binary codes, which outperform binary codes on channels with more than 1 bit/symbol per channel use~\cite{Declercq2007}. Since binary codewords can be encoded in the signs of Alice and Bob's correlated sequences, $\mathbf{X_{0}}$ and $\mathbf{Y_{0}}$, the reconciliation system can therefore be modelled as a BIAWGNC~\cite{Jouguet2011}.

\subsubsection{Reverse Reconciliation Algorithm for BIAWGNC}

A virtual model of the BIAWGNC can be induced from the physical parameters that characterize the quantum transmission~\cite{Jouguet2011}, where the optical fiber and homodyne detector losses are captured in the form of a signal-to-noise ratio with respect to the optical input signal, whose variance is normalized based on Alice's modulation variance $V_{A}$. Assuming that the BIAWGNC has a zero mean and noise variance of $\sigma_{Z}^2$, $Z \sim \mathcal{N}(0,\sigma_{Z}^2)$, the SNR can be expressed as $s=1/\sigma_{Z}^2$. In order to perform key reconciliation, Alice and Bob now construct two new correlated Gaussian sequences from their sifted correlated sequences $\mathbf{X_0}$ and $\mathbf{Y_0}$ of length $N_\text{quantum}$. Alice and Bob first select a subset of $n$ elements from $\mathbf{X_0}$ and $\mathbf{Y_0}$, where $n < N_\text{quantum}$. Here, $n$ is chosen to be equivalent to the LDPC code block length. Alice and Bob then normalize their subset of $n$ elements by the modulation variance $V_A$, such that Alice and Bob now share correlated Gaussian sequences $\mathbf{X}$ and $\mathbf{Y}$, each of length $n$, where $X \sim \mathcal{N}(0, 1)$, $Y \sim \mathcal{N}(0, 1+\sigma_{Z}^2)$, and the property $\mathbf{Y}~=~\mathbf{X}+~\mathbf{Z}$ holds~\cite{Jouguet2011}. 

Bob uses a quantum random number generator to generate a uniformly-distributed random binary sequence $\mathbf{S}$ of length $k$, where $S_{i} \in \{0,1\}$. He then performs a computationally inexpensive LDPC encoding operation to generate an LDPC codeword $\mathbf{C}$ of length $n$, where $C_{i} \in \{0,1\}$, by appending $(n-k)$ redundant parity bits to $\mathbf{S}$ based on a binary LDPC parity-check matrix $\mathbf{H}$ that is also known to Alice. Eve may also have access to $\mathbf{H}$, however, the QKD security proof still holds since Eve is assumed to have infinite mathematical genius. Bob prepares his classical message to Alice, $\mathbf{M}$, by modulating the signs of his correlated Gaussian sequence $\mathbf{Y}$ with the LDPC codeword $\mathbf{C}$, such that $M_{i} = (-1)^{C_{i}}Y_{i}$ for $i=1, 2, \dots, n$. The symmetry in the uniform distribution of Bob's random binary sequence $\mathbf{S}$ ensures that the transmission of $\mathbf{M}$ over the authenticated classical public channel does not reveal any additional information about the secret key to Eve~\cite{Lodewyck2007}. 

Assuming error-free transmission over the classical channel, Alice attempts to recover Bob's codeword using her correlated Gaussian sequence $\mathbf{X}$ based on the following division operation:
\begin{equation} \label{eq:Ri} \begin{split}
R_{i} & = \frac{M_{i}}{X_{i}} = \frac{  (-1)^{C_{i}}Y_{i} }{ X_{i} } = \frac{(-1)^{C_{i}}(X_{i}+Z_{i})}{X_{i}} \\
& = (-1)^{C_{i}} + (-1)^{C_{i}}\frac{Z_{i}}{X_{i}},
\end{split} \end{equation}
for $i=1, 2, \dots, n$. Here, Alice observes a virtual channel with binary input ($\pm 1$) and additive noise $(-1)^{C_{i}}\frac{Z_{i}}{X_{i}}$. In this case, the division operation in the noise term represents a fading channel, however, since Alice knows the value of each $X_{i}$, the norm of $\mathbf{X}$ is revealed and the overall channel noise remains Gaussian with zero mean and variance $\sigma_{Ni}^2 = \sigma_{Z}^2 / ||X_{i}||$ for each $i=1, 2, \dots, n$~\cite{Jouguet2011}. Alice then attempts to reconstruct $\mathbf{S}$ by applying the computationally intensive Sum-Product belief propagation algorithm for LDPC decoding, to remove the channel noise from her received vector $\mathbf{R}$. The LDPC decoding algorithm requires the channel noise variance $\sigma_{Ni}^2$ to be known for each $i=1, 2, \dots, n$. By discarding the $(n-k)$ parity bits from the decoded codeword, Alice can build an estimate $\mathbf{\hat S}$ of Bob's original binary sequence $\mathbf{S}$ for further post-processing in the next privacy amplification step to asymptotically reduce Eve's knowledge about the secret key~\cite{GG02}. LDPC decoding is successful if $\mathbf{\hat S} = \mathbf{S}$, whereas a frame error is said to have occurred when $\mathbf{\hat S} \neq \mathbf{S}$.

\subsubsection{Multi-Dimensional Reconciliation}

Up until this point, the discussion has assumed a 1-dimensional reconciliation scheme in $\mathbb{R}$, with $\pm 1$ binary inputs on the virtual BIAWGNC. Leverrier~et~al.\ showed that the quantum transmission can be extended to longer distances with proven security by employing multi-dimensional reconciliation schemes constructed from spherical rotations in $\mathbb{R}^2$, $\mathbb{R}^4$, and $\mathbb{R}^8$, where the multiplication and division operators are defined~\cite{Leverrier2008, Leverrier2009}. These spaces are commonly referred to as the set of complex numbers $\mathbb{C}$, the quaternions $\mathbb{H}$, and the octonions $\mathbb{O}$, respectively. As shown in Eq.~\ref{eq:Ri}, the division and multiplication operations must be defined for the reverse reconciliation procedure. By Hurwitz's theorem of composition algebras, normed division is only defined for four finite-dimensional algebras: the real numbers $\mathbb{R}$ ($\mathbb{R}^{d=1}$), the complex numbers $\mathbb{C}$ ($\mathbb{R}^{d=2}$), the quaternions $\mathbb{H}$ ($\mathbb{R}^{d=4}$), and the octonions $\mathbb{O}$ ($\mathbb{R}^{d=8}$)~\cite{Hurwitz1898}. Hence, the remainder of this discussion considers only the $d=1,2,4,8$ dimensions.

The multi-dimensional approach is a further reformulation of the reduction of the physical Gaussian channel to a virtual BIAWGNC at low SNR. For $d$-dimensional reconciliation, $d \in \{1,2,4,8\}$, each consecutive group of $d$ quantum coherent-state transmissions from Alice to Bob can be mapped to the same virtual BIAWGNC. As a result, the channel noise variance among all $d$ virtual channels is uniform. For the $d=1$ case, each $R_{i}$ defined in Eq.~\ref{eq:Ri} has a unique channel noise variance defined by $\sigma_{Ni}^2 = \sigma_{Z}^2/||X_{i}||$ for $i~=~1, 2, \dots, n$. For the $d=2$ case, the reconciliation is performed over successive $(R_{2i-1},R_{2i})$ pairs: $(R_{1},R_{2}), (R_{3},R_{4}), \dots, (R_{n-1},R_{n})$, which are constructed from the quadrature transmission of successive $(M_{2i-1},M_{2i})$ pairs for $i=1,2,\dots,n/2$ in $\mathbb{R}^{d=2}$. Similar to $d=1$, each $i$th received value is still comprised of a $\pm 1$ binary input and a noise term, such that $R_{2i-1} = (-1)^{C_{2i-1}} + N_{2i-1}$ and $R_{2i} = (-1)^{C_{2i}} + N_{2i}$ for $i=1,2,\dots,n/2$. While the real and imaginary noise components, $N_{2i-1}$ and $N_{2i}$, are not equal, the variance of the channel noise is uniform over both dimensions, such that $\sigma_{N(2i-1)}^2 = \sigma_{N(2i)}^2$ for each $(R_{2i-1},R_{2i})$ pair. This can be extended to the $d=4$ and $d=8$ cases, where each $d$-tuple of successive $R_{i}$ values has a unique channel noise for each dimensional component, but the channel noise variance remains equal over all $d$ dimensions. For example, for $d=4$, each received 4-tuple, $(R_{4i-3},R_{4i-2},R_{4i-1},R_{4i})$ for $i=1,2,\dots,n/4$, has a unique noise term for each of its four components, but the channel noise variance over all four dimensions remains uniform. 

The following derivation extends Alice's message reconstruction calculation presented in Eq.~\ref{eq:Ri} to $d$-dimensional vector spaces, $d \in \{1,2,4,8\}$, where the multiplication and division operators are defined. The derivation of the channel noise for $d=2,4,8$ is much more rigorous than for $d=1$, however, the procedure can be simplified by applying associative and distributive algebraic properties that hold true for the complex, quaternion, and octonion vector spaces. It follows then that
\begin{align}
\mathbf{R} 
& = \mathbf{M} \mathbf{X}^{-1} \nonumber \\
& = (\mathbf{U}\mathbf{Y}) \mathbf{X}^{-1} \nonumber \\
& = (\mathbf{U}(\mathbf{X} + \mathbf{Z})) \mathbf{X}^{-1} \nonumber \\
& = (\mathbf{U}\mathbf{X} + \mathbf{U}\mathbf{Z}) \mathbf{X}^{-1} \text{ by right-linearity } a(b+c)=ab+ac \nonumber \\
& = \mathbf{U}\mathbf{X}\mathbf{X}^{-1} + \mathbf{U}\mathbf{Z}\mathbf{X}^{-1} \text{ by left-linearity } (b+c)a=ba+ca \nonumber \\
& = \mathbf{U} + \mathbf{U}\mathbf{Z}\mathbf{X}^{-1} \text{ by right-cancellation } abb^{-1} = a \nonumber \\
& = \mathbf{U} + \mathbf{U}\mathbf{Z}\dfrac{\mathbf{X}^{*}}{||\mathbf{X}||^{2}}.  
\end{align}
Here, $\mathbf{R}$, $\mathbf{M}$, $\mathbf{X}$, $\mathbf{Y}$, and $\mathbf{Z}$ are $d$-dimensional vectors, and $\mathbf{U}$ is the $d$-dimensional vector comprised of $(-1)^{C_{i}}$ components. For example, for $d=2$, $\mathbf{U}=[(-1)^{C_{2i}}, (-1)^{C_{2i-1}}]$, while for $d=4$, $\mathbf{U}=[(-1)^{C_{4i-3}}, (-1)^{C_{4i-2}}, (-1)^{C_{4i-1}}, (-1)^{C_{4i}}]$. The multi-dimensional noise for a virtual BIAWGNC is given by the term $\mathbf{N}=(\mathbf{\mathbf{U}\mathbf{Z}\mathbf{X}^{*}})/||\mathbf{X}||^{2}$, such that $\mathbf{R}~=~\mathbf{U}~+~\mathbf{N}$. The Cayley-Dickson construction can then be applied to complete the derivation of the multi-dimensional noise $\mathbf{N}$ for $d=2,4,8$. Since the noise is uniformly distributed in each dimension, $\mathbf{C}$ can be assumed to be the all-zero codeword, i.e. $C_i=0$ for all $i=1,2,\dots,n$, to further simplify the derivation. For $d=2$, the channel noise of both the real and imaginary components can be expressed as $N_{2i-1}=a_i Z_{2i-1}+b_i Z_{2i}$ and $N_{2i}=a_i Z_{2i}- b_i Z_{2i-1}$, where $a_i~=~\dfrac{X_{2i-1}+X_{2i}}{X_{2i-1}^2+X_{2i}^2}$ and $b_i=\dfrac{X_{2i}-X_{2i-1}}{X_{2i-1}^2+X_{2i}^2}$ for $i~=~1,2,\dots,n/2$. It follows then that the channel noise variance for $d=2$ is given by $\sigma_{N(2i-1)}^2 = \sigma_{N(2i)}^2 = (a_i^2 + b_i^2)\sigma_{Z}^2$. The noise derivation for $d=4$ and $d=8$ is much longer and not included here.

\subsubsection{Reconciliation Efficiency}
\label{sec:RecEff}

The reverse reconciliation algorithm for the BIAWGNC can be reduced to an asymmetric Slepian-Wolf source-coding problem with input $\mathbf{M}$ and side information $\mathbf{X}$, where Alice and Bob observe correlated Gaussian sequences $\mathbf{X}$ and $\mathbf{Y}$, respectively~\cite{Bloch2006, Elkouss2010}. Since Alice must discard $(n-k)$ parity bits from the linear block code after LDPC decoding, it follows then that the efficiency of the reverse reconciliation algorithm is given by $\beta = \frac{R_\text{code}}{I(X;Y)}$, where $I(X;Y)$ is the mutual information between $\mathbf{X}$ and $\mathbf{Y}$, and $R_\text{code}$ is the LDPC code rate defined as $R_\text{code} = k/n$ from the $n$-length codeword $\mathbf{C}$ and $k$-length random information string $\mathbf{S}$~\cite{Bloch2006, Jouguet2011}. The mutual information $I(X;Y)$ corresponds to the Shannon capacity of the quantum channel, hence the reconciliation efficiency can be expressed more simply as: 
\begin{equation}
\beta = \frac{R_\text{code}}{C(s)} = \frac{R_\text{code}}{R_\text{max}}, 
\label{eq:beta-def}
\end{equation}
where $C(s) = \frac{1}{2} \log_{2}(1+s)$ is the Shannon capacity and $s$ is the SNR of the BIAWGNC. The Shannon capacity defines the maximum achievable code rate $R_\text{max}$ for a given SNR, and thus, the $\beta$-efficiency characterizes how close the reconciliation algorithm operates to this fundamental limit~\cite{Richardson2001_Shokrollahi}. 

The reconciliation efficiency plays a crucial role in the performance of CV-QKD. The $\beta$-efficiency at a particular SNR operating point determines the code rate, and ultimately, the number of parity bits discarded in each message. Assuming that the LDPC coding scheme has been optimized for a particular SNR operating point such that the code rate $R_\text{code}$ is fixed, the reconciliation efficiency then depends solely on the SNR of the quantum channel, which is a function of Alice's coherent-state modulation variance and the physical transmission losses in the optical fiber. Hence, for a fixed optical transmission distance between Alice and Bob, the reconciliation efficiency can be optimized by tuning Alice's modulation variance $V_{A}$, and designing an optimal error-correction scheme for a target SNR. Sections \ref{sec:LDPCforRecon} and \ref{sec:GPU} explore how changes in the $\beta$-efficiency affect the reconciliation distance and  maximum achievable secret key rate.

\subsection{Privacy Amplification}

%Since Eve may have collected sufficiently enough information during her observations of the quantum and classical channels, Alice and Bob reduce Eve's knowledge of the key by applying a shared universal hashing function to their respective binary strings $\mathbf{\hat{S}}$ and $\mathbf{S}$, in order to generate a unique symmetric key~\cite{GG02, Lodewyck2007}. Alice and Bob can then use their shared secret key to encrypt messages with perfect secrecy using the one-time pad technique~\cite{Alleaume2014}. A complete discussion of the privacy amplification step is beyond the scope of this work. 

Since Eve may have collected sufficient information during her observations of the quantum and classical channels, Alice and Bob must asymptotically reduce Eve's knowledge of the key by independently applying a shared universal hashing function on a concatenated block of their independent binary strings $\mathbf{\hat{S}}$ and $\mathbf{S}$, in order to generate a unique symmetric key~\cite{GG02, Lodewyck2007}. Alice first discards her erroneously decoded $\mathbf{\hat S}$ messages, and informs Bob as to which messages she discarded. Bob then discards his original $\mathbf{S}$ messages that correspond to the $\mathbf{\hat S}$ messages that were discarded by Alice. Alice concatenates all of her correctly decoded $\mathbf{\hat S}$ messages to construct a long secret key block of length $N_\text{privacy}=mk$ bits, where $k$ is the length of the LDPC-decoded message $\mathbf{\hat S}$, and $m$ is some large non-zero integer. Bob also concatenates his corresponding $\mathbf{S}$ messages to construct a long secret key, also of length $N_\text{privacy}$ bits. Alice and Bob then independently perform universal hashing on their independent secret key blocks to reduce Eve's knowledge of the key. Alice and Bob can use the resulting symmetric key to encrypt and decrypt messages with perfect secrecy using the one-time pad technique~\cite{Alleaume2014}. 

The speed of privacy amplification is an active area of research, with published results showing maximum speeds of 100Mb/s for a block size of $N_\text{privacy}=10^8$ bits~\cite{Takahashi2016}. The computational complexity of universal hashing can be reduced from $O(n^2)$ to $O(n\log_2n)$ by applying the fast Fourier transform (FFT) or number theoretical transform (NTT) on a Toeplitz matrix~\cite{Hayashi2016}. Estimation of security parameters is also performed during privacy amplification using $(N_\text{quantum} - N_\text{privacy})$ bits, however, a complete discussion of parameter estimation and privacy amplification is beyond the scope of this work.

\subsection{Maximizing Secret Key Rate with Collective Attacks}
\label{sec:MaximizingSecretKeyRate}

The primary metric that defines the performance of a QKD system is the maximum rate at which Alice and Bob can securely generate and reconcile keys over a fixed-distance optical fiber in the presence of an eavesdropper that has access to both the quantum and classical channels. The maximum secret key rate must be proven secure against a collective Gaussian attack, the most optimal man-in-the-middle attack, where Eve first prepares an ancilla state to interact with each one of Alice's coherent states during the quantum transmission, and then listens to the public communication between Alice and Bob during the reconciliation step in order to perform the most optimal measurement on her collected ancillae to reconstruct the classical messages transmitted by Bob~\cite{Lodewyck2007}. Assuming perfect error-correction during the reconciliation step, the maximum theoretical secret key rate for a CV-QKD system with one-way reverse reconciliation can be defined as 
\begin{equation}
\label{eq:Kopt}
K_\text{opt}=\beta I_{AB} - \chi_{BE} \ \ \ \text{(bits/pulse)},
\end{equation}
where $I_{AB}$ is the mutual information between Alice and Bob, $\beta$ is the previously defined reconciliation efficiency, and $\chi_{BE}$ is the Holevo bound on the information leaked to Eve. Here, $I_{AB}$ is equivalent to the Shannon channel capacity, and is defined as 
\begin{equation}
\label{eq:IAB}
I_{AB} = \frac{1}{2} \log_{2}(1+s) = \frac{1}{2} \log_{2} \bigg( { \frac{V + \chi_\text{total}}{1 + \chi_\text{total}} } \bigg),
\end{equation}
where $V = V_{A} + 1$, $V_{A}$ is Alice's adjustable modulation variance, and $\chi_\text{total}$ is the total noise between Alice and Bob. The Holevo bound is defined as
\begin{equation}
\resizebox{230pt}{!}{$
\chi_{BE} = G\bigg( { \dfrac{\lambda_{1}-1}{2} } \bigg) + G\bigg( { \dfrac{\lambda_{2}-1}{2} } \bigg) - G\bigg( { \dfrac{\lambda_{3}-1}{2} } \bigg) - G\bigg( { \dfrac{\lambda_{4}-1}{2} } \bigg),
$} 
%\nonumber
\end{equation}
where $G(x) = (x+1)\log_{2}(x+1) - x\log_{2}x$, and the Eigenvalues $\lambda_{1,2,3,4}$ are given by $\lambda_{1,2}^2 = \frac{1}{2} ( A \pm \sqrt{A^2 - 4B} )$ and $\lambda_{3,4}^2 = \frac{1}{2} (C \pm \sqrt{C^2 - 4D} )$, where
\begin{align*}
A &= V^2(1-2T) + 2T + T^2(V+\chi_\text{line})^2 \\
B &= T^2(V\chi_\text{line}+1)^2\\
C &= \frac{V\sqrt{B}+T(V+\chi_\text{line}) + A\chi_\text{hom}}{T(V+\chi_\text{total})}\\
D &= \sqrt{B} \frac{V + \sqrt{B}\chi_\text{hom}}{T(V+\chi_\text{total})}.
\end{align*}
%using align* removes equation numbering

Here, $K_\text{opt}$ represents the asymptotic limit on the secret key rate based on ideal theoretical security models, and does not consider the imperfections of a practical CV-QKD system, which might enable additional side-channel attacks~\cite{Jouguet2012}. Such imperfections include the finite-size effects~\cite{Leverrier2010, Curty2014, Diamanti2015}, excess electronic and phase noise from uncalibrated optical equipment, as well as discretized Gaussian modulation with finite bounds on the distribution and randomness~\cite{Jouguet2012}. Leverrier proved that CV-QKD with coherent states provides composable security against collective attacks~\cite{Leverrier2015}, however, extending the information-theoretic security proofs from collective attacks to general attacks in the finite-size regime of CV-QKD is currently an active area of research~\cite{Diamanti2015,Usenko2016}. At the time of writing, the highest CV-QKD key rates can be achieved using coherent states and homodyne detection with security against collective attacks and some finite-size effects~\cite{Jouguet2013}. The motivation of this work is to show that the key reconciliation (error correction) algorithm can be accelerated such that the throughput of LDPC decoding is higher than the asymptotic secret key rate achievable using realistic quantum channel parameters and optical equipment available today. The finite-size effects on secret key rate are considered later in this section, while the other imperfections of a practical CV-QKD system are beyond the scope of this work.

\begin{figure}[t!]
\centering
\includegraphics[trim=0in 0in 0in 0in, width=0.48\textwidth]{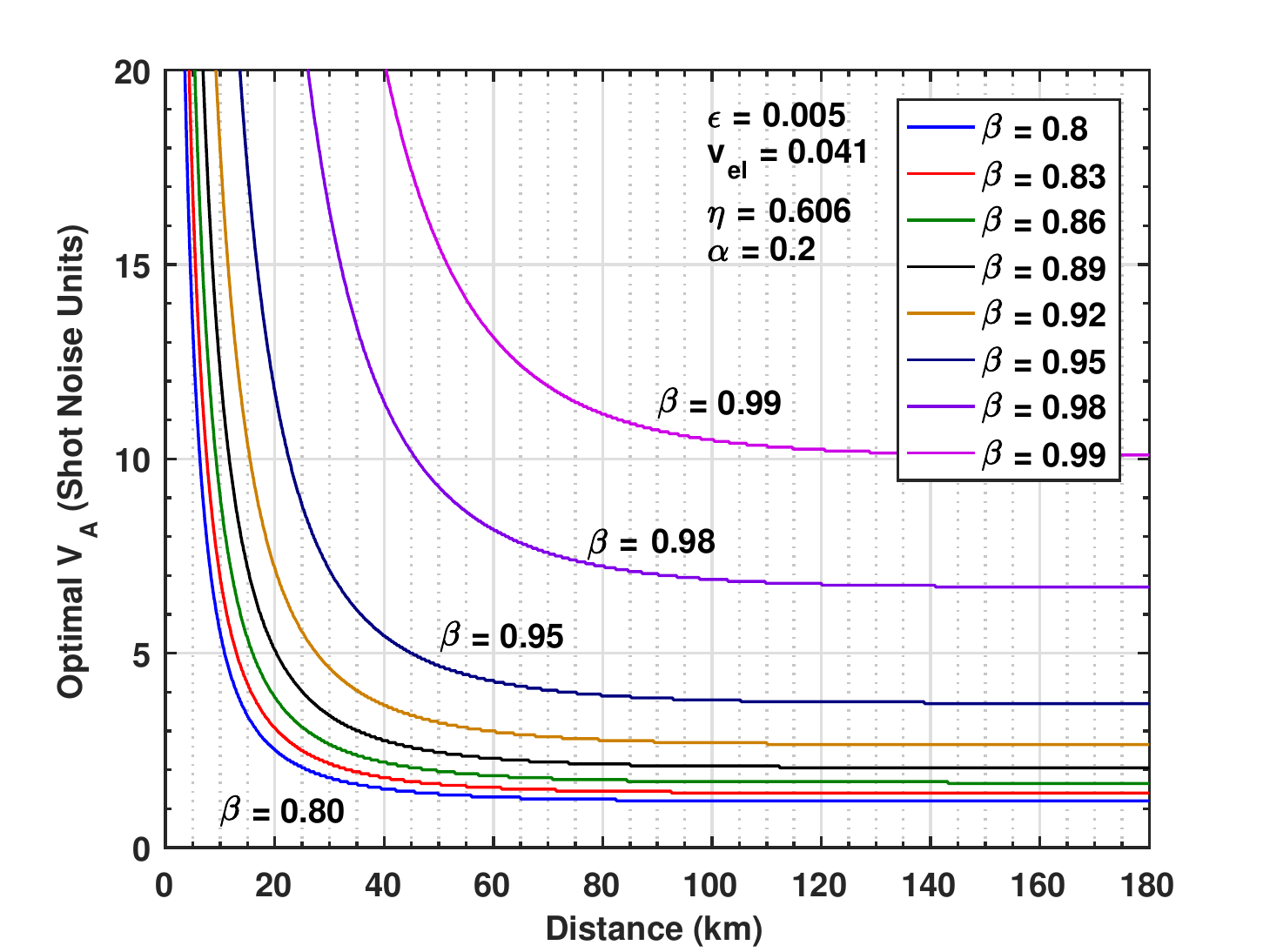}
\caption{Optimal $V_{A}$ vs. transmission distance for maximum theoretical secret key rate, from $\beta=0.8$ to $\beta=0.99$, based on the assumed physical operating parameters of the quantum channel.}
\label{fig:OptimalVA}
\end{figure}

\begin{figure}[t!]
\centering
\includegraphics[trim=0in 0in 0in 0in, width=0.48\textwidth]{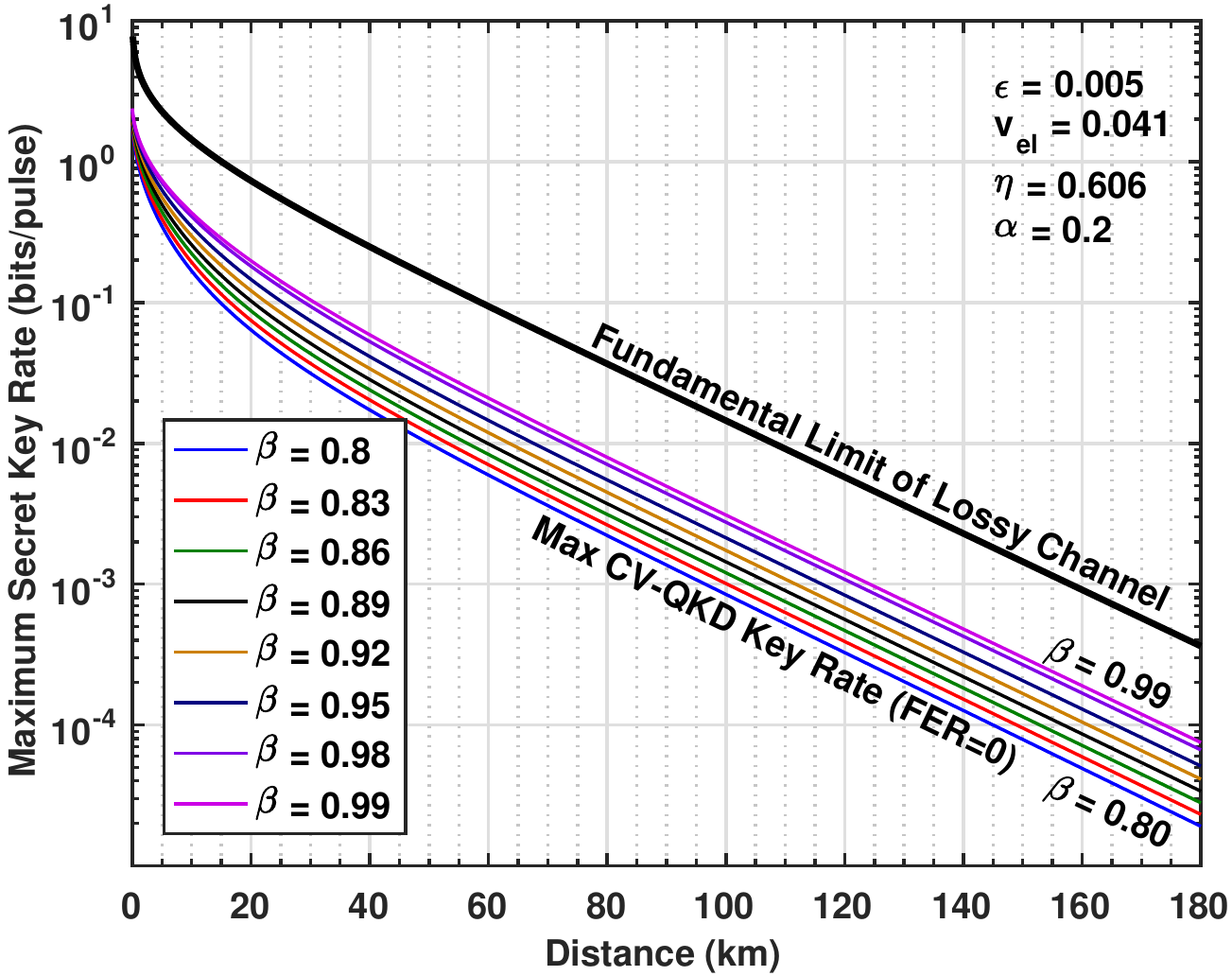}
\caption{Maximum theoretical secret key rates vs. transmission distance. The maximum CV-QKD key rate is defined by $K_\text{opt}$ from $\beta=0.8$ to $\beta=0.99$ based on the optimal $V_A$. The fundamental limit for a lossy channel is defined by $K_\text{lim}= -\log_2(1-T)$.}
\label{fig:MaxSKR}
\end{figure}

Optimizing Alice's modulation variance for each quantum transmission distance ensures a maximum SNR on the BIAWGNC~\cite{Jouguet2011}, and thus, a maximum achievable secret key rate $K_\text{opt}$ for a particular $\beta$-efficiency. Figure~\ref{fig:OptimalVA} presents the optimal modulation variance $V_{A}$ as a function of $\beta$ for quantum transmission distances up to 180km, assuming perfect error-correction in the reconciliation step. Figure~\ref{fig:MaxSKR} shows the corresponding maximum theoretical secret key rate $K_\text{opt}$ for CV-QKD based on the computed optimal $V_{A}$ at each distance. 

Pirandola~et~al. recently showed that there exists an upper bound on the secret key rate for a lossy channel~\cite{Pirandola2015}. This fundamental limit is determined by the transmittance $T$ of the fiber-optic channel, and is given by
\begin{equation}
K_\text{lim} = -\log_2(1-T) \ \ \text{(bits/pulse).}
\label{eq:Klim}
\end{equation}
The transmittance was previously defined as $T = 10^{-\alpha d/10}$, where the distance $d$ is expressed in kilometers and the standard loss of a fiber optic cable is assumed to be $\alpha = 0.2$dB/km. The upper bound is plotted in Fig.~\ref{fig:MaxSKR}. 

The BIAWGNC model for long-distance CV-QKD under investigation in this work has also been proven secure against collective attacks, thus the expression for the asymptotic secret key rate $K_\text{opt}$ still holds~\cite{Leverrier2008, Jouguet2011}. At long distances, $I_{AB}$ and $\chi_{BE}$ are nearly equal, thus in order to maximize the secret key rate, it would appear that the reconciliation efficiency $\beta$ must also be maximized. However, this is not necessarily true since $K_\text{opt}$ only provides an expression for the maximum achievable secret key rate and does not consider the speed of reconciliation, nor the uncorrectable errors. The frame error rate (FER) of the reconciliation algorithm must also be considered.

\subsection{Frame Error Rate for Reverse Reconciliation}

\begin{figure}[h!]

\tikzstyle{level 1}=[sibling distance=4cm]
\tikzstyle{level 2}=[sibling distance=3cm]
\tikzstyle{level 3}=[sibling distance=2cm]
\scalebox{0.98}{
\begin{tikzpicture}[level distance=1.4cm, grow=down]

					\node {Decoded Codeword $\mathbf{\hat C}$} 
            		child{
            			node {\begin{tabular}{c} $\mathbf{\hat C H^\top = 0}$ \\ Parity Check Pass \\ $\mathbf{\hat C}$ Valid \end{tabular}}            				
            				child{
            					node {CRC Pass}
            					child{
            						node {\begin{tabular}{c} $\mathbf{\hat S = S}$ \\ No Error \end{tabular}}
            					}
            					child{
            						node {\begin{tabular}{c}$\mathbf{\hat S \neq S}$ \\ Undetected \\ Error \end{tabular}}
            					}
            				} 
            				child{
            					node {CRC Fail}
            					child{
            						node {\begin{tabular}{c}$\mathbf{\hat S \neq S}$ \\ Detected \\ Error \end{tabular}}
            					}            					
            				}
            		}
            		child{
            			node {\begin{tabular}{c} $\mathbf{\hat C H^\top \neq 0}$ \\ Parity Check Fail \\ $\mathbf{\hat C}$ Error \end{tabular}}   
            			child{
            				node {\begin{tabular}{c} Skip CRC \end{tabular}}
             				child{
            					node {\begin{tabular}{c} $\mathbf{\hat S \neq S}$ \\ Detected \\ Error \end{tabular}}
            				}           				
            			}
            		};

\end{tikzpicture}

}

\caption{Possible decoding scenarios and error detection techniques.}
\label{fig:PerrorTree}
\end{figure}
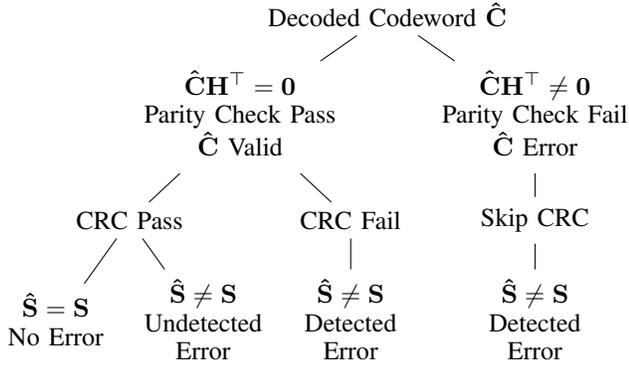

In reverse reconciliation, Alice attempts to construct a decoded estimate $\mathbf{\hat S}$ of Bob's original message $\mathbf{S}$ in order to perform privacy amplification and build a secret key. The tree diagram in Fig.~\ref{fig:PerrorTree} highlights four possible decoding scenarios for generating $\mathbf{\hat S}$ from Alice's decoded codeword $\mathbf{\hat C}$.

After LDPC decoding, Alice performs a parity check $\mathbf{\hat C H^\top}$ to validate that her decoded codeword $\mathbf{\hat C}$ is valid. When the parity check fails, i.e. $\mathbf{\hat C H^\top \neq 0}$, Alice knows that a decoding error has occurred and the frame is discarded since it can not be used to generate a secret key. However, when the parity check passes, i.e. $\mathbf{\hat C H^\top = 0}$, Alice knows that $\mathbf{\hat C}$ is a valid codeword, however, she does not yet know if $\mathbf{\hat C}$ is equal to Bob's original encoded codeword $\mathbf{C}$. 

For any binary linear block code, the number of possible codewords is $2^k=2^{nR_\text{code}}$. Thus, for codes with a long block-length $n$, the number of possible codewords grows exponentially, and it is possible for the decoder to converge to a valid codeword where the decoded message is incorrect, i.e. $\mathbf{\hat S \neq S}$. In coding theory, this is referred to as an undetected error. This scenario is problematic for secret key generation where both parties must share the same message after decoding in order to perform universal hashing in the next privacy amplification step. 

\begin{figure}[t!]
\centering
\includegraphics[trim=0.0in 0in 0in 0in, width=0.48\textwidth]{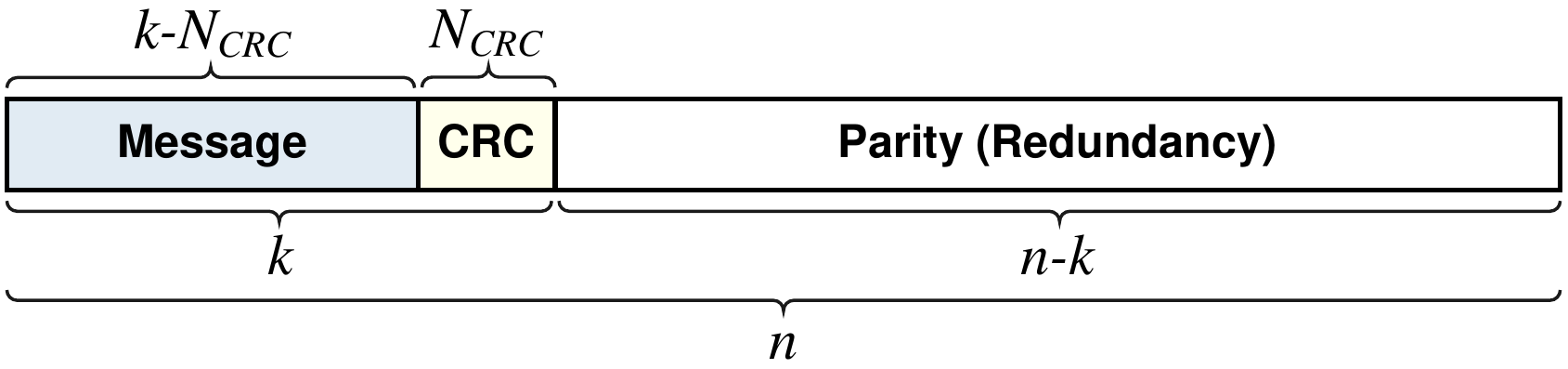}
\caption{LDPC frame components: message, CRC, and parity bits.}
\label{fig:LinearBlockCode}
\end{figure}

In order to detect invalid decoding errors when $\mathbf{C \hat H^\top=0}$, a cyclic redundancy check (CRC) of Bob's original message $\mathbf{S}$ can be transmitted as part of the frame, and then verified against the computed CRC of Alice's decoded message $\mathbf{\hat S}$. Figure~\ref{fig:LinearBlockCode} presents the components of an LDPC-encoded frame, where $k$ information bits are comprised of $(k-N_\text{CRC})$ message bits and $N_\text{CRC}$ CRC bits, followed by $(n-k)$ parity bits to be discarded after LDPC decoding. If the CRC results of $\mathbf{S}$ and $\mathbf{\hat S}$ are equal, then the decoding is successful and $\mathbf{\hat S}$ can be used to distill a secret key, otherwise Alice knows that a decoding error has occurred and $\mathbf{\hat S}$ is discarded. The CRC needs to be performed only when the parity check passes, otherwise the frame is known to contain an error and the CRC is skipped. A truly undetected error occurs when both the parity check and CRC pass, but $\mathbf{\hat S \neq S}$. 

A frame error is said to have occurred when $\mathbf{\hat S} \neq \mathbf{S}$, i.e. when the decoding fails to reproduce the original message. Both detected and undetected errors contribute to the overall FER. The probability of frame error is defined as follows:
\begin{equation}
P_\text{e} = P_\text{detected error} + P_\text{undetected error}.
\label{eq:Pe}
\end{equation}
From Fig.~\ref{fig:PerrorTree}, it follows then that the detected and undetected error probabilities are defined as
\begin{equation}
%\resizebox{240pt}{!}{$
P_\text{detected error} = P(\mathbf{\hat C H^\top \neq 0}) + P(\mathbf{\hat C H^\top = 0} \ \cap \ \text{CRC Fail})
%$}
\nonumber
\end{equation}
\begin{equation}
%\resizebox{240pt}{!}{$
P_\text{undetected error} = P(\mathbf{\hat C H^\top = 0} \ \cap \ \text{CRC Pass} \ \cap \ \mathbf{\hat S \neq S}).
%$}
\nonumber
\end{equation}

There exists a rare case not shown in Fig.~\ref{fig:PerrorTree}, where the parity check passes and CRC fails, yet $\mathbf{\hat S = S}$. Although the decoded message is correct, it will be discarded by the decoder due to the failed CRC check. As a result, there is a rare chance that this frame will be lost and the secret key rate will be reduced. However, this case is not considered by convention in communication theory.

\subsection{Impact of Reconciliation Error and Efficiency on Key Rate}

The remainder of this work investigates the trade-offs in error-correction performance, reconciliation efficiency, reconciliation distance, and secret key rate, by assuming that the physical parameters of the quantum channel are fixed, and that Alice's modulation variance $V_{A}$ has been optimally set for each transmission distance and desired $\beta$-efficiency. In practice, the asymptotic secret key rate $K_\text{opt}$ is scaled by the FER since decoded frames with known error can not be used to generate a secret key and must therefore be discarded. As such, the effective secret key rate of a practical CV-QKD system is given by
\begin{equation}
\resizebox{227pt}{!}{$
K_{\text{eff}}=\Big(1-P_{\text{detected error}}\Big) \Big((1-P_{\text{undetected error}})\beta I_{AB} - \chi_{BE}\Big).
$}
\label{eq:Keff}
\end{equation}
Alice and Bob can discard frames with detected error, while frames with undetected error further reduce the mutual information $I_{AB}$ between Alice and Bob. In Section~\ref{sec:LDPCforRecon}, it is empirically shown that $P_{\text{undetected error}}=0$ using a 32-bit CRC code, thus the total decoding FER can be expressed more simply as $P_e = P_\text{detected error}$. This simplified expression for the FER is assumed for the remainder of this work, and thus the effective secret key rate expression given by Eq.~\ref{eq:Keff} can be reduced to
\begin{equation}
K_\text{eff} = (1-P_\text{e})(\beta I_{AB} - \chi_{BE}).
\label{eq:Keffsimple}
\end{equation}

Up until this point, the $\beta$-efficiency has been assumed to be independent of the reconciliation algorithm, however, as shown in Eq.~\ref{eq:Keff}, the effective secret key rate $K_\text{eff}$ is dependent on both $\beta$ and FER. Given the set of optimal $V_{A}$ values and assuming that the physical operating parameters of the quantum channel remain constant, the virtual BIAWGNC channel can be induced and described solely in terms of the SNR at a particular distance with an effective secret key rate $K_\text{eff}$. As described further in Section~\ref{sec:LDPCforRecon}, there exists a trade-off between reconciliation distance and effective secret key rate, such that for a single SNR, one of the following two operating conditions is possible: (1) long  distance with a low secret key rate, or (2) short distance with a high secret key rate. In fact, for a fixed LDPC code rate $R_\text{code}$, the SNR only depends on the reconciliation efficiency and is independent of transmission distance. From Eq.~\ref{eq:beta-def}, the SNR of a virtual BIAWGNC can be expressed as a function of $\beta$ such that
\begin{equation}
\label{eq:snr(beta)}
s(\beta)=2^{2R_\text{code}/ \beta}-1.
\end{equation}
From a code design perspective then, an optimal rate $R_\text{code}$ LDPC code can be designed to achieve a target FER at a particular SNR. Since Alice and Bob remain stationary once deployed in the field, their transmission distance remains fixed, and thus an optimal LDPC code can be designed to achieve the maximum operating secret key rate over a range of distances by providing the optimal trade-off between $\beta$ and FER.

\subsection{Secret Key Rate with Finite-Size Effects}

The security of the CV-QKD protocol must account for the finite length of the secret key, which is generated via universal hashing in the privacy amplification step using a block of length of $N_\text{privacy}$ bits. Alice constructs her privacy amplification block from her correctly decoded $\mathbf{\hat S}$ messages, while Bob constructs his privacy amplification block from his original corresponding $\mathbf{S}$ messages. Due to the finite block size, the secret key rate is reduced by an offset coefficient $\Delta(N_\text{privacy})$ and scaling coefficient $N_\text{privacy}/N_\text{quantum}$, where $N_\text{quantum}$ is the number of symbols sent from Alice to Bob during the first quantum transmission step. The secret key rate, accounting for finite-size effects, is given by
\begin{equation}
\resizebox{226pt}{!}{$
K_\text{finite} = \bigg(\dfrac{N_\text{privacy}}{N_\text{quantum}}\bigg)\bigg(1-P_e\bigg)\bigg(\beta I_{AB} - \chi_{BE} - \Delta(N_\text{privacy}) \bigg).
$}
\label{eq:Kfinite}
\end{equation}
Leverrier~et~al.\ showed that $N_\text{quantum}$ can be arbitrarily chosen as $N_\text{quantum}=2N_\text{privacy}$~\cite{Leverrier2010}, and that when $N_\text{privacy}>10^4$, the finite-size offset factor $\Delta(N_\text{privacy})$ can be approximated as 
\begin{equation}
\Delta(N_\text{privacy}) \approx 7 \sqrt{\dfrac{\log_2(2/\epsilon)}{N_\text{privacy}}},
\end{equation}
where a conservative choice for the security parameter is $\epsilon=10^{-10}$~\cite{Leverrier2010}. The LDPC block length $n$ is not directly included in this expression, however, the LDPC block length does affect the reconciliation efficiency $\beta$ and FER $P_e$. The impact of $\beta$ and $P_e$ on reconciliation distance is discussed in greater detail in the next section of this paper.

The next section presents a study of the optimal block size $N_\text{privacy}$ for privacy amplification, as well as an overview of the design and application of LDPC codes for reverse reconciliation in long-distance CV-QKD systems on the virtual BIAWGNC.

\section{LDPC Codes for Reconciliation}
\label{sec:LDPCforRecon}

The design of efficient reconciliation algorithms is one of the central challenges of long-distance CV-QKD~\cite{Jouguet2011}. Early reconciliation algorithms failed to achieve efficiencies above 80$\%$~\cite{Assche2004}, while more advanced algorithms that now achieve 95$\%$ efficiency suffer from computational complexity~\cite{Jouguet2013, Jouguet2014}. LDPC codes are highly suitable for low-SNR reconciliation in CV-QKD due to their near-Shannon limit error-correction performance and absence of patent licensing fees~\cite{Richardson2001_Shokrollahi}. However, designing efficient LDPC codes with block lengths on the order of 10$^{6}$ bits remains a challenge.

This section introduces a complexity-reduction technique for LDPC code design that has been widely adopted in hardware-based LDPC decoders, namely through the implementation of architecture-aware codes~\cite{Mansour2003}. A popular class of such codes are quasi-cyclic codes, whose parity-check matrices are constructed from an array of cyclically-shifted identity matrices that provide a sufficient degree of randomness and enable computational decoding speedup as a result of their highly parallelizable structure, which provides a simple mapping to hardware~\cite{Mansour2003, Fossorier2004}. Previous independent works by Martinez-Mateo and Walenta have explored the application of existing QC-LDPC codes from the IEEE 802.11n standard for DV-QKD, however, these works were not able to demonstrate reliable reconciliation beyond 50km~\cite{Martinez-Mateo2013, Walenta2014}. While this distance may have been a limitation of DV-QKD, the short block lengths of such existing QC-LDPC codes (on the order of 10$^{3}$ bits) remain unsuitable for long-distance CV-QKD. Recently, Bai et al. theoretically showed that rate 0.12 QC codes with block lengths of 10$^6$ bits can be constructed using progressive edge growth techniques, or by applying a QC extension to random LDPC codes with block lengths of 10$^5$ bits~\cite{Bai2017}. However, the reported QC codes target an SNR of -1dB, and are thus not suitable for long-distance CV-QKD beyond 100km. At the time of writing, there has not been any reported investigation of the construction of QC codes for multi-edge LDPC codes targeting low-SNR channels below -15dB for long-distance CV-QKD.  This work shows that by applying the structured QC-LDPC code construction technique to the random multi-edge LDPC codes previously explored by Jouguet~et~al.\ for long-distance CV-QKD~\cite{Jouguet2011}, it is possible to construct codes that achieve sufficient error-correction performance while enabling the acceleration of the computationally-intensive LDPC decoding algorithm such that the reconciliation step is no longer the bottleneck for secret key distillation beyond 100km.

\begin{figure}[t!]
\centering
\includegraphics[trim=0.
0in 0in 0in 0in, width=0.48\textwidth]{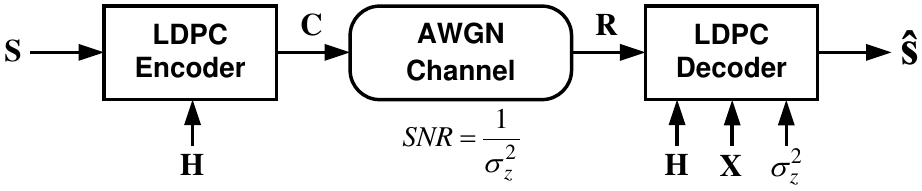}
\caption{Simplified model of BIAWGNC with LDPC encoding and decoding.}
\label{fig:LDPCchannelmodel}
\end{figure}

As previously described, there exists a trade-off between reconciliation distance and maximum secret key rate for a given $\beta$-efficiency. Once the target FER and operating SNR are known, an optimal LDPC code, i.e.\ parity-check matrix $\mathbf{H}$, can be designed independent of other CV-QKD system parameters. The reverse reconciliation problem can thus be reduced to the simpler model shown in Fig.~\ref{fig:LDPCchannelmodel} as a result of the BIAWGNC approximation at low SNR. The variables shown in Fig.~\ref{fig:LDPCchannelmodel} are the same variables described in Section \ref{sec:Background}, where $\mathbf{S}$ is Bob's random binary sequence, $\mathbf{C}$ is Bob's binary LDPC-encoded codeword, the Gaussian channel is described by $Z \sim \mathcal{N}(0,\sigma_{Z}^2)$, $\mathbf{R}$ is Alice's received channel value, $\mathbf{X}$ is Alice's correlated Gaussian sequence, and $\mathbf{\hat S}$ is Alice's LDPC-decoded estimate of Bob's original random binary sequence. Alice and Bob also share a predefined parity-check matrix $\mathbf{H}$ for encoding and decoding.

This work demonstrates the application of multi-edge QC-LDPC codes for long-distance CV-QKD through the design of several rate 0.02 binary parity-check matrices with block lengths on the order of 10$^{6}$ bits. While a complete QKD system would offer multi-rate code programmability for various operating channels, this work focuses on the design of a single, low-rate code for a large range of transmission distances to fully study the effects of $\beta$-efficiency and FER on the maximum achievable secret key rate and reconciliation distance. While some works have explored the use of rate-adaptive or repetition codes to achieve high-efficiency decoding with multiple code rates~\cite{Jouguet2011}, the exploration of multi-rate code design for long-distance CV-QKD is beyond the scope of this work. The remainder of this section describes the construction of LDPC codes, the belief propagation decoding algorithm, the error-correction performance of the designed rate 0.02 codes, and the achievable secret key rates for multiple $\beta$-efficiencies beyond 100km.

\subsection{General Construction of LDPC Codes}

\begin{figure}[htbp]
\centering
\includegraphics[trim=0.
0in 0in 0in 0in, width=0.48\textwidth]{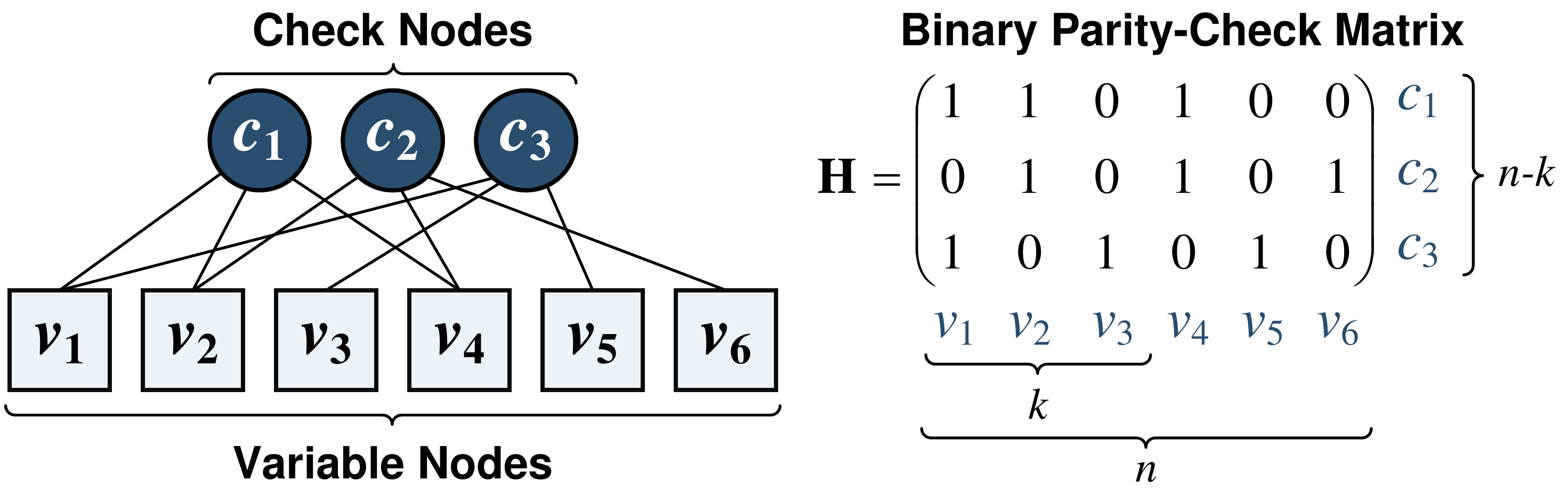}
\caption{LDPC Tanner graph and corresponding binary parity-check matrix.}
\label{fig:LDPCtannermatrix}
\end{figure}

LDPC codes are a class of linear block codes defined by a sparse parity-check matrix $\mathbf{H}$ of size $(n-k)\times n,\,k\leq n$, and code rate $R_\text{code} = k/n$~\cite{Fossorier1999}. Given $\mathbf{H}$, an equivalent definition of the code is given by its Tanner graph~\cite{Tanner1981}. A Tanner graph $\mathcal{G}$ is a bipartite graph with two independent vertex sets, known as check nodes (CNs) and variable nodes (VNs), which correspond to the rows and columns of $\mathbf{H}$, respectively. As shown in Fig.~\ref{fig:LDPCtannermatrix}, an edge between VN $v_i$ and CN $c_j$ belongs to $\mathcal{G}$ if and only if $\mathbf{H}(j,i)=1$. 

An LDPC code of length $n$ is fully specified by the number of variable and check nodes, and their respective degree distributions. The number of edges connected to a vertex in $\mathcal{G}$ is called the degree of the vertex. The degree distribution of $\mathcal{G}$ is a pair of polynomials $\omega(x)=\sum_i\omega_ix^i$ and $\psi(x)=\sum_i\psi_ix^i$, where $\omega_i$ and $\psi_i$ respectively denote the number of variable and check nodes of degree $i$ in $\mathcal{G}$. The performance of tree-like Tanner graphs can be analyzed using a technique called density evolution~\cite{Richardson2001}. As $n\rightarrow\infty$, the error-correction performance of Tanner graphs with the same degree distribution is nearly identical~\cite{Richardson2001}. Hence, the variable and check node degree distributions can be normalized to $\Omega(x)=\sum_i\frac{\omega_i}{n}x^i$ and $\Psi(x)=\sum_i\frac{\psi_i}{n-k}x^i$, respectively. 

The design of binary LDPC codes of rate $R_\text{code}$ and block length $n$ consists of a two-step process. First, find the normalized degree distribution pair $(\Omega(x),\Psi(x))$ of rate $R_\text{code}$ with the best performance. Then, if $n$ is large, randomly sample a Tanner graph $\mathcal{G}$ that satisfies the degree distribution defined by $\omega(x)$ and $\psi(x)$ (up to rounding error), and find the corresponding parity-check matrix $\mathbf{H}$. Unfortunately, this method is non-trivial in the design of low-rate codes that approach Shannon capacity at low SNR. 

\subsection{Multi-Edge LDPC Codes}

Multi-edge LDPC codes, first introduced by Richardson and Urbanke, provide two advantages over standard LDPC codes: (1) near-Shannon capacity error-correction performance for low-rate codes, and (2) low error-floor performance for high-rate codes~\cite{Richardson2002}. The latter is not a significant concern for long-distance CV-QKD where the reconciliation FER is on the order of $10^{-1}$, however, the design of a high-performance low-rate code is crucial to achieving high $\beta$-efficiency~\cite{Jouguet2011}.  

The multi-edge framework can be applied to both regular and irregular LDPC codes with uniform and non-uniform vertex degree distributions, respectively, by introducing multiple edge types into the Tanner graph specifying the code~\cite{Richardson2002}. In a standard LDPC code, the polynomial degree distributions are limited to a single edge type, such that all variable and check nodes are statistically interchangeable. In order to improve performance, multi-edge LDPC codes extend the polynomial degree distributions to multiple independent edge types with an additional edge-type matching condition~\cite{Richardson2002}. 

To describe the design of multi-edge LDPC codes, let the potential connections of a variable or check node be called its sockets. Let the vector $\mathbf{d}=(d_1,d_2,\dots,d_t)$ be a multi-edge node degree of $t$ types. A node of degree $\mathbf{d}$ has $d_1$ sockets of type 1, $d_2$ sockets of type 2, etc. When generating a Tanner graph, only sockets of the same type can be connected by an edge of that type. Multi-edge normalized degree distributions are straightforward generalizations based on multi-edge degrees $\Omega(x_1,x_2,\dots,x_t) = \sum_{\mathbf{d}}\Omega_{\mathbf{d}}x_1^{d_1}x_2^{d_2}\cdots x_t^{d_t}$ and $\Psi(x_1,x_2,\dots,x_t) ~= \sum_{\mathbf{d}}\Psi_{\mathbf{d}}x_1^{d_1}x_2^{d_2}\cdots x_t^{d_t}$, where $\Omega_{d_1,d_2,\dots,d_t}$ and $\Psi_{d_1,d_2,\dots,d_t}$ are the respective fractions of variable and check nodes with $d_1$ edges of type 1, $d_2$ edges of type 2, etc. The rate of a multi-edge LDPC code is then defined as $R_\text{code} = \Omega(\mathbf{1}) - \Psi(\mathbf{1})$, where $\mathbf{1}$ denotes the all-ones vector with implied length~\cite{Richardson2002}.

The multi-edge LDPC code used in this work is rate 0.02 with normalized degree distribution
\begin{equation}
\Omega(x_1,x_2,x_3) = \frac{9}{400}x_1^{2}x_2^{57}x_3^{0} + \frac{7}{400}x_1^{3}x_2^{57}x_3^{0} + \frac{24}{25}x_1^{0}x_2^{0}x_3^{1}
\nonumber
\end{equation}
\begin{equation} 
\begin{split}
\Psi(x_1,x_2,x_3) = &\frac{3}{320}x_1^{3}x_2^{0}x_3^{0} + \frac{17}{1600}x_1^{7}x_2^{0}x_3^{0} \\&+ \frac{3}{5}x_1^{0}x_2^{2}x_3^{1} + \frac{9}{25}x_1^{0}x_2^{3}x_3^{1}.
\end{split}
\nonumber
\end{equation}
\noindent
This degree distribution was designed by Jouguet~et~al.\ by modifying a rate 1/10 multi-edge degree structure introduced by Richardson and Urbanke~\cite{Jouguet2011, Richardson2002}. Structurally, the sub-graph induced by edges of type 1 corresponds to a high-rate LDPC code with variable degree 3 and check degree 7 or 8. Parity checks are added to this code, each checking two or three variable nodes as specified by edges of type 2. The resulting parity bits are degree-1 variable nodes specified by edges of type 3. For the BIAWGNC, the minimum SNR for which the tree-like Tanner graph with this multi-edge degree distribution is error free is $2.863\times10^{-2}$ or -15.47dB~\cite{Jouguet2011}.

The LDPC parity-check matrices in this work were generated by randomly sampling Tanner graphs that satisfied the multi-edge degree distribution defined by $\omega(x)$ and $\psi(x)$, and the edge-type matching condition. The random sampling technique does not degrade code performance in this case, since the operating FER is known to be high ($P_\text{e} \approx 10^{-1}$). At such high FER, the error-floor phenomenon is not a significant concern as the code is strictly designed to operate in the waterfall region in order to achieve high $\beta$-efficiency~\cite{Richardson2003}. The rate 0.02 LDPC codes explored in this work target a block length of $n=1\times10^{6}$ bits in order to achieve near-Shannon capacity error-correction performance. As a result, the parity-check matrix $\mathbf{H}$ has dimensions $n-k=n(1-R_{code})=9.8\times10^{5}$ by $n=1\times10^{6}$. Due to the low code rate and large block length, the random parity-check matrix construction does introduce LDPC decoder implementation complexity, which directly affects decoding latency and maximum achievable secret key rate. The LDPC decoder implementation complexity for such a code can be reduced with minimal degradation in error-correction performance by imposing a quasi-cyclic structure to the parity-check matrix. 

\subsection{Quasi-Cyclic LDPC Codes}

While purely-random LDPC codes have been shown to achieve near-Shannon capacity error-correction performance under belief propagation decoding~\cite{Chung2001}, the hardware-based implementation of decoders for random codes is a challenge with large block lengths, especially on the order of $10^{6}$ bits. The bottleneck stems from the complex interconnect network between CN and VN processing units that execute the belief propagation algorithm~\cite{Mohsenin2010, Kim2011}. 

In a traditional LDPC decoder, variable and check node processing units iteratively exchange messages across an interconnect network described by a Tanner graph, as shown in Fig.~\ref{fig:LDPCtannermatrix}. Purely-random parity-check matrices introduce unstructured interconnect between the variable and check nodes, resulting in unordered memory access patterns and complex routing, which limit scalability in ASIC or FPGA implementations. Quasi-cyclic codes alleviate such decoder complexities by imposing a highly-regular matrix structure with a sufficient degree of randomness~\cite{Fossorier2004}. This work extends the design of low-rate, mutli-edge LDPC codes to QC codes for hardware realization.

\begin{figure}[t!]
\centering
\includegraphics[trim=0.
0in 0in 0in 0in, width=0.48\textwidth]{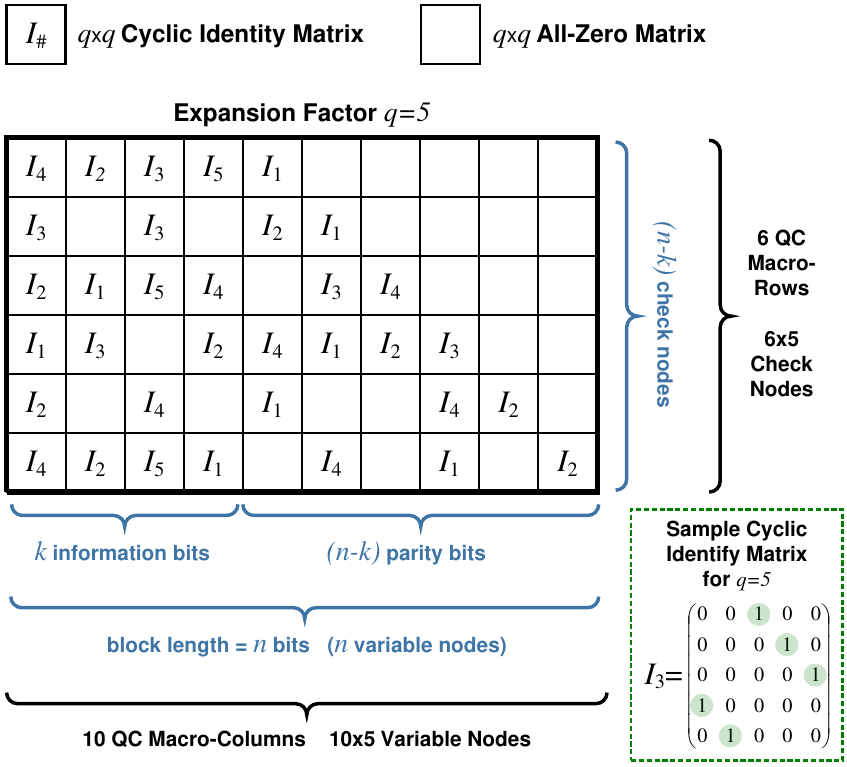}
\caption{Sample quasi-cyclic binary parity-check matrix for $q=5$ constructed from uniformly-sized ($q \times q$), cyclically-shifted identity matrices and all-zero matrices.}
\label{fig:ParityCheckMatrix}
\end{figure}

QC codes are defined by a parity-check matrix constructed from an array of $q\times q$ cyclically-shifted identity matrices or $q\times q$ zero matrices~\cite{Fossorier2004}. As shown in Fig.~\ref{fig:ParityCheckMatrix}, the tilings evenly divide the $(n-k)\times n$ parity-check matrix $\mathbf{H}$ into $n/q$ QC macro-columns and $(n-k)/q$ QC macro-rows. The expansion factor $q$ in a QC parity-check matrix determines the trade-off between decoder implementation complexity and error-correction performance. For small $q$, the parity-check matrix exhibits a high degree of randomness, which improves error-correction performance, while a large $q$ reduces decoder complexity with some performance degradation. 

\begin{figure}[htbp]
    \centering
    \begin{subfigure}[t]{0.225\textwidth}
       \centering
       \includegraphics[width=1\textwidth]{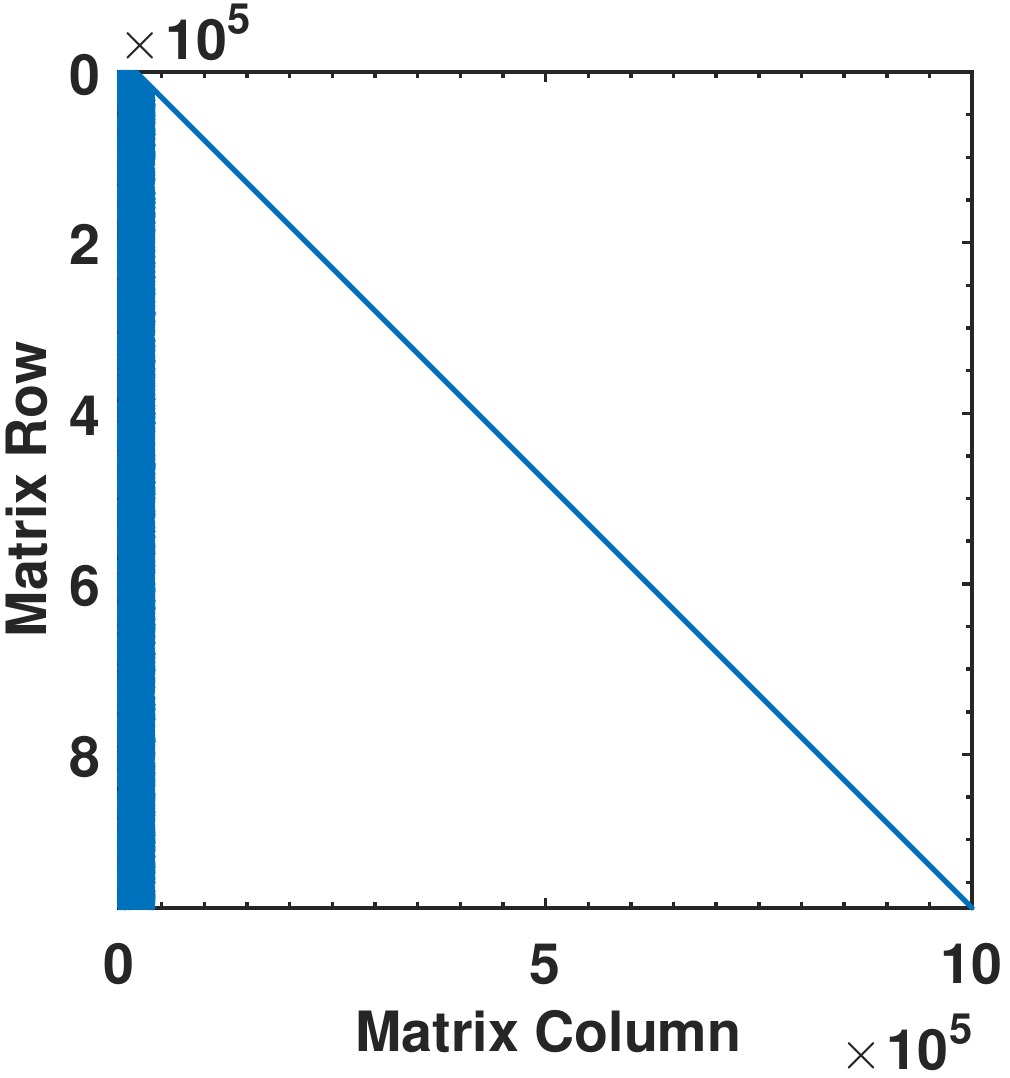}{}
       \captionsetup{font=small}      
       \caption{Full $q=50$ QC parity-check matrix structure with $1\times10^6$ columns and $9.8\times10^5$ rows. Empty space represents zeros.}
    \end{subfigure} \hfill   
    \begin{subfigure}[t]{0.225\textwidth}
        \centering
        \includegraphics[width=1\textwidth]{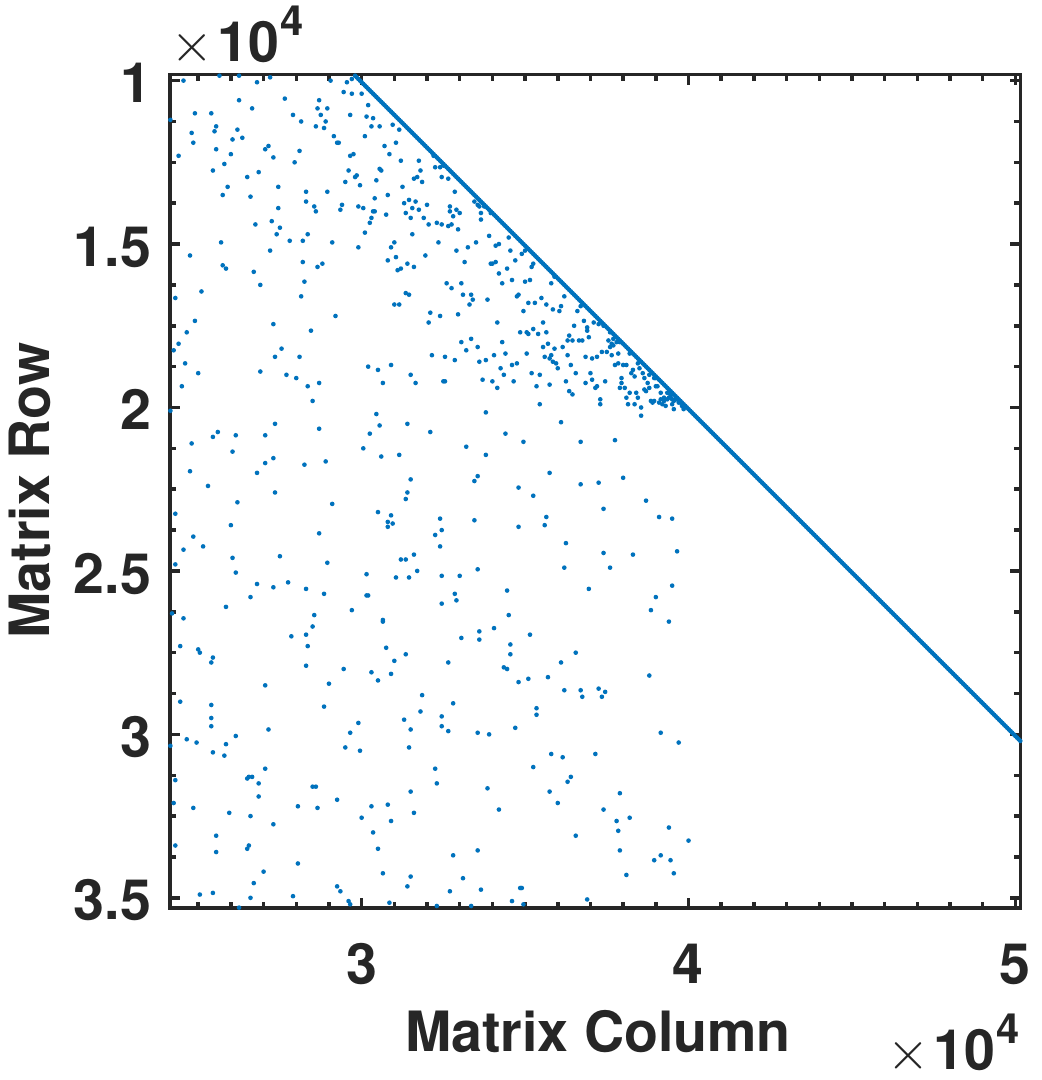}{}
        \captionsetup{font=small} 
        \caption{Zoom-in of top left corner of $q=50$ QC matrix shown in (a). Each dot represents a $50\times50$ cyclic identity matrix.}
    \end{subfigure}
    \caption{Structure of designed parity-check matrices.}
    \label{fig:spy_plots}
\end{figure}

To design a multi-edge QC-LDPC code, repeat the random multi-edge sampling process using $n/q$ as the block length instead of $n$ to obtain a base Tanner graph $\mathcal{G}_B$. The base parity-check matrix $\mathbf{H}_B$ is obtained from $\mathcal{G}_B$ by populating each non-zero entry by a random element of the set $\{1,2,\dots,q\}$. Let $I_i$ be the circulant permutation matrix obtained by cyclically shifting each row of the $q\times q$ identity matrix to the right by $i-1$. The QC parity-check matrix $\mathbf{H}$ is obtained from $\mathbf{H}_B$ by replacing each non-zero entry of value $i$ by $I_i$, and each zero entry by the $q\times q$ all-zeros matrix. 

In this work, multi-edge QC-LDPC parity-check matrices of rate 0.02 were generated for expansion factors $q\in\{21,50,100,500,1000\}$. Under belief propagation decoding, the error-correction performance of the $q\in\{100,500,1000\}$ QC codes was significantly worse in comparison to a random multi-edge code with the same degree distribution. Thus, only the $q\in\{21,50\}$ codes are presented in the remainder of this study. In order to maintain the same degree distributions, the block length for the $q=21$ code with rate $R_\text{code}=0.02$ was adjusted to $n=1.008\times 10^{6}$ bits. Similarly, the $q=50$ code has a block length of $n=1\times 10^{6}$ bits and rate $R_\text{code}=0.01995$.

Figure~\ref{fig:spy_plots} shows the structure of the parity-check matrices designed in this work. Both the purely-random and QC matrices have a similar structure, which contains a dense area of 1s or cyclic identity matrices on the left, and a long diagonal of degree-1 VNs to the right. The starting point of the diagonal is determined by the VN degree distribution of the multi-edge matrix. In the case of the QC codes, no cyclic shifts are implemented along the diagonal, thus all submatrices are $q\times q$ $I_1$ identity matrices. This matrix structure greatly improves the decoding speed as degree-1 VNs along the diagonal need to pass VN-to-CN messages only in the first decoding iteration, while CN-to-VN messages need to be passed to degree-1 VNs only if the early-termination condition is enabled. The degree-1 VNs along the diagonal correspond to the majority (but not all) of the $(n-k)$ parity bits that are discarded after decoding, thus the VN update computation needs to be performed in these degree-1 VNs only if a decision needs to be made when early termination is enabled. A small fraction of the $(n-k)$ parity bits correspond to VNs with more than one CN connection in the denser area of the matrix to the left of the diagonal. These VNs must perform the VN update computation in each iteration along with the first $k$ VNs, which correspond to the $k$ information bits of the block. 

The parity component of $\mathbf{H}$, $i>k$ in $\mathbf{H}(j,i)$, is lower-triangular for both the purely-random and QC parity-check matrices designed in this study. An example of this type of construction is shown in Fig.~\ref{fig:ParityCheckMatrix}, and is also illustrated in Fig.~\ref{fig:spy_plots}.  While the lower-triangular construction does not necessarily impact decoding complexity or error-correction performance, it does simplify the LDPC encoding procedure, which can be performed via forward substitution if $\mathbf{H}$ is of this form. Further investigation of LDPC encoding complexity for such large codes is beyond the scope of this work.

\subsection{LDPC Decoding: Belief Propagation Algorithm}

LDPC decoding is performed using belief propagation, an iterative message-passing algorithm commonly used to perform inference on graphical models such as factor graphs~\cite{Kschischang2001}. In the context of LDPC, the decoding procedure attempts to converge on a valid codeword by iteratively exchanging probabilistic updates between variable and check nodes along the edges of the Tanner graph until the parity-check condition is satisfied or the maximum number of iterations is reached. The Sum-Product algorithm is the most common variant of belief propagation~\cite{Kschischang2001}, and is described in Algorithm~\ref{SPA} for $d=1$ dimensional reconciliation on a BIAWGNC. 

\begin{algorithm}[htbp]
\caption{Sum-Product algorithm for $d=1$ scheme}
\label{SPA}
\textbf{Input:} $\mathbf{X}, \mathbf{R}, \sigma_{Z}^2$; \hspace{3mm} \textbf{Output:} $\mathbf{\hat S}$
\begin{algorithmic}[0]
	\STATE \textbf{Step 1:} LLR initialization at each VN, $v=1,2,\dots,n$
	\STATE $\sigma_{N_{v}}^2 \gets \dfrac{\sigma_{Z}^2}{||X_{v}||}, \ Q_{v} \gets \ln\bigg({ \dfrac{P(C_{v}=0|R_{v})}{P(C_{v}=1|R_{v})} }\bigg) = \dfrac{2R_{v}}{\sigma_{N_{v}}^2}$
	\STATE $L_{vc}^{(t=0)} \gets Q_{v}$,\ \ $\forall c \in M(v)$ for first iteration 
 	\FOR {Iteration $t=1$ to Max Iterations}
		\STATE \textbf{Step 2:} Check node update (CN-to-VN messages)\\	
	$\operatorname{sgn}(m_{cv}^{(t)}) \gets \prod_{v' \in N(c) \backslash v} \operatorname{sgn}(L_{v'c}^{(t-1)})$\\	
	$\big|m_{cv}^{(t)}\big| \gets \Phi^{-1} \bigg(\sum_{v' \in N(c) \backslash v} \Phi\big(\big|L_{v'c}^{(t-1)}\big|\big) \bigg)$\\	
	$m_{cv}^{(t)} \gets \operatorname{sgn}(m_{cv}^{(t)}) \times \big|m_{cv}^{(t)}\big|$	
		\STATE \textbf{Step 3:} Variable node update (VN-to-CN messages)\\
	$L_{vc}^{(t)} \gets Q_{v} + \sum_{c' \in M(v) \backslash c} m_{c'v}^{(t)}$\\
	$L_{v}^{(t)} \gets Q_{v} + \sum_{c \in M(v) } m_{cv}^{(t)}$
		\STATE \textbf{Step 4:} Hard decision and early termination check\\
	$\hat C_{v}^{(t)} \gets \begin{cases} 0,&L_{v}^{(t)} \geq 0\\ 1,&\text{otherwise} \end{cases}$	
		\STATE \textbf{if} $\mathbf{\hat C  H^\top=0}$ (mod 2) \textbf{then} Go to \textbf{Step 5}
 	\ENDFOR
	\STATE \textbf{Step 5:} Discard parity bits $\ \hat S_{v} \gets \hat C_{v}, \ v \leq k$
\end{algorithmic}
\end{algorithm}

In Algorithm~\ref{SPA}, Step 1 prepares the $Q_v$ log-likelihood ratio (LLR) input values at each variable node based on Alice's correlated Gaussian sequence $\mathbf{X}$ and the channel noise variance $\sigma_Z^2$. All VN-to-CN messages from VN $v$ are initialized to the received channel LLR $Q_v$ before the first message-passing iteration. As previously discussed in Section~\ref{sec:Background}, the expression for the channel noise variance $\sigma_{N_{v}}^2$ is different for $d=2,4,8$ reconciliation schemes, but the remainder of the algorithm is the same. Steps 2 to 5 specify the message-passing interaction between the CNs and VNs until the codeword syndrome defined by $\mathbf{\hat C H^\top}$ is equal to zero, or the maximum predetermined number of decoding iterations is reached. In Step 2, $m_{cv}^{(t)}$ is the message from CN $c$ to VN $v$ in iteration $t$, and $\Phi(x) = \Phi^{-1}(x) = -\ln(\tanh(x/2))$. In Step 3, $L_{vc}^{(t)}$ is the message from VN $v$ to CN $c$, and $L_{v}^{(t)}$ is the updated LLR belief of bit $v$ in the frame, whose decision is given by $\hat C_{v}^{(t)}$ in Step 4. In Step 2, the set of VNs connected to CN $c$ is defined as $N(c)=\{v|v\in\{1,2,\dots,n\}\wedge H_{vc}=1 \}$, where the notation $v' \in N(c) \backslash v$ refers to all VNs in the set $N(c)$ excluding VN $v$. Similarly, in Step 3, the set of CNs connected to VN $v$ is defined as $M(v)=\{c|c\in\{1,2,\dots,n-k\}\wedge H_{vc}=1 \}$, where $c' \in M(v) \backslash c$ refers to all CNs in the set $M(v)$ excluding CN $c$. As previously described, LDPC decoding is successful if the decoded message $\mathbf{\hat S}$ is equal to the original message $\mathbf{S}$. 

Due to the non-linearity of the $\tanh(x)$ function in the Sum-Product algorithm, most hardware-based LDPC decoders instead implement variants of the Min-Sum algorithm~\cite{Mohsenin2010, Kim2011}, which provides an acceptable approximation to Sum-Product decoding without the need for complex lookup tables. Despite the benefit of computational speedup, Min-Sum does not perform well at low SNR~\cite{Anastasopoulos2001}, and is thus not suitable for long-distance CV-QKD. The results presented in the remainder of this work were achieved using Sum-Product decoding.

\subsection{Error-Correction Performance of Multi-Edge QC Codes}

\begin{figure}[t!]
\centering
\includegraphics[trim=0in 0in 0in 0in, width=0.48\textwidth]{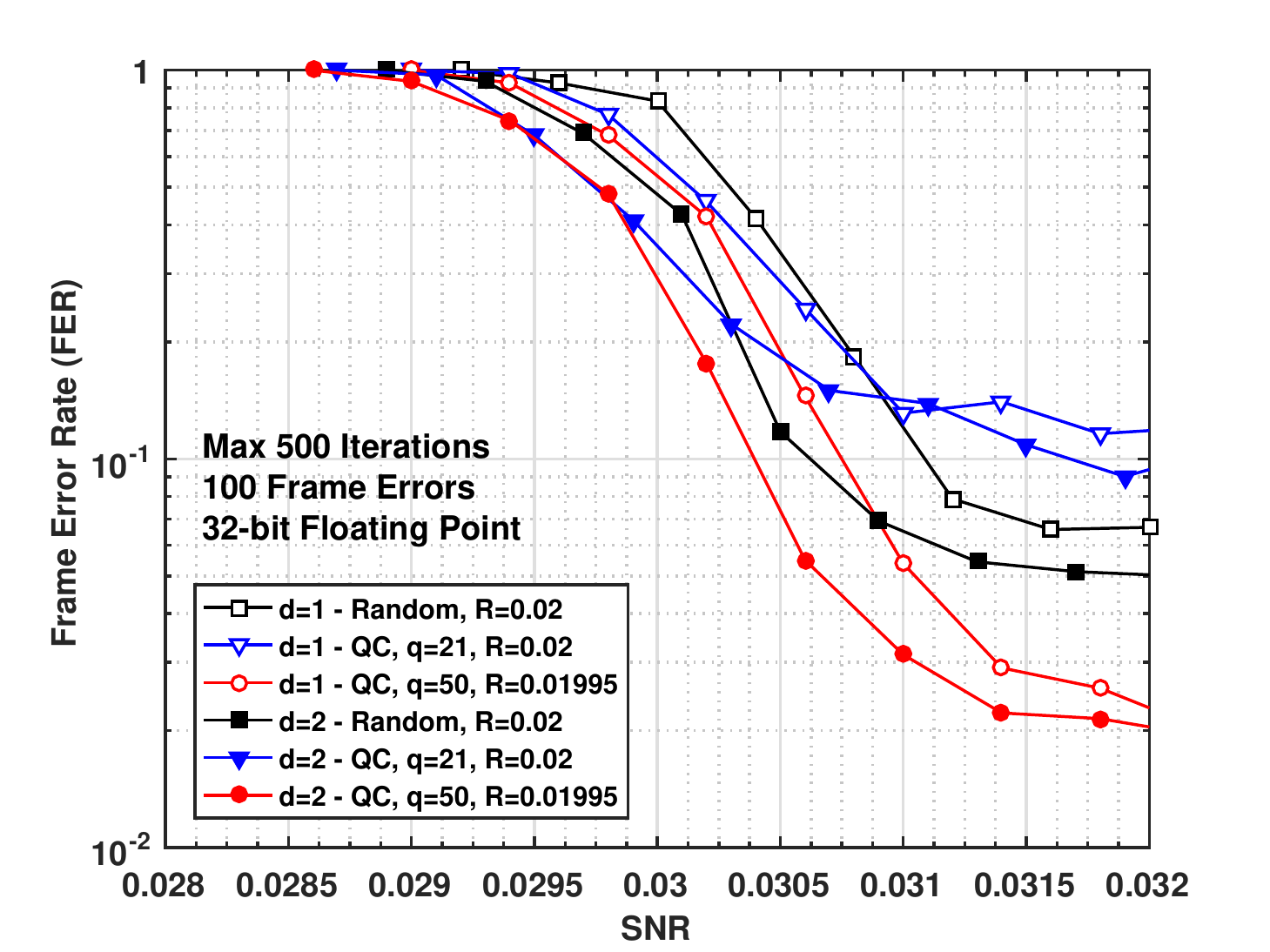}
\caption{FER vs. SNR for Sum-Product decoding with $d=1$ and $d=2$ dimensional reconciliation on BIAWGNC.}
\label{fig:FERvsSNR-1D}
\end{figure}

\begin{figure}[t!]
\centering
\includegraphics[trim=0in 0in 0in 0in, width=0.48\textwidth]{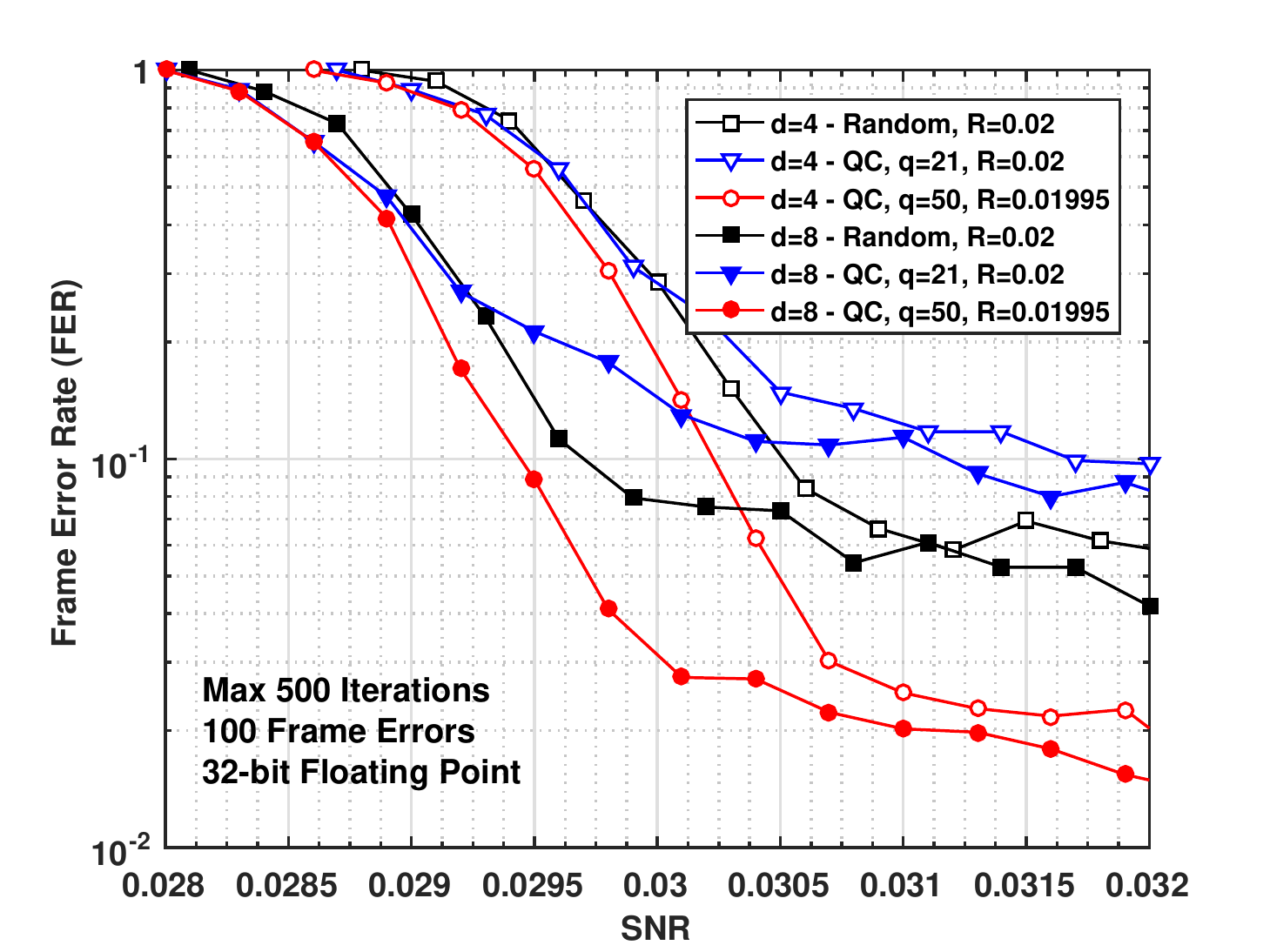}
\caption{FER vs. SNR for Sum-Product decoding with $d=4$ and $d=8$ dimensional reconciliation on BIAWGNC.}
\label{fig:FERvsSNR-8D}
\end{figure}

The multi-edge LDPC codes designed in this work achieve similar FER performance on the BIAWGNC compared to those developed by Jouguet~et~al.\ for long-distance CV-QKD with multi-dimensional reconciliation~\cite{Jouguet2011}. Table~\ref{table:matrices} summarizes the parameters of the three codes designed in this work, and Figures \ref{fig:FERvsSNR-1D} and \ref{fig:FERvsSNR-8D} present their FER vs. SNR error-correction performance under Sum-Product decoding for $d=1,2,4,8$ reconciliation dimensions. FER simulations were performed for the complete linear SNR range corresponding to the range of efficiencies between $\beta = 0.8$ and $\beta = 0.99$, as defined by Eq.~\ref{eq:snr(beta)}. This range of $\beta$-efficiency values was chosen to illustrate the trade-off between distance and finite secret key rate. For clarity, however, Figures \ref{fig:FERvsSNR-1D} and \ref{fig:FERvsSNR-8D} present the FER results only for the SNR range corresponding to $\beta$-efficiencies between $\beta = 0.88$ and $\beta = 0.99$.

\begin{center}
\begin{table}[t!]
%\caption{Multi-Edge LDPC Codes}
\caption{Designed rate 0.02 multi-edge LDPC codes}
\label{table:matrices}
\scalebox{0.88}{
    \begin{tabular}{c c c c}
    %\toprule[1.2pt]    
    \Xhline{3\arrayrulewidth} 
    \multirow{2}{*}{\bf Structure} & \bf Expansion & \bf Code Rate & \bf Block Length (Bits) \\
	& \bf Factor $q$ & $R_\text{code}$ & $n$ \\   \hline
	\Xhline{3\arrayrulewidth} 
	Random & N/A & 0.02 & $1 \times 10^6$ \\    
	Quasi-Cyclic & 21 & 0.02 & $1.008 \times 10^6$ \\    
	Quasi-Cyclic & 50 & 0.01995 & $1 \times 10^6$ \\
	%\bottomrule[1.2pt]
	\Xhline{3\arrayrulewidth}   \\
    \end{tabular}
}
\end{table}
\end{center}

Despite their identical degree distributions, the $q=50$ QC code achieves the best overall FER performance over $d=1,2,4,8$ dimensions in comparison to the purely-random and $q=21$ QC codes, due to its slightly lower code rate. At low SNR where $\beta$ is high, the $q=21$ QC code also performs better than the purely-random code over all dimensions, likely due to the longer block length. At higher SNR though, the purely-random code achieves a lower error-floor than the $q=21$ QC code due to higher randomness in the parity-check matrix. 

In general, the $d=2$, $d=4$, and $d=8$ reconciliation schemes achieve approximately 0.04dB, 0.08dB, and 0.2dB of coding gain, respectively, over the $d=1$ scheme in the waterfall region for all three codes. As previously mentioned, FER performance in the waterfall region is of particular interest for long-distance CV-QKD since it corresponds to the high $\beta$-efficiency region of operation at low SNR close to the Shannon limit. The error-floor region beyond the waterfall is not of practical use in CV-QKD as it corresponds to the low $\beta$-efficiency region where transmission distance is limited.

As previously discussed in Section~\ref{sec:Background}, for any binary linear block code, the number of possible codewords is $2^k=2^{nR_\text{code}}$. In this case, when $n=1 \times 10^6$ bits and $R_\text{code}=0.02$, the number of possible valid codewords for the decoder to choose from is approximately $4 \times 10^{6020}$. In order to detect invalid decoding errors when the parity check $\mathbf{\hat C H^\top = 0}$ but $\mathbf{\hat S \neq S}$, a 32-bit CRC code is included in each LDPC frame. In this work, $N_{CRC}=32$ bits were sufficient to detect all invalid decoded messages without sacrificing information throughput. Having full control of the simulation environment, it was also empirically found that $P_\text{undetected error}=0$ using a 32-bit CRC code. 

\begin{figure}[t!]
\centering
\includegraphics[trim=0in 0in 0in 0in, width=0.48\textwidth]{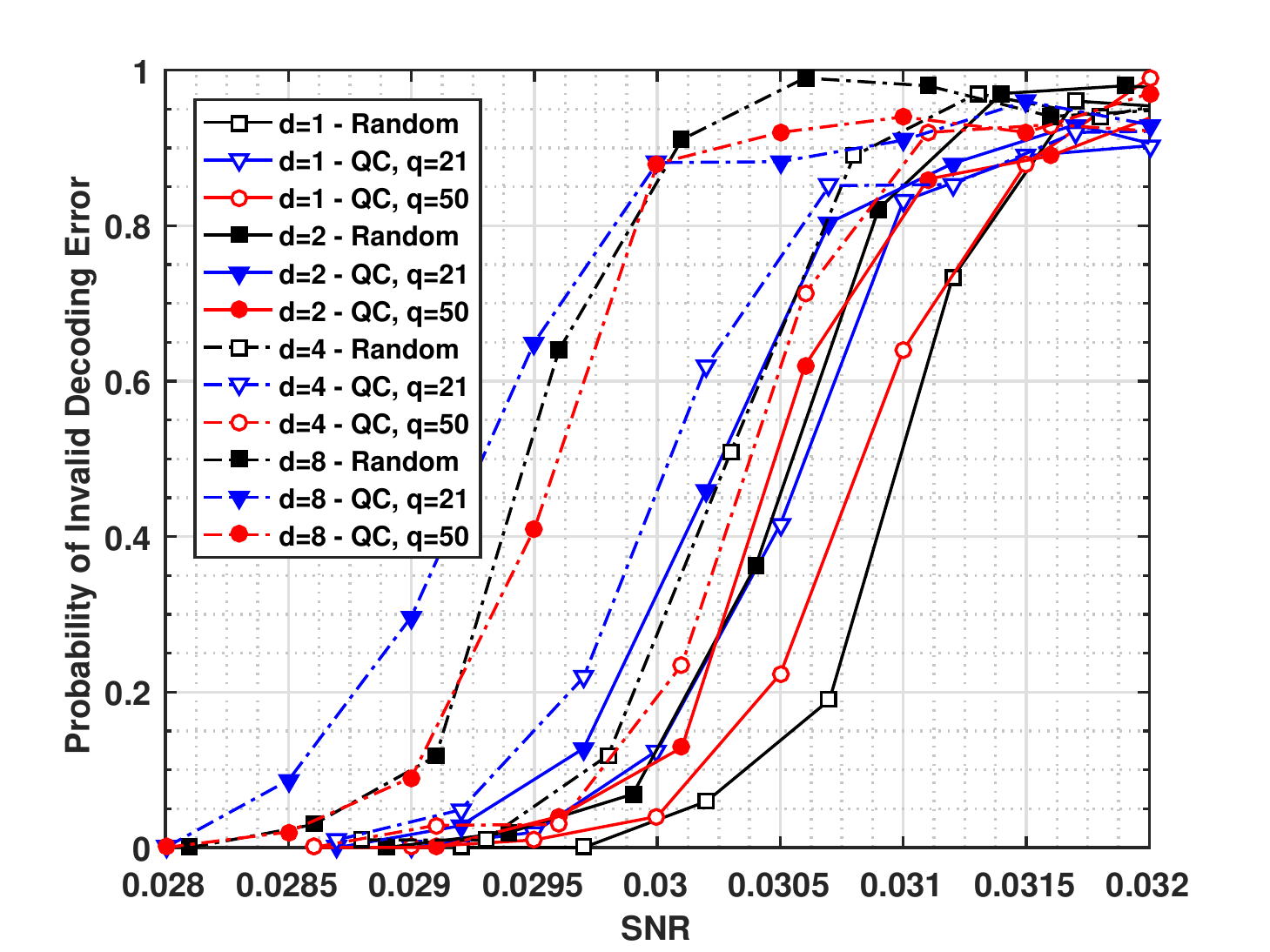}
\caption{Probability of invalid decoding error vs. SNR for Sum-Product decoding with $d=1,2,4,8$ dimensional reconciliation on BIAWGNC. Probability of error is computed for invalid messages that are correctly decoded but CRC fails.}
\label{fig:InvalidDecodingRate}
\end{figure}

The probability of an invalid decoding error is given by
\begin{equation}
%\resizebox{0.46\textwidth}{!}{$
P( \mathbf{\hat C H^\top=0} \ \cap \ \text{CRC Fail} ) = \dfrac{\text{Number of CRC Errors}}{  \text{Total Number of Frame Errors} }.
%$}
\nonumber
\end{equation}
Figure~\ref{fig:InvalidDecodingRate} shows the probability of an invalid decoding error over the SNR range of interest for $d=1,2,4,8$ reconciliation dimensions on the BIAWGNC for the three LDPC codes designed in this work. In general, the probability of invalid decoding increases with SNR and becomes the main source of frame error, particularly in the error-floor region as a result of the large block length and low code rate. In the region of operation for long-distance CV-QKD where the FER $P_\text{e} \approx 1$, invalid decoding convergence still contributes to nearly 10\% of all frame errors. A concatenated higher-rate code was not included as part of the message component to correct residual errors~\cite{Lodewyck2007, Jouguet2011}.

\begin{figure}[htbp]
\centering
\includegraphics[trim=0in 0in 0in 0in, width=0.48\textwidth]{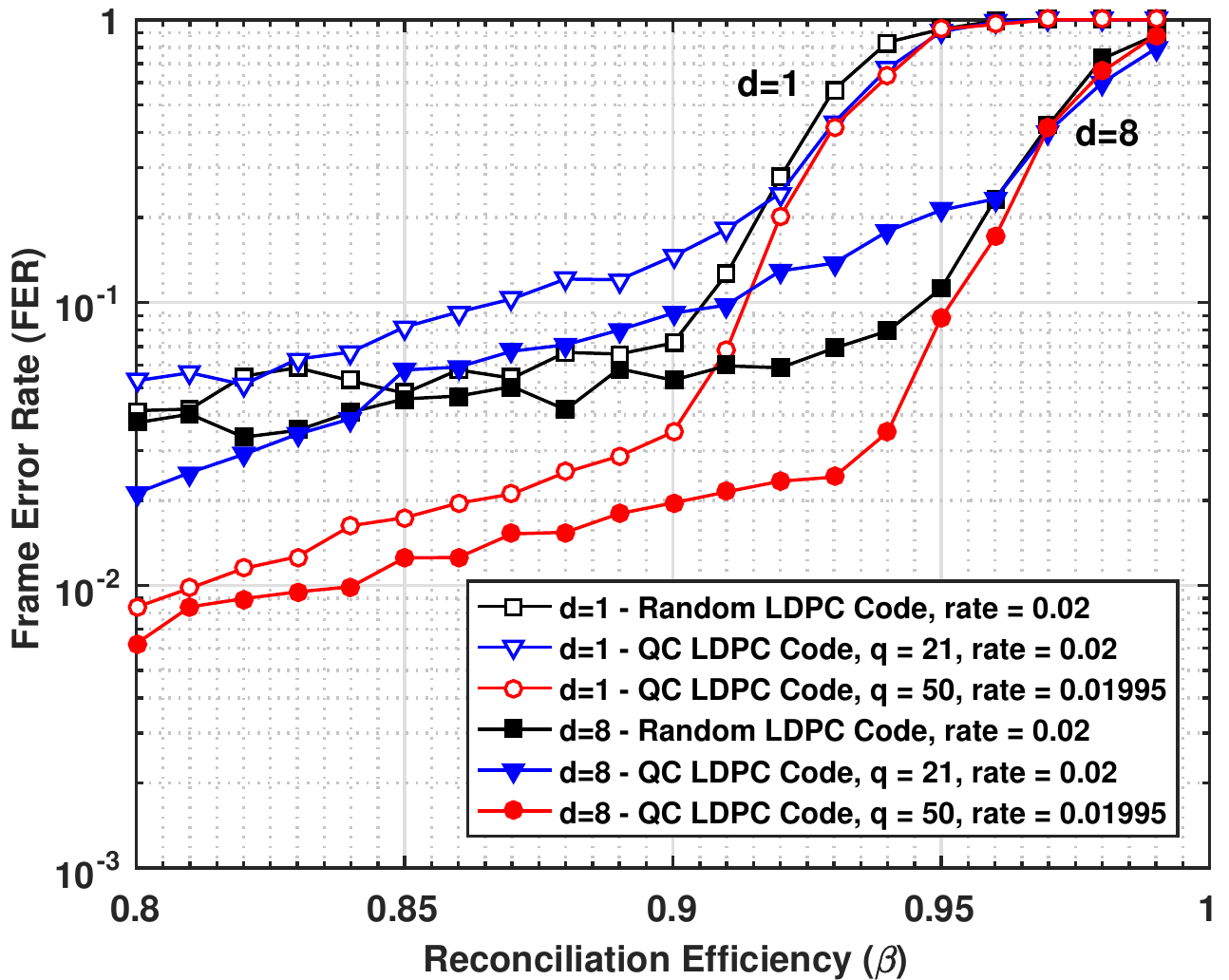}
\caption{FER vs. reconciliation efficiency for Sum-Product decoding with $d=1$ and $d=8$ dimensional reconciliation on BIAWGNC. FER values are derived from the FER vs. SNR curves based on Eq.~\ref{eq:snr(beta)}.}
\label{fig:FERvsBeta}
\end{figure}

Up until this point, the performance of the reconciliation algorithm has been presented as a coding theory problem, where an optimal LDPC code was designed to achieve a particular FER at a given SNR operating point. The SNR was considered as an abstraction of the virtualized BIAWGNC in order to demonstrate fixed-rate code performance, independent of other CV-QKD system parameters such as modulation variance, transmission distance, and physical losses. Assuming that the transmission distance and physical parameters of the quantum channel are fixed, Alice's modulation variance can be optimally tuned such that the effective secret key rate is then solely determined by the FER and $\beta$-efficiency of the LDPC-decoding reconciliation algorithm. 

Figure~\ref{fig:FERvsBeta} shows that for each fixed-rate LDPC code, there exists a unique FER-$\beta$ pair, where each $\beta$ corresponds to a particular SNR operating point based on Eq.~\ref{eq:snr(beta)}. While it may appear from Eq.~\ref{eq:Keffsimple} that maximizing $\beta$ would produce a higher effective secret key rate, Fig.~\ref{fig:FERvsBeta} shows that $\beta$ and FER are positively correlated, such that there exists an optimal trade-off between $\beta$ and FER where $K_\text{eff}$ is maximized for a fixed transmission distance. To achieve key reconciliation at long distances, the operating point must be chosen in the waterfall region where $\beta$ is high, despite the high FER. 

The results presented in this section showed that higher reconciliation schemes, namely $d=4$ and $d=8$, extend code performance to lower SNR where the FER $P_\text{e}>0$ and $\beta \rightarrow 1$. As such, the $d=8$ scheme would be most preferred for long-distance reconciliation. The next section examines the impact of reconciliation dimension, $\beta$-efficiency, and FER on the effective secret key rate over a range of transmission distances for the LDPC codes designed in this work.

\subsection{Finite Secret Key Rate}

This section extends the discussion of the effective secret key rate to include finite-size effects. Key reconciliation for a particular $\beta$-efficiency is only achievable over a limited range of distances where the finite secret key rate $K_\text{finite}>0$. In general, for a single FER-$\beta$ pair, LDPC decoding can achieve either (1) a high secret key rate at short distance, or (2) a low secret key rate at long distance. For long-distance CV-QKD beyond 100km, key reconciliation is only achievable with high $\beta$-efficiency at the expense of low secret key rate. This section provides an overview of the maximum achievable finite secret key rates and reconciliation distances for the three LDPC codes designed in this work. Results are presented for the $d=1$ and $d=8$ reconciliation dimensions in order to demonstrate the effectiveness of higher-order dimensionality on reconciliation distance. The results also consider the finite size of the privacy amplification block. The $d=1$ and $d=8$ secret key rate results are shown for privacy amplification blocks of $N_\text{privacy}=10^{10}$ and $N_\text{privacy}=10^{12}$ bits to demonstrate the impact of block size on the maximum transmission distance.

The range of transmission distances for each $\beta$ is limited by the total noise between Alice and Bob. From Eq.~\ref{eq:IAB}, the total noise can be expressed as a function of $\beta$, such that 
\begin{equation}
\label{eq:Xtotal(d,beta)}
\mathbf{\chi^{\prime}_{total}}(\beta)= \frac{\mathbf{V_{A}^{opt}}(\beta)}{s(\beta)}-1,
\end{equation}
where $\mathbf{V_{A}^{opt}}(\beta)$ is a vector of Alice's optimal modulation variances for a particular $\beta$-efficiency from Fig.~\ref{fig:OptimalVA}, and the virtual SNR $s(\beta)$ is given by Eq.~\ref{eq:snr(beta)} for a fixed-rate LDPC code. From the expression for the total channel noise, $\chi_\text{total} = \chi_\text{line} + \frac{\chi_\text{hom}}{T}$, a set of transmission distance points for a particular $\beta$ can then be described by the vector
\begin{equation}
\label{eq:dprime(beta)}
\mathbf{d^{\prime}}(\beta) = \frac{10}{\alpha} \log_{10} \Bigg( \frac{\eta(\mathbf{\chi^{\prime}_{total}}(\beta)-\epsilon+1)}{1+V_\text{el}} \Bigg), 
\end{equation}
in order to compute the maximum theoretical finite secret key rate based on Eq.~\ref{eq:Kfinite}. 

\begin{figure}[!t]
\centering
\includegraphics[trim=0in 0in 0in 0in, width=0.48\textwidth]{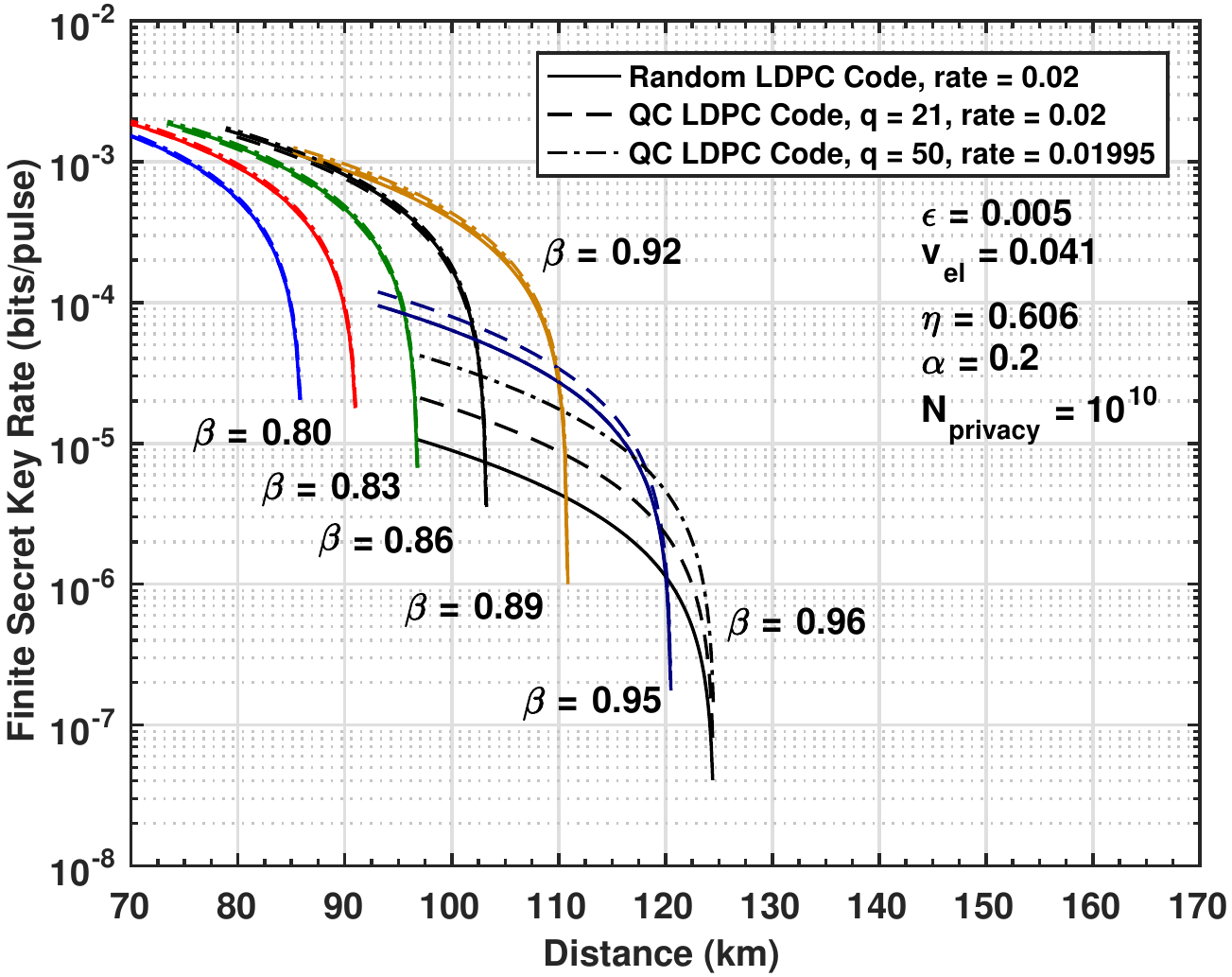}
\caption{$d=1$ dimensional reconciliation with $N_\text{privacy}=10^{10}$ bits: finite secret key rate $K_\text{finite}$ vs. distance for collective attacks on BIAWGNC with Sum-Product decoding.}
\label{fig:1D-ESKR-1E10}
\end{figure}

\begin{figure}[!t]
\centering
\includegraphics[trim=0in 0in 0in 0in, width=0.48\textwidth]{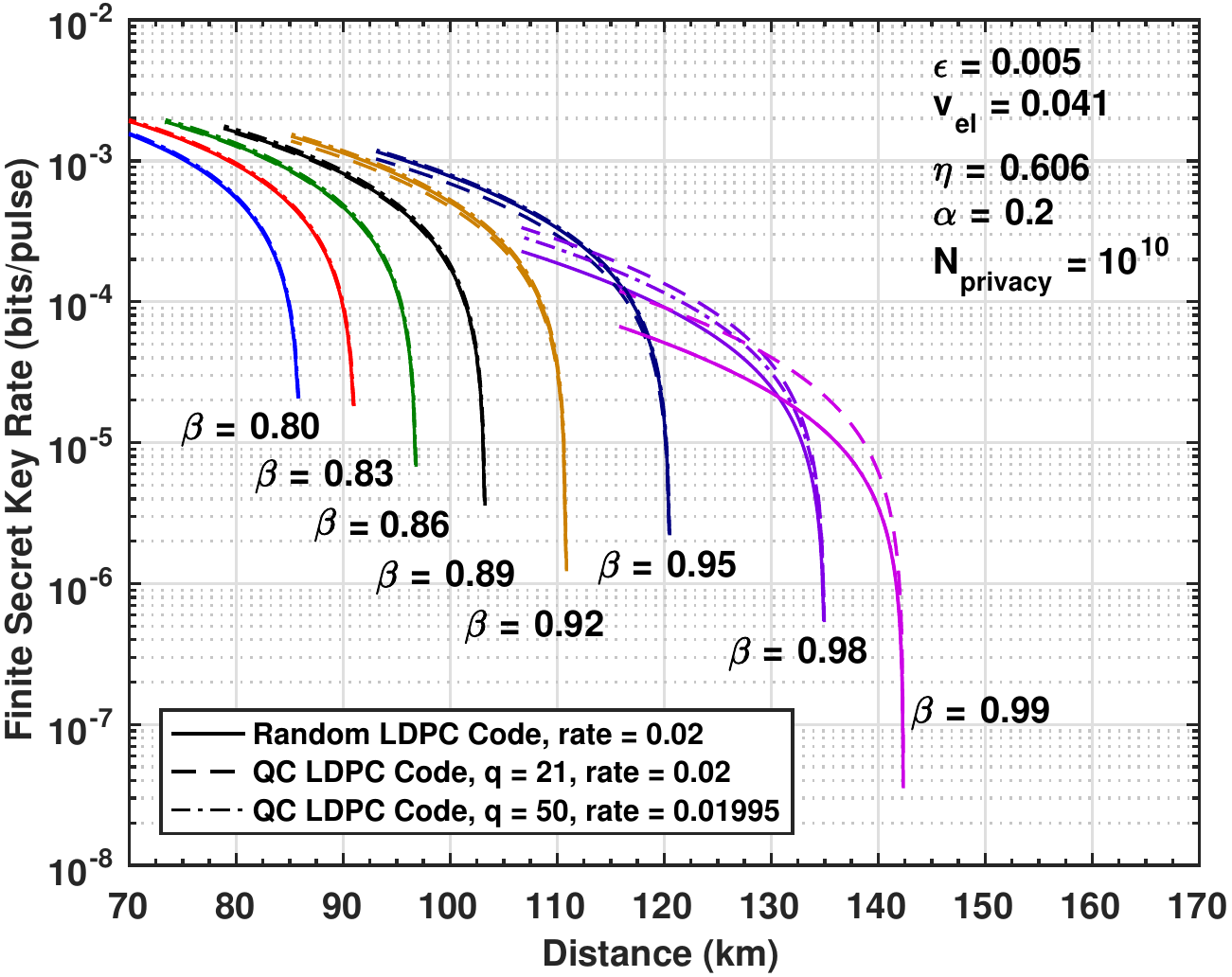}
\caption{$d=8$ dimensional reconciliation with $N_\text{privacy}=10^{10}$ bits: finite secret key rate $K_\text{finite}$ vs. distance for collective attacks on BIAWGNC with Sum-Product decoding.}
\label{fig:8D-ESKR-1E10}
\end{figure}

\begin{figure}[!t]
\centering
\includegraphics[trim=0in 0in 0in 0in, width=0.48\textwidth]{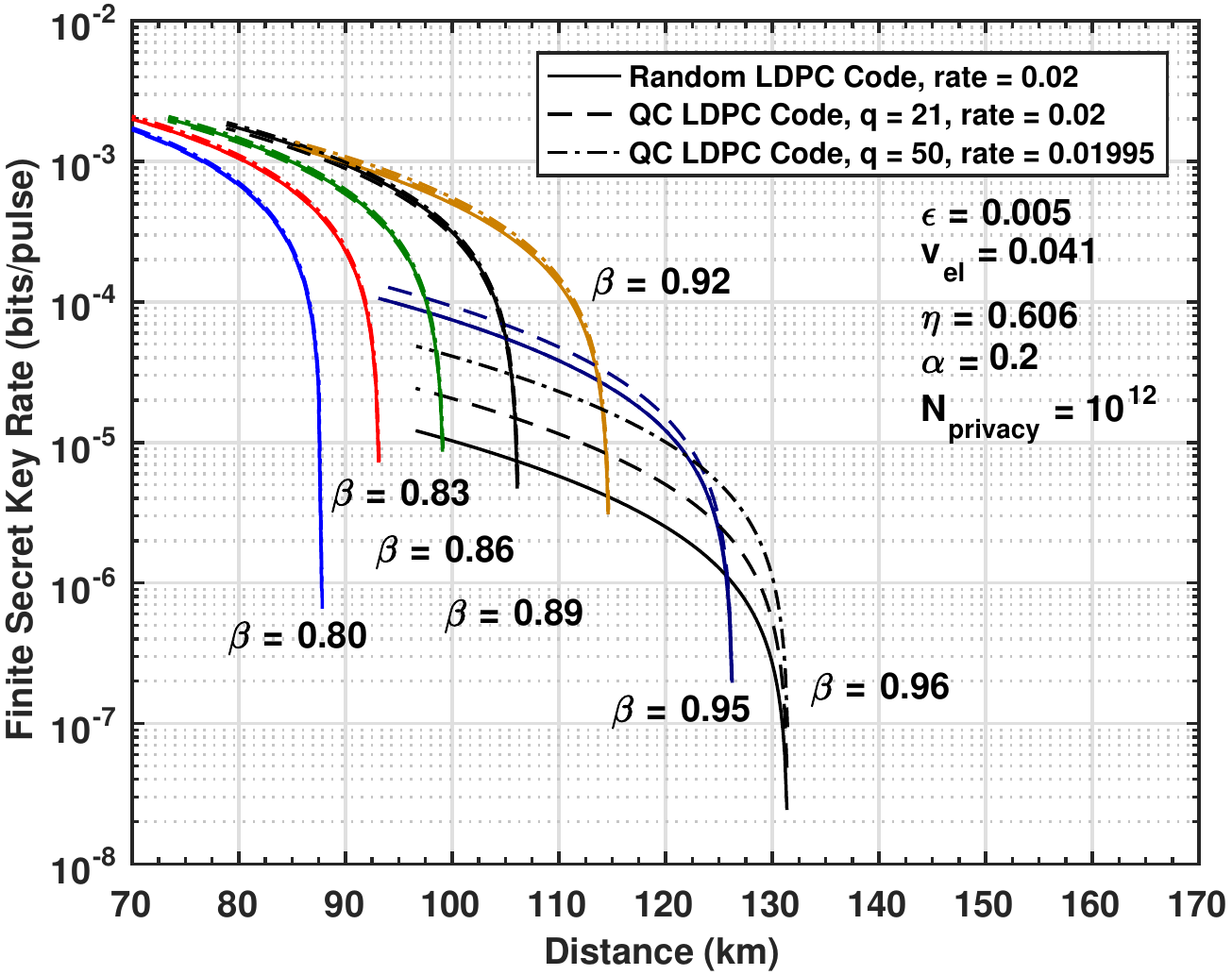}
\caption{$d=1$ dimensional reconciliation with $N_\text{privacy}=10^{12}$ bits: finite secret key rate $K_\text{finite}$ vs. distance for collective attacks on BIAWGNC with Sum-Product decoding.s}
\label{fig:1D-ESKR-1E12}
\end{figure}

\begin{figure}[!t]
\centering
\includegraphics[trim=0in 0in 0in 0in, width=0.48\textwidth]{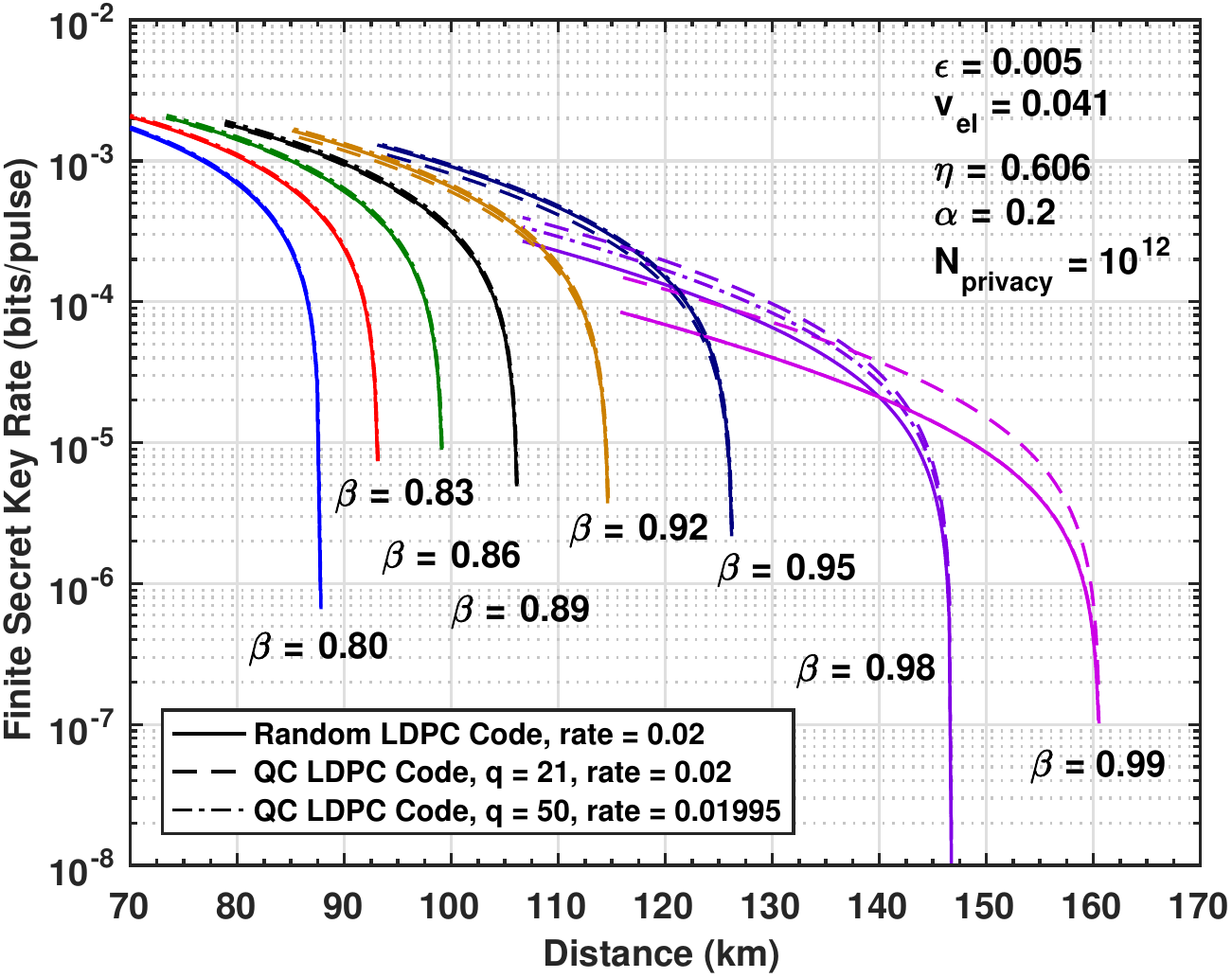}
\caption{$d=8$ dimensional reconciliation with $N_\text{privacy}=10^{12}$ bits: finite secret key rate $K_\text{finite}$ vs. distance for collective attacks on BIAWGNC with Sum-Product decoding.}
\label{fig:8D-ESKR-1E12}
\end{figure}

Figures \ref{fig:1D-ESKR-1E10} and \ref{fig:8D-ESKR-1E10} present the finite secret key rate results for the three LDPC codes over the transmission distance range of interest with $N_\text{privacy}=10^{10}$ bits based on the $d=1$ and $d=8$ reconciliation dimensions, respectively. Each $\beta$-efficiency curve in Figures \ref{fig:1D-ESKR-1E10} and \ref{fig:8D-ESKR-1E10} represents a FER-$\beta$ pair where the FER and SNR are constant over the entire transmission distance range, while $V_A$ is optimally chosen to achieve the maximum secret key rate at each distance point. When $\beta$ is high, the FER $P_e \rightarrow 1$, and thus $K_{\text{finite}} \rightarrow 0$ as erroneous frames are discarded after decoding. As a result, the maximum reconciliation distance is limited by the error-correction performance of the LDPC code. Figures \ref{fig:1D-ESKR-1E12} and \ref{fig:8D-ESKR-1E12} present the finite secret key rate results with $N_\text{privacy}=10^{12}$ bits over $d=1$ and $d=8$ reconciliation dimensions, respectively. When $N_\text{privacy}=10^{12}$ bits, the maximum transmission distance is extended by 18km over the result with $N_\text{privacy}=10^{10}$ bits for $d=8$ reconciliation with $\beta=0.99$ efficiency. This demonstrates the importance of selecting a large block size for privacy amplification.

In each of the $N_\text{privacy}=10^{10}$ and $N_\text{privacy}=10^{12}$ cases, the three LDPC codes achieve similar finite secret key rates and reconciliation distances with both $d=1$ and $d=8$ schemes for $\beta \leq 0.92$, since the codes are operating close to their respective error floors. However, for $\beta>0.92$, the FER becomes a limiting factor to achieving a non-zero secret key rate. The $d=1$ scheme achieves a maximum efficiency of $\beta=0.96$, where the maximum distance is limited to 124km with $N_\text{privacy}=10^{10}$ bits, and 132km with $N_\text{privacy}=10^{12}$ bits. For $\beta > 0.96$, the FER $P_\text{e}=1$, thus $K_\text{finite}=0$. The $d=8$ scheme operates up to $\beta=0.99$ efficiency, with a maximum distance of 142km with $N_\text{privacy}=10^{10}$ bits, and 160km with $N_\text{privacy}=10^{12}$ bits. Furthermore, the $d=8$ scheme achieves higher secret key rates for all three LDPC codes at $\beta=0.95$ and $\beta=0.96$ in comparison to the $d=1$ scheme since the code FER performance is higher. While the complete results are not shown here, the $d=2$ and $d=4$ schemes both achieve a maximum efficiency of $\beta=0.97$, at 129km with $N_\text{privacy}=10^{10}$ bits, and 138km with $N_\text{privacy}=10^{12}$ bits.

The finite secret key rate $K_\text{finite}$ results presented in this section were normalized to the pulse rate, without consideration of the light source repetition rate $f_\text{rep}$. By considering the pulse rate, the complete operating secret key rate of the CV-QKD system can be defined as 
\begin{equation}
\label{eq:Kfiniteprime}
K_\text{finite}^\prime= f_\text{rep}K_\text{finite} \ \ \ \text{(bits/s)}.
\end{equation}
The next section presents an overview of a GPU-based LDPC decoder implementation where the information throughput for the three LDPC codes designed in this work is compared to the upper bound on secret key rate at the maximum reconciliation distance points.

%-----------------------------------------------------------------------------------
%
% GPU-Accelerated Decoding
% 
% Note for thesis: draw GPU structure picture so readers understand thread/mem 
% structure, and difference between shared/global memory
%-----------------------------------------------------------------------------------

\section{GPU-Accelerated LDPC Decoding}
\label{sec:GPU}

GPUs are a highly suitable platform for the implementation of LDPC decoders that target high information throughput with long block-length codes. Computational acceleration of the belief propagation algorithm is achieved by parallelizing the check and variable node update operations across thousands of single-instruction multiple-thread (SIMT) cores, which provide floating-point precision, high-bandwidth read/write access to on-chip memory, and intrinsic mathematical libraries for the logarithmic functions of the Sum-Product algorithm~\cite{Kang2012, Wang2013, Lin2014, Wang2011, Gal2014}. 

This section provides an overview of the GPU-based LDPC decoder implementation in this work. GPU throughput results are presented for the maximum CV-QKD distances under $d=1,2,4,8$ dimensional reconciliation, and also compared to the maximum achievable secret key rates for  reconciliation efficiencies $\beta > 0.85$. Finally, the implementation is compared to previous work by Jouguet and Kunz-Jacques for an LDPC code with block length of $2^{20}$ bits~\cite{Jouguet2014}, as well as other non-LDPC codes. The GPU decoding throughput results presented in this section quantitatively highlight the computational speedup that can be achieved using quasi-cyclic LDPC codes for long-distance CV-QKD.

\subsection{GPU-based LDPC Decoder Implementation}

\begin{figure}[htbp]
\centering
\includegraphics[trim=0in 0in 0in 0in, width=0.48\textwidth]{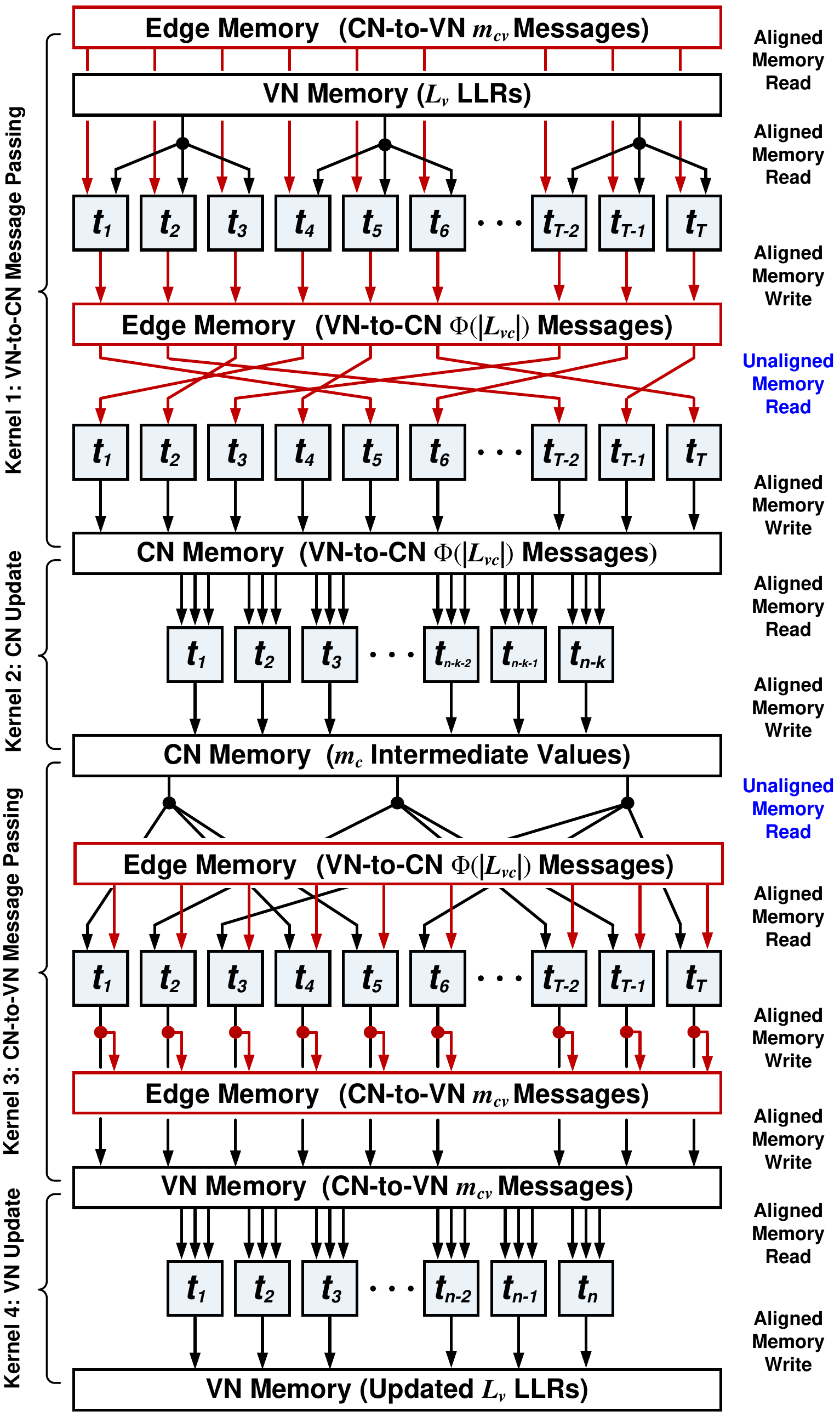}
\caption{GPU implementation of LDPC decoder showing four multi-threaded compute kernels and data flow from top to bottom for one decoding iteration. Coalesced memory access patterns and message variables are indicated. Thread $i$ is denoted by $t_i$, where $T$ in kernels 1 and 3 represents the maximum number of connections between all CNs and VNs, $(n-k)$ in kernel 2 is the number of CNs, and $n$ in kernel 4 is the number of VNs. Early termination is not shown.}
\label{fig:gpu-ldpc-imp}
\end{figure}

The LDPC decoder was implemented on a single NVIDIA GeForce GTX 1080 (Pascal Architecture) GPU with 2560 CUDA cores using the NVIDIA CUDA C++ application programming interface. Figure~\ref{fig:gpu-ldpc-imp} shows the data flow for a single decoding iteration of the parallelized Sum-Product algorithm, which is comprised of four multi-threaded compute kernels. Each kernel instantiates a different number of GPU threads depending on the level of parallelism for the operation. The individual compute operations of the Sum-Product algorithm are re-ordered to exploit the maximum amount of thread-level parallelism in each kernel such that the latency per iteration is minimized. The overall throughput of the GPU-based LDPC decoder is then determined by the number of iterations, the latency per iteration, and the block length.

The complexity of an LDPC decoder implementation stems from the highly-irregular interconnect structure between CNs and VNs described by the code's Tanner graph. For codes with short block lengths, the permutation network complexity does not introduce significant GPU decoding latency~\cite{Wang2013, Kang2012, Lin2014}, however, for codes with block lengths on the order of $10^6$ bits as those designed in this work, data permutation and message passing constitute between 25\% to 50\% of GPU runtime per decoding iteration, as shown in Table~\ref{table:GPU-LDPC-latency-performance-results}. While arithmetic operations are relatively inexpensive on a GPU, addressing global memory is very costly in terms of compute time. The most expensive GPU operation is addressing unordered memory, i.e.\ accessing non-consecutive memory locations, as multiple transactions are required to perform the unordered memory read or write, and all kernel threads must be stalled~\cite{Wang2013}. On the contrary, coalesced memory addressing, i.e.\ accessing consecutive memory locations, can be performed in a single transaction and allows for concurrent thread execution, which reduces the runtime of the kernel. Furthermore, uncoalesced memory writes are more expensive than uncoalesced memory reads. Thus, the throughput of a GPU-based LDPC decoder is highly dependent on memory access patterns.

\begin{table}[t!]
\begin{center}
\captionsetup{justification=centering}
\caption{GPU-based LDPC decoding latency and error-correction performance for rate 0.02 multi-edge codes}
\label{table:GPU-LDPC-latency-performance-results}
\scalebox{0.775}{
    \begin{tabular}{ !{\vrule width 3\arrayrulewidth} C{1.22in} !{\vrule width 3\arrayrulewidth} C{0.85in} | C{0.85in} | C{0.85in} !{\vrule width 3\arrayrulewidth} }
    \Xhline{3\arrayrulewidth}   
  
\bf LDPC Code &	\bf Random Multi-Edge & \bf $q=21$ QC Multi-Edge & \bf $q=50$ QC Multi-Edge \\ \hline
\Xhline{3\arrayrulewidth}

\bf Block Length (Bits) & $1 \times 10^6$ & $1.008 \times 10^6$ & $1 \times 10^6$ \\ \hline
\bf Code Rate & 0.02 & 0.02 & 0.01995 \\ \hline
\bf Connections in Parity Matrix & 3,337,494 & 160,185 & 66,747 \\ \hline
\Xhline{3\arrayrulewidth}

\multicolumn{4}{!{\vrule width 3\arrayrulewidth} c !{\vrule width 3\arrayrulewidth}}{\bf Latency by Kernel with Percent Breakdown for 1 Decoding Iteration} \\ \hline
\Xhline{3\arrayrulewidth}
\bf Kernel 1 Runtime VN-to-CN  (ms) & 1.773 (50.3\%) & 0.446 (34.4\%)	&	0.391 (33.2\%) \\ \hline
\bf Kernel 2 Runtime CN Update (ms) & 0.197 (5.6\%) & 0.204 (15.7\%) & 0.198 (16.8\%) \\ \hline
\bf Kernel 3 Runtime CN-to-VN (ms) & 	1.240 (35.1\%) & 0.317 (24.4\%) & 0.303 (25.7\%)	 \\ \hline
\bf Kernel 4 Runtime VN Update (ms) & 0.318 (9.0\%) & 0.331 (25.5\%) & 0.286 (24.3\%)	 \\ \hline
\bf Total Latency Per Iteration (ms) & 3.528 & 1.296 & 1.177 \\ \hline
\Xhline{3\arrayrulewidth}

\multicolumn{4}{!{\vrule width 3\arrayrulewidth} c !{\vrule width 3\arrayrulewidth}}{\bf FER Performance and Decoding Throughput at $\beta=0.99$ and $d=8$} \\ 
\Xhline{3\arrayrulewidth}
\bf Max Iterations & 500	& 500 &	500	 \\ \hline	
\bf Average Iterations $^*$ & 470	& 451 &	470	 \\ \hline	
\bf FER & 0.883 & 0.792 & 0.883 \\ \hline
\bf $K_\text{GPU}^\text{raw}$ Raw Throughput (Mb/s) & 0.603	& 1.724 &	1.807	 \\ \hline
\bf $K_\text{GPU}^\prime$ Information Throughput (Kb/s) & 1.409	& 7.160 &	4.207 \\ \hline	

 \Xhline{3\arrayrulewidth} 
    \end{tabular} 
}
\end{center}

\scalebox{0.8}{
\begin{tabular}{p{4.25in}}
$^*$ Early-termination check is enabled only after the number of decoding iterations is equal to the average number of iterations, which is determined empirically through FER simulation and stored in a lookup table.
\end{tabular}
}

\end{table}

The operations of the Sum-Product algorithm presented in Algorithm~\ref{SPA} were re-ordered to avoid uncoalesced memory writes and to use the maximum amount of thread-level parallelism for arithmetic computations. For example, the VN-to-CN message-passing permutation in kernel 1 also performs the $\Phi(\cdot)$ computation from the next CN-update step in each thread. The CN-update kernel (2) does not fully compute the $m_{cv}$ messages from each CN to its connected VNs, but instead, the final CN-to-VN $m_{cv}$ messages are computed in the CN-to-VN message-passing kernel (3). Due to the Tanner graph structure and data permutation nature of the LDPC decoder, uncoalesced memory reads are still required when reading from edge memory in kernel 1 and reading from CN memory in kernel 3. However, the latency of these operations is negligible compared to the overall latency of an entire iteration. Fully-coalesced memory writes are enabled by the different ordering of connected edges in the VN-to-CN and CN-to-VN message-passing kernels (1 and 3). In the VN-to-CN message-passing kernel (1), the edge connectivity is ordered by consecutive VNs, while in the CN-to-VN message-passing kernel (3), the edges are ordered by consecutive CNs. Each CN-VN edge in the edge memory has a unique index that is addressed by both message-passing kernels (1 and 3). Several additional memory optimizations improve the overall GPU throughput. Shared memory is used in each thread to store local variables and to avoid expensive global memory accesses, while texture caches are used to store frequently-accessed static variables such as channel LLRs and the parity-check matrix.

As shown in Fig.~\ref{fig:gpu-ldpc-imp}, message-passing kernels (1 and 3) instantiate up to $T$ threads, where $T$ is the maximum number of edge connections between all CNs and VNs, kernel 2 instantiates $(n-k)$ threads equal to the total number of CNs in the matrix, and kernel 4 instantiates up to $n$ threads equal to the total number of VNs in the matrix. When early termination is enabled, $T$ threads are required in kernels 1 and 3, and $n$ threads are required in kernel 4. However, when early termination is disabled, the number of threads instantiated in kernels 1, 3, and 4 can be reduced due to the long-diagonal construction of the parity-check matrices in this work. The message-passing kernels (1 and 3) need only to instantiate threads that correspond to the CN-VN connections to the left of the long diagonal in the matrix structure shown in Fig.~\ref{fig:spy_plots}. Similarly, the VN-update kernel (4) needs only to instantiate threads that correspond to VNs to the left of the long diagonal. This reduction in the number of threads provides a marginal speedup in each iteration. 

While not shown in Fig.~\ref{fig:gpu-ldpc-imp}, the early-termination check is implemented via multiple kernels that perform a parallel reduction following the VN-to-CN message-passing kernel (1) in order to compute the parity at each CN. Additional computations and memory reads/writes are required in the message-passing and VN-update kernels (1, 3, and 4). The following additional operations must be performed to enable an early-termination check: send the decision bit from each VN to its connected CNs, send all $m_{cv}$ messages from each CN to its connected VNs (including those corresponding to connections along the long diagonal), and calculate the decision bit in each VN. To reduce overall decoding latency and maximize throughput, the early-termination check is performed only after a fixed number of decoding iterations. This fixed number of iterations corresponds to the average number of iterations required at each SNR point, and is pre-determined empirically through FER simulation for each code. The decoder uses a lookup table to decide after how many decoding iterations to enable the early-termination check based on the current SNR.

A quasi-cyclic matrix structure reduces data permutation and memory access complexity by eliminating random, unordered memory access patterns. In addition, QC codes require fewer memory lookups for message passing since the permutation network can be described with approximately $q$-times fewer terms, where $q$ is the expansion factor of the QC parity-check matrix, in comparison to a purely-random matrix for the same block length. Table~\ref{table:GPU-LDPC-latency-performance-results} presents a breakdown of the latency of each GPU kernel for the three LDPC codes designed in this work. While the CN and VN update kernels (2 and 4) have similar runtime for both random and QC codes, QC codes achieve faster runtime in data permutation kernels (1 and 3) due the approximately $q$-times fewer CN-VN edge connections in the parity-check matrix. Since the parity-check matrices designed in this work are sparse, a compressed data structure is used to store CN-VN edge connections to reduce memory read latency in the message-passing kernels.

Table~\ref{table:GPU-LDPC-latency-performance-results} also highlights the respective error-correction performance and GPU throughput of the three codes at the maximum $\beta=0.99$ efficiency with $d=8$ reconciliation. The raw GPU throughput (including parity bits) is given by 
\begin{equation}
K_\text{GPU}^\text{raw} = \dfrac{\text{Block Length}}{\text{Latency Per Iteration} \times \text{Iterations}} \ \   \text{(bits/s).}
\end{equation}
Similar to the finite secret key rate, the information throughput of the GPU decoder must be scaled by (1) the FER $P_\text{e}$ to account for discarded frames when decoding is unsuccessful, i.e.\ CRC does not pass or parity check fails, and (2) the code rate $R_\text{code}$ to account for the parity bits that must be discarded after decoding. The average GPU information throughput is then given by
\begin{equation}
K_\text{GPU}^\prime =  K_\text{GPU}^\text{raw} \, R_\text{code} \ (1-P_\text{e}). 
\end{equation}
Thus, for any LDPC code, the GPU throughput is determined by the latency per iteration and the number of decoding iterations. The latency per iteration depends on the LDPC code structure and the number of memory lookups, while the FER is bound by the maximum number of iterations.

\begin{table*}[!t]
\begin{center}
\captionsetup{justification=centering}
\caption{Overview of secret key rate and GPU throughput at maximum reconciliation distance\\ with rate 0.02 multi-edge codes and $N_\text{privacy}=10^{12}$ bits}
\label{table:maxDistanceRateComparison}

\renewcommand{\arraystretch}{1.2}
\scalebox{0.785}{
    \begin{tabular}{ !{\vrule width 3\arrayrulewidth} C{0.846in} | C{0.846in} | C{0.69in} | C{0.615in} |  C{1.2in} | C{1.28in} | C{0.725in} | C{0.725in} | C{0.7in} !{\vrule width 3\arrayrulewidth}}
    \Xhline{3\arrayrulewidth} 
    
    \bf Reconciliation Dimension & 
    \bf Maximum Reconciliation Efficiency & 
    \bf LDPC Code & 
    \bf Maximum Distance (km) &     
 	\bf Operating Secret Key Rate $K_\text{finite}^\prime$ at Max Distance with $\mathbf{f_\text{rep}=1}$MHz (bit/s) &    
    \bf Fundamental Key Rate Limit $K_\text{lim}^\prime$ at Max Distance with $\mathbf{f_\text{rep}=1}$MHz (Kbit/s) & 
    \bf GPU Raw Throughput $K_\text{GPU}^\text{raw} $ (Mbit/s) & 
    \bf GPU Info. Throughput $K_\text{GPU}^\prime$ (Kbit/s) &
	\bf $K_\text{GPU}^\prime$ Speedup Over $K_\text{lim}^\prime$ ($K_\text{GPU}^\prime / K_\text{lim}^\prime$)\\ 
	\Xhline{3\arrayrulewidth}

\multirow{3}{*}{$d=1$} & \multirow{3}{*}{$\beta = 0.96$} 
  & Random     & $131.38$  & $0.060$ & $3.405$  & $0.612$ & $0.111$ & $0.033\times$ \\
& & QC, $q=21$ & $131.38$  & $0.119$  & $3.405$  & $1.887$ & $0.686$  & $0.202\times$ \\
& & QC, $q=50$ & $131.43$  & $0.235$ & $3.397$  & $1.966$ & $1.426$  & $0.420\times$ \\
\hline

\multirow{3}{*}{$d=2$} & \multirow{3}{*}{$\beta = 0.97$}
  & Random     & $137.99$ & $0.051$ & $2.510$  & $0.612$ & $0.223$ & $0.087\times$ \\
& & QC, $q=21$ & $137.99$ & $0.203$ & $2.510$  & $1.856$ & $2.700$  & $1.076\times$ \\
& & QC, $q=50$ & $137.85$ & $0.050$ & $2.526$ & $1.983$ & $0.360$ & $0.142\times$ \\
\hline

\multirow{3}{*}{$d=4$} & \multirow{3}{*}{$\beta = 0.97$}
  & Random     & $137.99$  & $0.101$ & $2.510$ & $0.604$ & $0.439$ & $0.175\times$ \\
& & QC, $q=21$ & $137.99$  & $0.302$ & $2.510$ & $1.818$ & $3.938$ & $1.569\times$ \\
& & QC, $q=50$ & $137.85$  & $0.401$ & $2.526$  & $1.855$ & $2.692$  & $1.065\times$ \\
\hline

\multirow{3}{*}{$d=8$}& \multirow{3}{*}{$\beta = 0.99$} 
  & Random     & $160.47$ & $0.230$ & $0.891$  & $0.604$ & $1.409$  & $1.581\times$ \\
& & QC, $q=21$ & $160.47$ & $0.410$ & $0.891$  & $1.724$ & $7.160$  & $8.033\times$ \\
& & QC, $q=50$ & $160.52$ & $0.224$ & $0.889$  & $1.808$ & $4.207$ & $4.733\times$ \\     
\hline    
    
	\Xhline{3\arrayrulewidth}   
    \end{tabular} 
}

\end{center}

\end{table*}

\begin{figure}[t!]
\centering
\includegraphics[trim=0in 0in 0in 0in, width=0.48\textwidth]{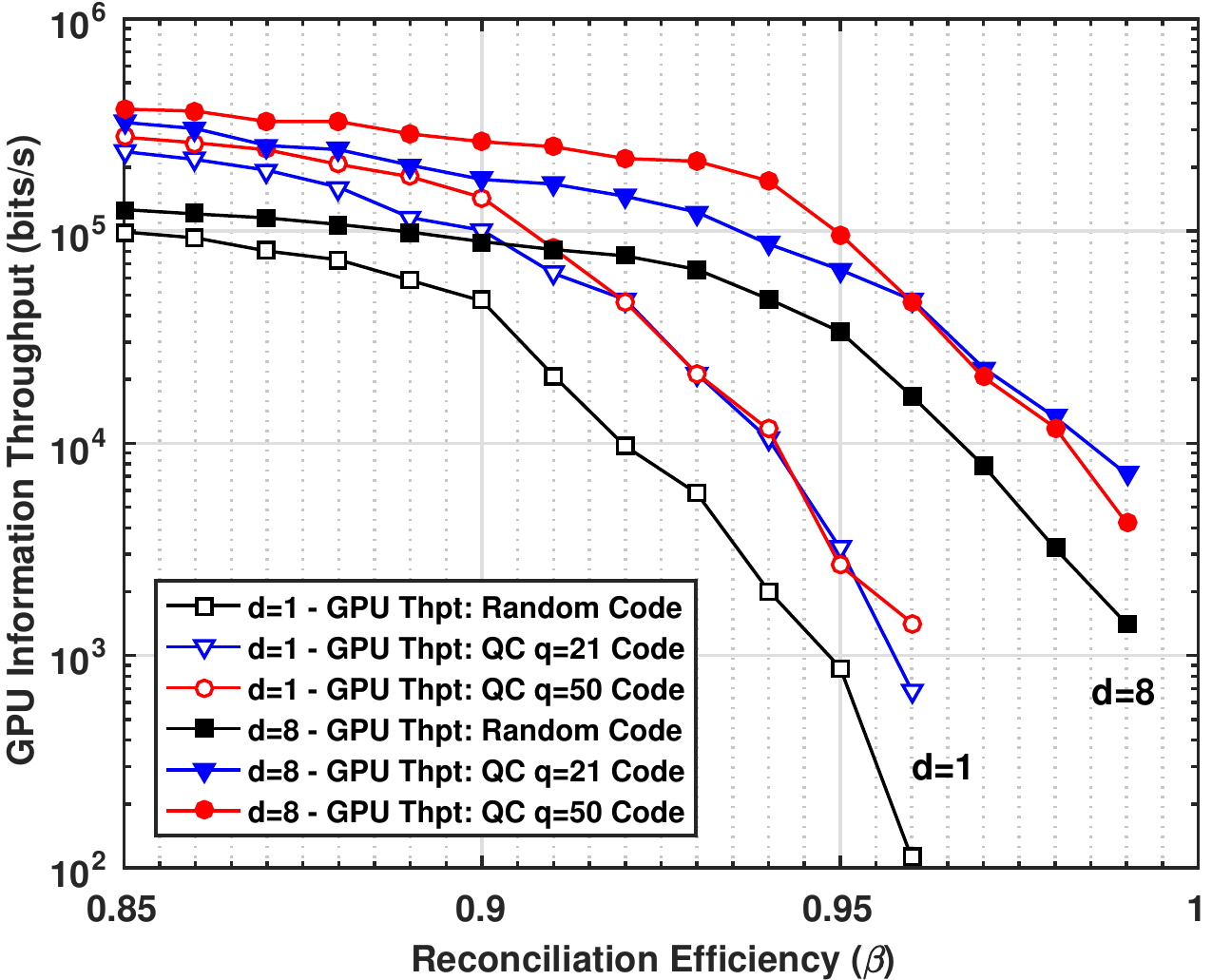}
\caption{Measured information throughput $K_\text{GPU}^\prime$ vs. reconciliation efficiency for $d=1$ and $d=8$ dimensional reconciliation. Each measurement point corresponds to a particular SNR operating point with a measured FER presented in Fig.~\ref{fig:FERvsBeta}.}
\label{fig:GPUThptvsBeta}
\end{figure}

\begin{figure}[t!]
\centering
\includegraphics[trim=0in 0in 0in 0in, width=0.48\textwidth]{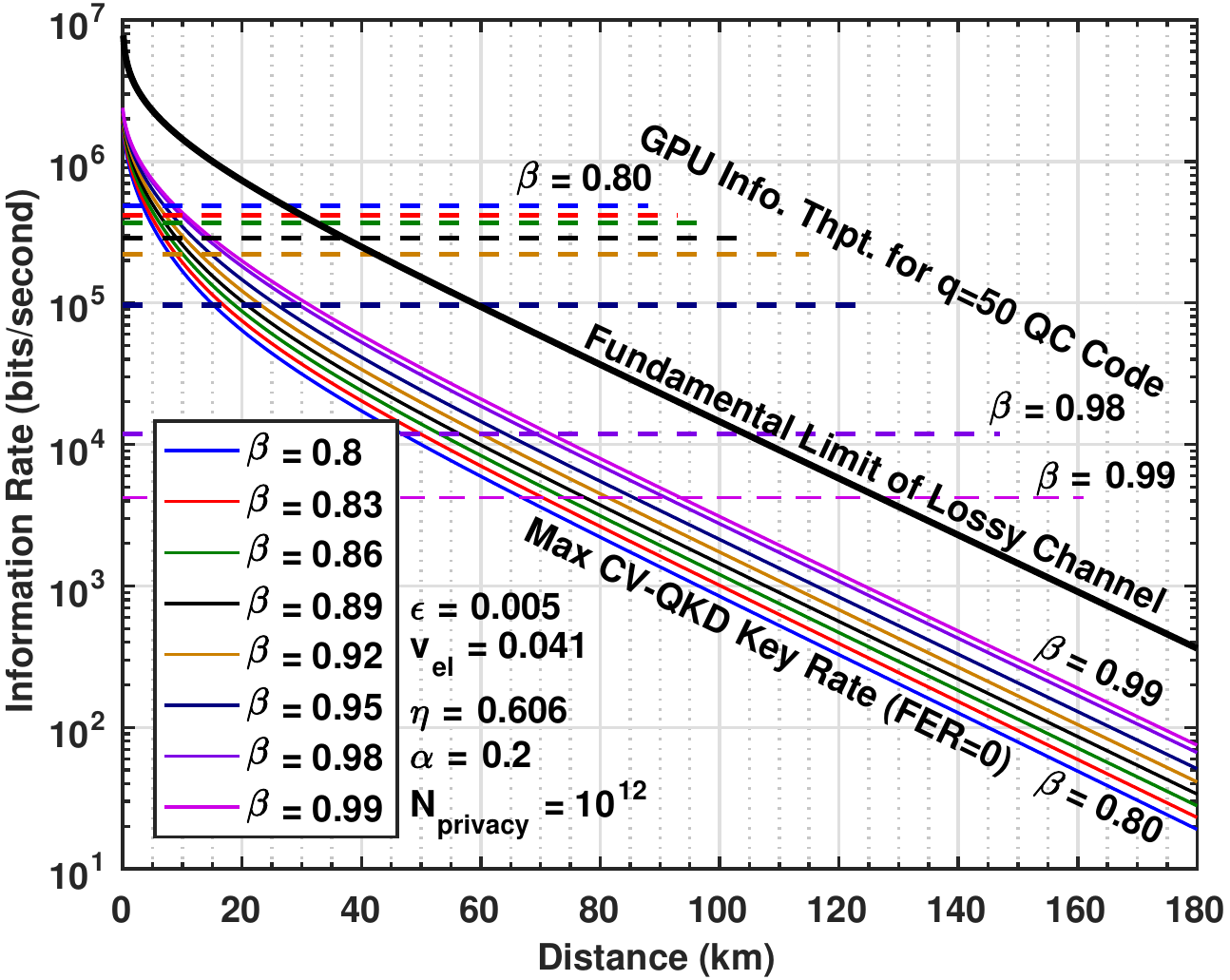}
\caption{GPU information throughput $K_\text{GPU}^\prime$ of the $q=50$ QC-LDPC code with $d=8$ dimensional reconciliation up to the maximum distance point for each $\beta$ efficiency with $N_\text{privacy}=10^{12}$ bits, maximum CV-QKD key rate with perfect reconciliation ($\text{FER}=0$) $K_\text{opt}^\prime$, and fundamental secret key rate limit for lossy channel $K_\text{lim}^\prime$ vs. distance. $K_\text{opt}^\prime$ and $K_\text{lim}^\prime$ are scaled by $f_\text{rep}=$~1MHz.}
\label{fig:newlimit_vs_8dGPU}
\end{figure}

Some GPU-based LDPC decoders use fixed-point number representations and/or frame-level parallelism to maximize computational speedup for codes with short block lengths ($n<10^5$ bits) in high-SNR regions above 0dB where the Min-Sum algorithm achieves sufficient error-correction performance~\cite{Wang2011, Kang2012, Wang2013, Gal2014, Lin2014}. This work, however, uses single-precision floating point to minimize FER with Sum-Product decoding at SNRs below -15dB. Due to the large block length ($n=10^6$ bits), all GPU threads are fully utilized, thus external (frame-level) parallelism does not provide additional speedup. Asynchronous data transfer to the GPU is another technique often employed to minimize overhead latency, however, this does not provide any significant performance boost as the Sum-Product computation dominates overall execution time due to the large number of iterations required for low-SNR decoding.

\subsection{Information Throughput Results}

Figure~\ref{fig:GPUThptvsBeta} presents the measured information throughput $K_\text{GPU}^\prime$ from the GPU decoder for all three LDPC codes at each $\beta$-efficiency point, which corresponds to a unique SNR-FER point in Figures~\ref{fig:FERvsSNR-1D}~and~\ref{fig:FERvsSNR-8D} for the $d=1$ and $d=8$ dimensional reconciliation cases, respectively. Table~\ref{table:maxDistanceRateComparison} compares the performance of the rate 0.02 random and QC codes at the maximum achievable distance for each reconciliation dimension. The $q=21$ and $q=50$ QC codes designed in this work achieve approximately $3\times$ higher raw decoding throughput $K_\text{GPU}^\text{raw}$ over the purely-random code with $d=1,2,4,8$ dimensional reconciliation at the maximum distance point for each $\beta$-efficiency. When scaled by the corresponding FER and code rate, the QC codes achieve between $5.1\times$ and $12.8\times$ higher information throughput $K_\text{GPU}^\prime$ over the purely-random code. Table~\ref{table:maxDistanceRateComparison} also presents the operating secret key rate $K_\text{finite}^\prime$ defined by Eq.~\ref{eq:Kfiniteprime}, and the fundamental secret key rate limit $K_\text{lim}$ for a lossy channel defined by Eq.~\ref{eq:Klim}. Here, the fundamental limit is scaled by the light source repetition rate $f_\text{rep}$, such that 
\begin{equation}
K_\text{lim}^\prime=f_\text{rep}K_\text{lim}.
\end{equation}
A realistic CV-QKD repetition rate of $f_\text{rep}=$~1MHz is assumed for the comparison~\cite{Jouguet2014directrecon, Jouguet2013, Huang2016}. For distances beyond 130km, the operating secret key rate $K_\text{finite}^\prime$ is between $2176\times$ and $57112\times$ lower than the fundamental limit $K_\text{lim}^\prime$, with $d=8$ and $d=1$ dimensional reconciliation, respectively.

The rightmost column in Table~\ref{table:maxDistanceRateComparison} ($K_\text{GPU}^\prime / K_\text{lim}^\prime$) presents the two key results of this work. First, it shows that the GPU decoder can achieve between $1.07\times$ and $8.03\times$ higher information throughput $K_\text{GPU}^\prime$ over the fundamental secret key rate limit $K_\text{lim}^\prime$ with a 1MHz source using QC-LDPC codes with $d=4$ and $d=8$ dimensional reconciliation. Since the decoder delivers an information throughput higher than the fundamental key rate limit, it can be concluded that LDPC decoding is no longer the post-processing bottleneck in CV-QKD, and thus, the secret key rate remains only limited by the physical parameters of the quantum channel. The second result is that $d=1$ and $d=2$ dimensional reconciliation schemes are not well-suited for long-distance CV-QKD since the $K_\text{GPU}^\prime$ speedup over $K_\text{lim}^\prime$ is less than $1\times$. In general, Table~\ref{table:maxDistanceRateComparison} shows that QC codes achieve lower decoding latency than the purely-random code at long distances, thereby making them more suitable for reverse reconciliation at high $\beta$ efficiencies. 

The results presented in Table~\ref{table:maxDistanceRateComparison} and Fig.~\ref{fig:newlimit_vs_8dGPU} assumed a light source repetition rate of $f_\text{rep}=$1MHz. While a higher source repetition rate such as $f_\text{rep}=$~100MHz or $f_\text{rep}=$~1GHz would raise the fundamental secret key rate limit $K_\text{lim}^\prime$ above the maximum GPU decoder throughput $K_\text{GPU}^\prime$, it would still not introduce a post-processing bottleneck for CV-QKD. The GPU decoder currently delivers an information throughput $K_\text{GPU}^\prime$ between 1868$\times$ and 18790$\times$ higher than the operating secret key rate $K_\text{finite}^\prime$ with a 1MHz light source at the maximum distance points for $d=1,2,4,8$ dimensional reconciliation schemes beyond 130km. Even with a source repetition rate of $f_\text{rep}=$~1GHz, the GPU information throughput $K_\text{GPU}^\prime$ would still exceed the operating secret key rate $K_\text{finite}^\prime$ for distances beyond 130km by 1.8$\times$ and 18.7$\times$, assuming the same quantum channel parameters. Therefore, GPUs remain a viable platform for the implementation of reconciliation algorithms for long-distance CV-QKD.

This is further illustrated in Fig.~\ref{fig:newlimit_vs_8dGPU} where the GPU information throughput $K_\text{GPU}^\prime$ of the $q=50$ QC-LDPC code with $d=8$ dimensional reconciliation is compared to the asymptotic secret key rate limit $K_\text{opt}$ for each $\beta$-efficiency with perfect reconciliation ($P_e=0$), as presented in Fig.~\ref{fig:MaxSKR}. Here, a 1MHz source is assumed, and the scaled $K_\text{opt}^\prime$ is given by
\begin{equation}
K_\text{opt}^\prime=f_\text{rep}K_\text{opt}.
\end{equation}
The result presented in Fig.~\ref{fig:newlimit_vs_8dGPU} also shows that $K_\text{GPU}^\prime$ is higher than the upper bound on secret key rate $K_\text{lim}^\prime$ on a lossy channel with a 1MHz source from $\beta=0.8$ to $\beta=0.99$.

\subsection{Comparison to Other CV-QKD Implementations}
\label{sec:Comparison}

While QKD has been well-studied over the past 30 years, the exploration of long-distance CV-QKD is still nascent, with very few published implementations in the low-SNR regime for optical transmission distances beyond 100km. Hardware-based implementations of DV-QKD and short-distance CV-QKD have previously been demonstrated using FPGAs and GPUs~\cite{Martinez-Mateo2013, Zbinden2013, Dixon2014, Walenta2014}, however, at the time of writing, there is only one reported CV-QKD implementation designed to operate in the low-SNR regime for long-distance reconciliation~\cite{Jouguet2014}.

Jouguet and Kunz-Jacques reported a GPU-based LDPC decoder implementation that achieves 7.1Mb/s throughput at $\text{SNR}=0.161$ ($\beta=0.93$) on the BIAWGNC~\cite{Jouguet2014}, for a random multi-edge LDPC code with a block length of $2^{20}$ bits based on the rate 1/10 multi-edge code designed by Richardson and Urbanke with an SNR threshold of 0.1556\cite{Richardson2002}. For throughput comparison purposes, two additional multi-edge codes with the same code rate, block length, and SNR threshold were designed in this work: a purely-random code and a $q=512$ QC code.

\begin{table}[t!]
\begin{center}
\captionsetup{justification=centering}
\caption{GPU LDPC decoding comparison at $\text{SNR}=0.161$ with $d=8$ on BIAWGNC targeting $\text{FER}=0.04$ with rate 1/10 codes}
\label{table:GPUcomparison}
\scalebox{0.78}{
    \begin{tabular}{ !{\vrule width 3\arrayrulewidth} C{1.18in} !{\vrule width 3\arrayrulewidth} C{0.28in} | C{0.28in} | C{0.385in} | C{0.385in} | C{0.85in} !{\vrule width 3\arrayrulewidth} }
    \Xhline{3\arrayrulewidth}   
  
\bf \multirow{2}{*}{Specification} & \multicolumn{4}{c|}{\bf This Work}	& \bf Jouguet \& Kunz-Jacques \\ 
& \multicolumn{4}{c|}{2016} & 2014~\cite{Jouguet2014} \\ \hline

\Xhline{3\arrayrulewidth}   

\bf Code Rate & \multicolumn{4}{c|}{$1/10$} &	$1/10$ \\ \hline
\bf Block Length (Bits)	&	\multicolumn{4}{c|}{$2^{20}$} &	$2^{20}$ \\ \hline
\bf SNR  &	\multicolumn{4}{c|}{0.161}	&	0.161 \\ \hline

\bf LDPC Code Structure &	\multicolumn{2}{C{0.75in}|}{Random Multi-Edge} &	\multicolumn{2}{C{0.92in}|}{$q=512$ QC Multi-Edge}	& Random Multi-Edge \\ \hline

\bf Connections in Parity Matrix & \multicolumn{2}{C{0.75in}|}{4,063,229} &	\multicolumn{2}{C{0.92in}|}{7,932}	& N/A \\ \hline

\bf Early Termination &	No	&	Yes	 &	No	&	Yes	&	No \\ \hline
\bf Max Iterations & 88 & 88 & 100	&	100	&	100 \\ \hline
\bf Average Iterations & 88 & 78 & 100	&	78	&	100 \\ \hline
\bf FER ${^{(1)}}$ & 	0.04 & 0.04 &	0.0243	&	0.0243	&	0.04 \\ \hline
\bf Latency Per Iteration (ms) ${^{(2)}}$ & 4.73 & 4.84 &	1.28	&	1.47	&	1.48 \\ \hline		
\bf $K_\text{GPU}^\text{raw}$ GPU Raw Throughput (Mb/s) & 2.52	& 2.78 &	8.21	&	9.17	&	7.1 \\ \hline	
\bf $K_\text{GPU}^\prime$ GPU Info. Throughput (Kb/s) & 242 & 267	&	801 	&	895	&	682 \\ \hline

	 \Xhline{3\arrayrulewidth} 
	 
\bf GPU Model	&	\multicolumn{4}{c|}{NVIDIA GeForce GTX 1080}   &	AMD Radeon HD 7970 \\ \hline
\bf CMOS Technology & \multicolumn{4}{c|}{16nm} & 28nm \\ \hline
\bf GPU Cores & \multicolumn{4}{c|}{2560} & 2048 \\ \hline
\bf GPU GFLOPS & \multicolumn{4}{c|}{8228} & 3789 \\ \hline
\bf GPU Memory Bus Width (Bits) & \multicolumn{4}{c|}{256} & 384\\ \hline
\bf GPU Memory Bandwidth (GB/s) & \multicolumn{4}{c|}{320} & 264\\ \hline
	
	\Xhline{3\arrayrulewidth}   
    \end{tabular} 
}
\end{center}
\scalebox{0.8}{
\begin{tabular}{p{4.25in}}
${^{(1)}}$ FER $P_e$ corresponds to the probability of detected error, since $P_\text{undetected}=0$ with 32-bit CRC. All three codes achieve a CV-QKD distance of 83.8km based on the quantum channel parameters assumed in this work.\\
${^{(2)}}$ Latency per iteration is an average for the full decoding of a single frame, and also includes the data transfer latency between the CPU and GPU.\\
\end{tabular}
}
\end{table}

Table~\ref{table:GPUcomparison} presents a performance comparison between the two designed rate 1/10 codes and the result achieved by Jouguet and Kunz-Jacques at $\text{SNR}=0.161$ on the BIAWGNC~\cite{Jouguet2014}. The two designed codes achieve a FER of approximately 0.04 under the same decoding conditions as the comparison work with $d=8$ dimensional reconciliation. Similar to the results presented in Tables \ref{table:GPU-LDPC-latency-performance-results} and \ref{table:maxDistanceRateComparison}, the $q=512$ QC code achieves approximately $3\times$ lower latency per iteration than the purely-random rate 1/10 code designed in this work. Rate 1/10 QC codes with expansion factors $q \in \{64,128,256\}$ were also designed, however, the $q=512$ QC code achieved the lowest latency per iteration due to the lower number of required memory accesses in the GPU message-passing kernels, as a result of the lower number of connections in the QC parity-check matrix. While the designed rate 1/10 random code achieves a maximum raw throughput of only 2.78Mb/s, the $q=512$ QC code delivers a maximum raw throughput of 9.17Mb/s with early termination enabled only in iterations greater than the average number of iterations, as determined empirically through FER simulation. The $q=512$ QC code achieves a 1.29$\times$ higher throughput than the 7.1Mb/s reported by Jouguet and Kunz-Jacques~\cite{Jouguet2014}, further demonstrating that the QC code structure offers computational speedup benefits for multi-edge codes operating in the high $\beta$-efficiency region at low SNR. Although the comparison work is from 2014, both GPU models have a similar memory bus width, which is the primary constraint that limits the latency per iteration. As previously discussed, GPU decoder performance is bound by the memory access rate, and not the floating point operations per second (FLOPS). Thus, a wider GPU memory allows for a higher memory access rate, which in turn, reduces the decoding latency.

Other types of error-correcting codes have been studied for application in the low-SNR regime of CV-QKD, such as polar codes, repeat-accumulate (RA) codes, and raptor codes. Polar codes require block lengths on the order of $2^{27}$ bits to achieve comparable FER performance to the rate 1/10 multi-edge LDPC codes designed in this work, however, they have been shown to achieve low decoding latency on generic x86 CPUs due to their recursive decoding algorithm~\cite{Jouguet2014}. A polar-code performance comparison is not available for the rate 0.02 multi-edge QC-LDPC codes designed in this work. Punctured and extended low-rate RA codes have been constructed from ETSI DVB-S2 codes with block lengths of 64,800 bits to achieve $\beta > 0.85$ efficiency over a wide range of SNRs~\cite{Johnson2016}, however, their performance has not been investigated beyond 70km and there is currently no hardware implementation to provide a sufficient throughput comparison. Lastly, raptor codes achieve high $\beta$-efficiency at low SNR and guarantee error-free decoding ($P_\text{e}=0$) by sending as many coded symbols as required by the receiver~\cite{Shirvanimoghaddam2016}. However, their decoding latency may be a limitation to high-throughput reconciliation, and at the time of writing, there is no known hardware implementation of raptor codes for long-distance CV-QKD. The demand for long-distance communication through applications such as CV-QKD motivates the need for continued research in high-efficiency codes and their hardware realizations.

% -----------------------------------------------------------------------------------
%
% CONCLUSION
%
% -----------------------------------------------------------------------------------

\section{Conclusion}
\label{sec:Conclusion}

This work introduced multi-edge quasi-cyclic LDPC codes to accelerate the reconciliation step in long-distance CV-QKD by means of a GPU-based decoder implementation and multi-dimensional reconciliation schemes. With an 8-dimensional reconciliation scheme, the GPU-based decoder delivers an information throughput up to $8.03\times$ higher than the upper bound on secret key rate for a lossy channel with a 1MHz source, thereby demonstrating that key reconciliation is no longer a computational bottleneck in long-distance CV-QKD. Furthermore, the low-rate LDPC codes extend the maximum distance of CV-QKD from the previously achieved 100km to 160km based on the quantum channel parameters assumed in this work. 

The LDPC codes and reconciliation techniques applied in this work can be extended to post-processing algorithms in two areas that show promise for the future of QKD: (1) free-space QKD using low-Earth orbit satellites as communication relays to extend the distance of secure communication beyond 200km without fiber-optic infrastructure, and (2) fully-integrated chip implementations~\cite{Diamanti2016}. Recent works have experimentally demonstrated terrestrial free-space QKD for distances up to 143km~\cite{Yin2012, Handsteiner2015}, while satellite-based QKD has been proposed as a practical near-term solution to achieving long-distance QKD on a global scale~\cite{Bourgoin2015, Vallone2015}. In August 2016, China launched the Quantum Experiments at Space Scale (QUESS) satellite to generate secret keys between ground stations in Beijing and Vienna by transmitting entangled photon pairs from an orbit altitude of 500km~\cite{Gibney2016}. Free-space fading channels for satellite QKD typically operate at SNRs above 0dB~\cite{Rarity2002}, however, quasi-cyclic code construction techniques can still be employed to achieve high secret key rates, while GPUs would allow for simple integration with other satellite equipment for rapid prototyping, in contrast to ASIC- or FPGA-based LDPC decoder implementations. This paper presented the computational speedup achievable on a single state-of-the-art GPU. Further acceleration can be achieved through architectural optimizations in the design of a monolithic QKD chip that combines both optical and post-processing circuits. Photonic chips have already been realized for QKD transmitters and receivers~\cite{Ma2016, Bunandar2016, Diamanti2016,Sibson2017}, and further integration of post-processing algorithms would provide a considerable reduction in system size and power consumption. A final key takeaway here is that the quasi-cyclic LDPC code construction and GPU architecture techniques presented in this work can also be applied to forward error-correction implementations for DV-QKD where reconciliation is performed over the binary symmetric channel (BSC) instead of the BIAWGNC as in CV-QKD. 

This work addressed the challenge of achieving high-speed, high-efficiency reconciliation for long-distance CV-QKD over fiber-optic cable. In addition to extending information-theoretic security to general attacks for finite key sizes, a major remaining hurdle to extending the secure transmission distance in CV-QKD is the reduction of excess noise in the optical quantum channel. While recent techniques have been demonstrated to control excess noise to within a tolerable limit~\cite{HuangHuangLinZeng2016}, future work may also investigate the security of CV-QKD in the presence of non-Gaussian noise sources, and in particular, the performance of LDPC decoding at low SNR with non-Gaussian noise. GPU-based decoder implementations with quasi-cyclic codes would provide a suitable platform for such investigations. Furthermore, reducing the latency of privacy amplification for large block sizes on the order of $N_\text{privacy} \geq 10^{12}$ bits is necessary in order to realize secret key exchange for distances beyond 100km.

\section*{Acknowledgment}

The authors would like to thank the Natural Sciences and Engineering Research Council of Canada (NSERC) for supporting this research, Dr. Christian Weedbrook and Dr. Xingxing Xing for their technical guidance related to CV-QKD, Professor Hoi-Kwong Lo at the University of Toronto for his insights on MDI-QKD and state-of-the-art implementations, Professor Stefano Pirandola at the University of York for introducing the upper bound on secret key rate for lossy channels, and Dr. Christoph Pacher at the Austrian Institute of Technology for his clarifications on finite-size effects.

\ifCLASSOPTIONcaptionsoff
  \newpage
\fi

\newpage

\begin{IEEEbiography}[{\includegraphics[width=1in,height=1.25in,clip,keepaspectratio]{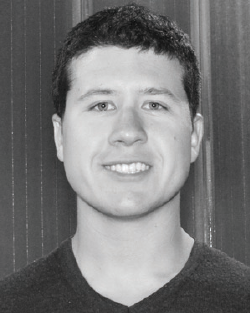}}]{Mario Milicevic}
received the B.A.Sc. degree in electrical engineering in 2010 from the University of Toronto, Canada, where he is currently pursuing the Ph.D. degree in electrical engineering under the supervision of Professor P.~Glenn~Gulak. His research interests include integrated circuit design for high-performance and low-power digital signal processing systems, cryptography, and security. He is an IEEE volunteer, and served as the student volunteer team lead at the IEEE International Solid-State Circuits Conference (ISSCC) from 2012 to 2016.
\end{IEEEbiography}

\begin{IEEEbiography}[{\includegraphics[width=1in,height=1.25in,clip,keepaspectratio]{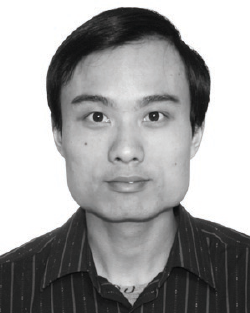}}]{Chen Feng} (S'08--M'14)
received the B.Eng.\ degree from the Department of Electronic and Communications Engineering, Shanghai Jiao Tong University, China, in 2006, and the M.A.Sc.\ and Ph.D.\ degrees from the Department of Electrical and Computer Engineering, University of Toronto, Canada, in 2009 and 2014, respectively. From 2014 to 2015, he was a Postdoctoral Fellow with Boston University, USA, and the  \'Ecole Polytechnique F\'ed\'erale de Lausanne (EPFL), Switzerland. He joined the School of Engineering, University of British Columbia, Kelowna, Canada, in July 2015, where he is currently an Assistant Professor. His research interests include data networks, coding theory, information theory, and network coding. Dr. Feng was a recipient of the prestigious NSERC Postdoctoral Fellowship in 2014. He was recognized by the \textsc{IEEE Transactions on Communications} (TCOM) as an Exemplary Reviewer in 2015. Since 2015, he served as an Associate Editor for the \textsc{IEEE Communications Letters}. He is a member of ACM and IEEE.
\end{IEEEbiography}

\begin{IEEEbiography}[{\includegraphics[width=1in,height=1.25in,clip,keepaspectratio]{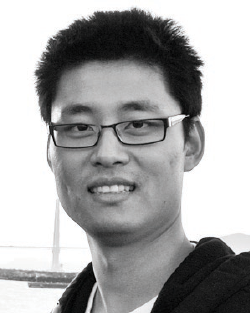}}]{Lei M. Zhang}
received his B.A.Sc. degree in Electrical Engineering from the University of Waterloo in 2009, his M.A.Sc. degree in Electrical Engineering from the University of Toronto in 2011, and his Ph.D. degree in Electrical Engineering from the University of Toronto in 2016 under the supervision of Professor Frank Kschischang. 
\end{IEEEbiography}

\begin{IEEEbiography}[{\includegraphics[width=1in,height=1.25in,clip,keepaspectratio]{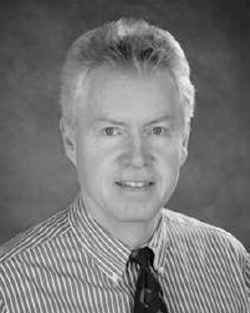}}]{P. Glenn Gulak}
is a Professor in the Department of Electrical and Computer Engineering at the University of Toronto. He is a Senior Member of the IEEE and a registered Professional Engineer in the Province of Ontario. His present research interests are in the areas of algorithms, circuits, and CMOS implementations of high-performance baseband digital communication systems and, additionally, in the area of CMOS biosensors. He has authored or co-authored more than 150 publications in refereed journals and refereed conference proceedings. In addition, he has received numerous teaching awards for undergraduate courses taught in both the Department of Computer Science and the Department of Electrical and Computer Engineering at the University of Toronto. 
%From Jan. 1985 to Jan. 1988 he was a Research Associate in the Information Systems Laboratory and the Computer Systems Laboratory at Stanford University. He has served on the ISSCC Signal Processing Technical Subcommittee from 1990 to 1999, ISSCC Technical Vice-Chair in 2000 and served as the Technical Program Chair for ISSCC 2001. From 2001 to 2003, he was CTO and VP Engineering of Valence Semiconductor located in Irvine, California.  He served on the Technology Directions Subcommittee for ISSCC from 2005 to 2008. 
He currently serves as the Vice-President of the Publications Committee for the IEEE Solid-State Circuits Society and an elected member of the IEEE Publication Services and Products Board (PSPB).
\end{IEEEbiography}

\end{document}